\documentclass[aps,prd,twocolumn,10pt,superscriptaddress,amssymb,amsmath,nofootinbib]{revtex4-2}

\usepackage[utf8]{inputenc}
\usepackage{lmodern}
\usepackage[T1]{fontenc}
\usepackage[colorlinks=true,allcolors=blue]{hyperref}
\usepackage{enumitem}
\usepackage{graphicx,color}
\usepackage{dcolumn}
\definecolor{alizarin}{rgb}{0.82, 0.1, 0.26}

 \usepackage{subcaption}
\usepackage{caption}
\usepackage{bm}
\usepackage{amsmath,amsfonts,amssymb}
\usepackage{acronym}
\usepackage{enumitem}
\usepackage{array}
\usepackage{multirow}
\usepackage{CJK}
\usepackage{array}
\usepackage{orcidlink}
\usepackage{xcolor}
\usepackage{booktabs}
\newlist{alphalist}{enumerate}{1}
\setlist[alphalist,1]{label=\textbf{\alph*.}}
\allowdisplaybreaks
\newcommand{\be}{\begin{equation}}
\newcommand{\ee}{\end{equation}}
\newcommand{\bea}{\begin{eqnarray}}
\newcommand{\eea}{\end{eqnarray}}

\newcommand{\IIT}{Indian Institute of Technology, Gandhinagar, Gujarat-382355, India}
\newcommand{\IITKGP}{Department of Physics, Indian Institute of Technology, Kharagpur, 721 302, India}
\newcommand{\Tsinghua}{Department of Astronomy, Tsinghua University, Haidian District, Beijing 100084, China}

\newcommand{\nocontentsline}[3]{}
\let\origcontentsline\addcontentsline
\newcommand\stoptoc{\let\addcontentsline\nocontentsline}
\newcommand\resumetoc{\let\addcontentsline\origcontentsline}

\begin{document}
%\begin{CJK*}{UTF8}{gbsn}

\title{{\Large {\bf 
Gravitational memory meets astrophysical environments: exploring a new frontier through osculations
}}}

\author{Rishabh Kumar Singh\orcidlink{0009-0007-6426-8136}}
\email{singhrishabh@iitgn.ac.in}
\affiliation{\IIT}

\author{Shailesh Kumar\orcidlink{0000-0001-7072-9452}}
\email{shaileshkumar.1770@gmail.com}
\affiliation{\IIT}
\affiliation{\IITKGP}

\author{Abhishek Chowdhuri\orcidlink{0000-0003-4474-790X}}
\email{chowdhuria@tsinghua.edu.cn}
\affiliation{\Tsinghua}

\author{Arpan Bhattacharyya\orcidlink{0000-0002-7933-6441}}
\email{abhattacharyya@iitgn.ac.in}
\affiliation{\IIT}
% \vspace{0.4cm}
\begin{abstract}
We study how dark matter environments influence nonlinear gravitational memory from intermediate-mass-ratio binaries. Incorporating environmental effects from the dark matter gravitational potential, dynamical friction, and accretion, we compute the leading-order nonlinear memory for both bound and unbound orbits under dark matter minispike profile. For quasi-circular inspirals in a minispike, we additionally include an empirical prescription for the time-dependent evolution of the dark matter profile, which gradually evolves along the inspiral and captures the cumulative environmental response. We find that dark matter can modify the orbital evolution and mode content of the memory relative to the vacuum case, with the cumulative effect depending sensitively on the density profile and on how the environment accelerates the inspiral. We use these waveforms to calculate mismatches in LISA space-based detector, highlighting where memory-driven differences may be large enough to warrant targeted parameter-estimation studies. Our results emphasize that astrophysical environments can leave a hereditary imprint on gravitational memory and provide a framework for connecting memory observables with dark matter dynamics.
\end{abstract}

\maketitle
%\tableofcontents

\section{Introduction}
Sources such as inspiralling binaries comprising black holes or compact objects are among the most prominent candidates for gravitational waves (GWs), providing ways to test the strong gravity regimes of black holes and astrophysical processes in extreme environments. GWs, directly detected by Earth-based observatories \cite{LIGOScientific:2016aoc, LIGOScientific:2016vlm, LIGOScientific:2016sjg, LIGOScientific:2017bnn, LIGOScientific:2018dkp}, have opened several new avenues in GW astronomy and further implicate the necessity of developing models for progressively complex sources with accurate precision. These models will not only provide a robust framework for testing general relativity (GR) but also will offer estimates on the detectability of GW signals by advanced Laser Interferometer Gravitational Wave Observatory (LIGO) and Laser Interferometer Space Antenna (LISA) detectors, particularly for binaries with small mass ratios. In addition to the GW signals emitted during distinct phases of binary evolution, these systems exhibit intriguing phenomena, such as gravitational/GW memory, and offer insights into environmental effects. Such aspects remain undetected in GW observations to date; however, their mutual interactions can significantly affect waveforms, enhancing their detectability through emerging observational approaches. 

In this line of endeavour, the present article examines the observational consequences of an interplay between gravitational memory and the astrophysical environment. The term \textit{gravitational memory} is a phenomenon in GR that refers to the permanent relative change in two nearby freely falling test masses after the passage of GW \cite{Zeldovich:1974gvh, Braginsky:1985vlg, 1987Natur.327..123B, Christodoulou:1991cr, PhysRevD.45.520, Favata:2008yd, Favata:2009ii, Favata:2010zu, Favata:2011qif, Tolish:2014bka, PhysRevD.89.084039, Hait:2022ukn, PhysRevD.106.064022}.  In other words, it describes a signal that settles at a nonzero finite value rather than decaying to zero. As the wave passes through the detector setup, it distorts the spacetime geometry, finally resulting in a permanent shift in the relative positions of the test masses/detectors. This intriguing area of study has recently attracted significant interest in GW physics, especially due to the potential for detection with future advanced LIGO or LISA detectors \cite{Lasky:2016knh, Talbot:2018sgr, Islo:2019qht, Boersma:2020gxx, Grant:2022bla, LISAConsortiumWaveformWorkingGroup:2023arg, Goncharov:2023woe, Sun:2022pvh}. There is various literature available discussing numerous aspects of GW memory ranging from the conventional displacement memory effect \cite{Zeldovich:1974gvh, Braginsky:1985vlg, 1987Natur.327..123B, strominger2018lecturesinfraredstructuregravity} to the kick memory by planar waves \cite{Zhang:2017rno, Zhang:2017geq, Zhang:2017jma, Chakraborty:2019yxn, BenAchour:2024ucn, Gaur:2024oms, Galoppo:2024vww}, spin memory \cite{Pasterski:2015tva, Mao:2018xcw}, memory effects in modified theories of gravity \cite{Tahura:2020vsa, Hou:2020tnd, Seraj:2021qja, Tahura:2021hbk, Heisenberg:2023prj, PhysRevD.105.024072}, and results from post-Newtonian (PN), self-force approaches \cite{Blanchet2014, Cunningham:2024dog} and worldline effective field theory \cite{Porto:2024cwd}. Further, recent developments have revealed a close link between the infrared structure of gravity and asymptotic symmetries that has led to the identification of the infrared triangle, which connects three fundamental ideas that were previously studied independently but now recognized to be related: \textit{gravitational memory}, \textit{asymptotic symmetries}, and \textit{soft theorems} \cite{Strominger:2014pwa, Hawking:2016sgy, Hawking:2016msc, strominger2018lecturesinfraredstructuregravity, Blanchet:2023pce, Solanki:2023wmv, Solanki:2024oci, DeLuca:2024cjl}. Since our investigation focuses solely on GW memory, readers are encouraged to look at a detailed review in \cite{strominger2018lecturesinfraredstructuregravity, Kumar:2021qrg} along with the references cited therein. The observational implications of the link between GW memory and asymptotic symmetries have also been examined from the perspective of numerical relativity \cite{Mitman:2020pbt, Mitman:2021xkq, Mitman:2022kwt, Mitman:2024uss}. GW memory is typically studied at future null infinity \cite{strominger2018lecturesinfraredstructuregravity}; however, lately, an analogous memory effect has been established near the horizon of black holes, revealing a connection with asymptotic symmetries akin to the conventional memory observed at future null infinity \cite{Donnay:2015abr, Donnay:2016ejv, Donnay:2018ckb, Blau:2015nee, Rahman:2019bmk, Bhattacharjee:2019jaf, Bhattacharjee:2020lgt, Bhattacharjee:2020vfb, Sarkar:2021djs, Bhat:2024cyq, Gasparotto:2023fcg}. {\color{black}It has also been shown that the nonlinear GW memory grows significantly during binary black hole mergers, with its magnitude and detectability influenced by the spin of black holes \cite{Pollney:2010hs}.} As a result, the field encompasses/offers a wide range of topics/opportunities to explore with promising possibilities of detection by future detectors.

On the other hand, the universe is permeated by a non-trivial matter medium, predominantly dark matter (DM), an enigmatic and pervasive component whose nature remains largely unknown and under active investigation. Indirect pieces of evidence support the presence of DM, albeit its exact properties continue to be unresolved and elude our understanding \cite{1990ApJ...356..359H, Bertone:2004pz, Navarro:1995iw, Clowe:2006eq, Corbelli:1999af, Navarro:1996gj, Gondolo:1999ef, Bertone:2005xv}. Intriguingly, in such a mystifying environment, dense matter clusters are expected to form halos, which may subsequently collapse to form black holes under the influence of gravity \cite{Navarro:1996gj, Gondolo:1999ef, Bertone:2005xv, Merritt:2002vj, Ullio:2001fb}. The existence of DM halos around galaxies and black holes at their centers suggests that DM can potentially impact black hole dynamics and GWs, inferring valuable insights into galactic properties. Recently, numerous studies have been conducted to analyze inspiralling and scattering binaries with distinct mass ratios in astrophysical environments and in modified gravity setups, which put forward the necessity of space-based detectors for possible detection of such an interaction from GW observations \cite{Cardoso:2021wlq, Destounis:2022obl, Cardoso:2022whc, Duque:2023seg, Shen:2023erj, Speeney:2024mas, Rahman:2023sof, PhysRevD.111.083041, Zhang:2024ugv, Macedo:2024qky, Ravanal:2024odh, Aurrekoetxea:2024cqd, Cheng:2024mgl,PhysRevD.109.124056,Bhattacharyya:2024kxj,Bhattacharyya:2024aeq,Usseglio:2025iwt, Cho:2021onr, Miller:2025yyx, PhysRevD.89.104059}.

Further, DM environments near SMBHs can become highly concentrated as the masses of SMBHs gradually increase from $10^{6}$\textendash $10^{9} M_{\odot}$, potentially forming compact structures, called \textit{DM spikes} \cite{Gondolo:1999ef}. While galaxy mergers and other events can disrupt DM spikes surrounding such supermassive binaries \cite{PhysRevLett.88.191301, PhysRevD.64.043504, PhysRevD.78.083506, PhysRevLett.113.151302, PhysRevLett.115.231302}, certain structures are formed around intermediate-mass black holes (IMBHs) with masses of $10^{2}$ \textendash $10^{5} M_{\odot}$, termed \textit{minispikes} \cite{PhysRevLett.95.011301, PhysRevD.72.103517}. GWs generated from binary black hole systems can be used to test the existence of DM minispikes as a result of their effects on the orbital dynamics of the inspiralling object \cite{Eda:2013gg, Eda:2014kra, PhysRevD.97.064003, PhysRevD.102.103022}. {\color{black}A stellar mass object} orbiting an IMBH, known as intermediate mass-ratio inspiral (IMRI), with a DM minispike is influenced by the gravity of the central IMBH, backreaction on the object due to GW emission, DM environment, and dynamical friction (DF) \cite{1943ApJ....97..255C, Ostriker_1999, Kim_2007, Eda:2014kra, PhysRevLett.110.221101, Dai:2021olt}. These systems also accrete, and the DF from such environments can considerably influence the evolution of IMBHs \cite{10.1093/mnras/104.5.273, 1983bhwd.book.....S, Macedo:2013qea}. Therefore, a comprehensive analysis of an inspiralling binary black hole system, such as an IMRI, requires incorporating the effects of accretion alongside gravitational forces, GW backreaction, and DF \cite{Eda:2014kra, PhysRevLett.110.221101, Yue:2018vtk, Nichols:2023ufs, Shadykul:2024ehz}. 
Studies indicate that such effects have nontrivial impacts on the phase and can yield detectable signals from missions such as LISA, Taiji, and TianQin \cite{Eda:2014kra, PhysRevD.97.064003}. Thus, detecting GWs from these systems could shed light on the properties of DM.

In this article, we consider elliptical, hyperbolic, and quasi-circular orbits of a stellar-mass object around the central massive black hole in an intermediate-mass-ratio (IMR) binary system immersed in a DM halo. Note that we analyze the quasi-circular case with the evolving DM minispike profile, following \cite{PhysRevD.105.043009}. Within these considerations, our main objective is to examine the interplay between gravitational memory and the DM environment, as defined by the DM minispike profile. {\color{black}Moreover, recently in \cite{Alnasheet:2025tpd}, it has been shown that astrophysical environments can induce transient features in GW tails and (linear) memory; however, the authors focus on linear scalar tails and memory sourced by a scalar charge, with the expectation that analogous features arise in the gravitational case. It is worth noting that our analysis adopts a distinct approach and examines the interplay between DM and nonlinear memory within the IMR framework, including the potential detectability of these two aspects by future detectors.}  We investigate such an interaction with the combined effects on the inspiralling object: radiation reaction, DF, and accretion phenomena. In other words, we focus on analyzing the effects of these on GW emissions resulting from orbital evolution and on the impacts of such environments on nonlinear GW memory. 

{\color{black} For further motivation, we wish to emphasize that the environmental effects considered in this work—DM gravity, DF, and accretion—also modify the oscillatory (non-memory) part of the GW signal, primarily through changes in the orbital phase evolution, and these effects are typically larger in magnitude than their imprint on the memory itself. In other words, these effects typically enter the phase at a lower PN order and are expected to be larger in magnitude than their imprint on the memory signal. However, the purpose of the present study is not to suggest that gravitational memory provides the dominant probe of DM environments, but rather that it offers a qualitatively distinct and complementary observable. However, we know that gravitational memory is a hereditary observable that depends on the cumulative energy flux emitted over the entire past history of the binary, rather than on the instantaneous orbital dynamics. As a result, environmental effects that act over long timescales can accumulate in the GW memory in ways that differ from their effects on the oscillatory phase and amplitude. This feature makes memory particularly sensitive to long-lived astrophysical environments, such as DM minispikes surrounding IMBHs. From this perspective, studying environmental effects on memory provides a complementary probe of such effects around IMBHs. Alternatively, the presence of environments can enhance the GW memory signal relative to the vacuum case, potentially increasing its detection prospects through IMRIs. Note that a detailed analysis of parameter estimation and degeneracies associated with environmental effects in both the oscillatory waveform and memory is deferred to future work. The present analysis provides an estimate on the interplay between nonlinear memory and environmental effects.}

Let us now outline the structure of the paper. In section (\ref{sec1}), we introduce the conceptual grounds of DM profiles and gravitational memory. Section (\ref{sec2}) provides equations of motion for computing the nonlinear memory for elliptical, hyperbolic, and quasi-circular orbits, including the combined effects of DM gravity, DF, and accretion. Section (\ref{sec3}) explicitly presents numerical results and the possible detectability of the interplay between memory and the DM environment with next-generation detectors. Finally, in section (\ref{dscn}), we summarize the key findings of the study and outline the future prospects of our work. We set the fundamental constants, gravitational constant ($G$) and speed of light ($c$) to unity.
 \\
\\
% {\color{black}{\it Notation}: We set the fundamental constants, gravitational constant ($G$) and speed of light ($c$) to unity.}

\section{Setup: Astrophysical environment and GW memory}\label{sec1}

This section provides a brief overview of relevant DM profiles under consideration and the basics of GW memory. We then motivate the study using DM density models and describe the computational framework used to examine the interplay between GW memory and these environments.

\subsection{Astrophysical environment}\label{environment}

Astrophysical studies suggest that DM constitutes a major component of the universe and may form overdense regions with steep power-law profiles\textemdash DM spikes\textemdash around black holes \cite{Gondolo:1999ef}. The existence and configuration of these profiles strongly depend on the black hole's formation history, particularly post-merger events \cite{Eda:2014kra}. Although such structures around supermassive black holes get depleted with time due to mergers of galaxies and other dynamical processes, they may persist around IMBHs that have not undergone any merger events in their evolutionary history \cite{Eda:2014kra}. Therefore, we consider central IMBHs, with masses in the range $(10^2 - 10^4) M_{\odot}$, embedded in DM environments. To comprehensively assess the impact of environmental conditions on GW memory, we model the surrounding DM distributions using DM minispike profile and analyze their effects on the system's dynamics.

%\subsubsection*{DM minispike profile}
%%%%%%%%%%%%%%%%%%%%%%%%%%%%%%%%%%%%%
\begingroup
\let\oldaddcontentsline\addcontentsline
\renewcommand{\addcontentsline}[3]{} % disables TOC entry
\subsubsection*{DM minispike profile}
\endgroup
%%%%%%%%%%%%%%%%%%%%%%%%%%%%%%%%%%%%%%

The DM minispike profile arises from the adiabatic growth of an IMBH within the DM halo. It is given by \cite{Eda:2014kra}: 
\begin{equation}
\rho_{\textup{DM}} (r) = 
\begin{cases} 
\rho_{\textup{sp}} \big( \frac{r_{\textup{sp}}}{r} \big)^{\alpha}, &  r_{\textup{min}} \leq r \leq r_{\textup{sp}} \\
0, &  r < r_{\textup{min}}\,, \\
\end{cases}
\label{eq:minispike_density}
\end{equation}  
where $\rho_{\textup{sp}}$ is the normalization constant and $r_{\textup{sp}}$ is empirically defined as $r_{\textup{sp}} \sim 0.2 r_h$ ($r_h$ tells the gravitational influence of central IMBH), $\alpha$ is the DM minispike exponent, $r$ is the distance from the center of IMBH and $r_{\textup{min}} = \frac{6 m_1}{c^2}$ is the innermost stable circular orbit (ISCO) distance for IMBH ($m_1$). The minispike exponent $\alpha = (9-2\alpha_{\textup{in}})/(4 - \alpha_{\textup{in}})$ is related to the initial inner-cusp density profile parameter $\alpha$. For a minispike, $0 \leq \alpha_{\textup{in}} \leq 2$ implies $2.25 \leq \alpha \leq 2.5$ \cite{Gondolo:1999ef}. The parameter values adopted for the DM minispike profile in this work are summarized in Table~\ref{tab:1}.

% \subsubsection*{DM NFW profile}
%%%%%%%%%%%%%%%%%%%%%%%%%%%%%%%%%%%%%
\begingroup
\let\oldaddcontentsline\addcontentsline
\renewcommand{\addcontentsline}[3]{} % disables TOC entry

Given that DM pervades the universe, assuming binary dynamics in a vacuum is an idealization that overlooks critical environmental effects. The presence of DM influences the evolution of compact binaries and, consequently, GW memory through mechanisms such as accretion and gravitational drag, leading to deviations from the dynamics predicted under vacuum conditions. Several studies have explored these effects, typically assuming a static DM density profile \cite{Macedo:2013qea, PhysRevD.103.023015, Dai:2021olt}. Recently, a few attempts have been made to incorporate the dynamic density behaviour of DM using N-body simulation \cite{karydas2024sharpeningdarkmattersignature}. In this paper, we have considered the evolution of compact binaries for a static DM profile in the case of eccentric and hyperbolic orbits; however, for quasi-circular orbits, we have empirically incorporated the dynamical nature of DM density \cite{PhysRevD.105.043009}. We will see that the environment significantly affects the nonlinear memory. 

\begin{widetext}
\begin{table*}
%\centering
%\captionsetup{width=0.65\textwidth}
\caption{Value of the parameters of DM Minispike profile for central IMBH $m_1= 10^3 M_{\odot}$ and total DM mass $M_{\textup{DM}} = 10^6 M_{\odot}$ \cite{Eda:2014kra}.}\vspace{0.2cm}
\renewcommand{\arraystretch}{1.5} % Increase row height
\large % Increase font size
\begin{tabular}{|c|c|c|c|c|c|}
\hline
$\alpha$ & $r_{\textup{sp}}$ & $\rho_{\textup{sp}}$ & $r_{\textup{min}}$& $M_{\odot}$ & \text{pc} \\
\hline\hline
$[2.25, 2.5]$ & $0.54\,\text{pc}$ & $226\,M_{\odot}/\text{pc}^3$& $2.87 \times 10^{-10}\,\text{pc}$& $1.99 \times 10^{30}\, \text{Kg}$ & $3.08\times 10^{16}\, \text{m}$\\
\hline
\end{tabular}
\label{tab:1}
\end{table*}
\end{widetext}

\subsection{GW memory}\label{memory}

To reiterate, GW memory manifests itself as a lasting change in the relative separation of test masses in a detector, persisting after the GW has passed through the setup. Quantitatively, it is defined as the difference between the late and early time values of the GW polarizations: 
\begin{equation}
\Delta h_{+,\times}^{\text{(mem)}} = \lim_{t \to + \infty} h_{+,\times} (t) -  \lim_{t \to - \infty} h_{+,\times} (t)\,,
\label{eq:gw_memory}
\end{equation}
where, $h_{+,\times}$ are two GW polarizations and $t$ is the time in the detector frame. There are two types of GW memory: (1) \textit{Linear memory} that arises from the non-oscillatory motion of the source. It appears in unbound systems such as hyperbolic binary encounters or supernova explosions \cite{Favata_2009}. (2) \textit{Nonlinear memory}, which originates from previously emitted GWs, which are always unbound and therefore contribute to the memory effect. This type of memory is present in all GW sources and is also known as \textit{Christodoulou memory} \cite{Favata_2009, PhysRevD.100.024009, PhysRevD.46.4304, Blanchet:1997jj, Blanchet_2017, Ivanov:2025ozg, 3jhf-vdjz, Bini:2025vuk}. For eccentric orbits, nonlinear memory affects the GW signal at leading-order (0PN), while for hyperbolic and parabolic orbits, it contributes at (2.5PN) order \cite{Favata:2011qif}. In general, we know that the GW polarizations can be expressed as the sum of multipole modes in the following way,
\begin{equation}
    h_{+}-i h_{\times}=\sum^{\infty}_{l=2}\sum^{l}_{m=-l}h^{lm} \;_{-2}\mathcal{Y}^{lm}(\Theta,\Phi)\,,
\label{eq:gw_polarization_decomposition}
\end{equation}
where
\begin{equation}
    h^{lm}=\frac{G}{\sqrt{2}R c^{l+2}}[\mathcal{U}^{lm}(T_{R})-\frac{i}{c}\mathcal{V}^{lm}(T_{R})]\,,
\label{eq:hlm_mode_in_terms_of_radiated_moments}
\end{equation}
$\mathcal{U}^{lm}$ and $\mathcal{V}^{lm}$ are spherical harmonic representations of the radiative mass and current multipoles, respectively; $_{-2}\mathcal{Y}^{lm}(\Theta,\Phi)$ are spin-weighted spherical harmonics; $T_R$ is the retarded time, and $(R,\Theta,\Phi)$ are the distance and angles that point from the source to the observer.

In leading order, the radiative mass and the current multipoles depend on the source mass ($\mathcal{I}_{lm}$) and current ($\mathcal{J}_{lm}$, see Eq. (2.22) of \cite{Favata:2008yd}) multipole moments, respectively. At higher PN order, these radiative moments receive additional contributions from tail terms and other nonlinear couplings. As we are focusing on the leading-order nonlinear memory contribution, we only take into account $\mathcal{U}^{lm}$ (there is no nonlinear memory contribution to the $\mathcal{V}^{lm}$, thus we neglect all current multipole moments) and neglect the higher-order terms. Under these assumptions, the waveform modes (leading-order nonlinear memory piece) become \cite{Favata:2011qif}:
\begin{equation}
    h_{lm}\simeq \frac{1}{\sqrt{2}R}\mathcal{I}^{(l)}_{lm}+h^{\textup{(mem)}}_{lm}\,,
\label{eq:hlm_mode_in_terms_of_mass_quad_moments_and_memory}
\end{equation}
where $\mathcal{I}^{(l)}_{lm}$ is $l^{\text{th}}$ time derivative of $\mathcal{I}_{lm}$ and $h^{\text{mem.}}_{lm}$ is the nonlinear memory piece, given as \cite{Favata:2011qif}:

\begin{equation}
\begin{aligned}
h^{\textup{(mem)}}_{lm}
&= \frac{16\pi}{R}\sqrt{\frac{(l-2)!}{(l+2)!}}
   \int^{T_{R}}_{-\infty} dt
   \int d\Omega \,
   \frac{dE_{\text{GW}}}{dt\, d\Omega} (\Omega)
   \mathcal{Y}^{*}_{lm} (\Omega)
\\
&= R\sqrt{\frac{(l-2)!}{(l+2)!}}
   \sum_{l' = 2}^{\infty}
   \sum_{l'' = 2}^{\infty}
   \sum_{m' = -l'}^{l'}
   \sum_{m'' = -l''}^{l''} (-1)^{m+m''}
\\
&\quad \times 
   \mathcal{G}^{2,-2,0}_{l' l'' l\, m' -m'' -m}
   \int^{T_{R}}_{-\infty} dt \,
   \big\langle
   \dot{h}_{l'm'} \dot{h}^{*}_{l''m''}
   \big\rangle\,.
\end{aligned}
\label{eq:hlm_mode_of_nonlinear_memory}
\end{equation}

% \begin{equation}
% \begin{aligned}
%      h^{\textup{(mem)}}_{lm}=& \frac{16\pi}{R}\sqrt{\frac{(l-2)!}{(l+2)!}}\int^{T_{R}}_{-\infty}dt\int d\Omega \frac{dE_{\text{GW}}}{dt d\Omega} (\Omega) \mathcal{Y}^{*}_{lm} (\Omega)\,, \\  
%      =& R\sqrt{\frac{(l-2)!}{(l+2)!}} \sum^{\infty}_{l^{'} = 2}  \sum^{\infty}_{l^{''} = 2} \sum^{l^{'}}_{m^{'} = -l^{'}} \sum^{l^{''}}_{m^{''} = -l^{''}} (-1)^{m+m^{''}}     \mathcal{G}^{2,-2,0}_{l^{'}l^{''} l m^{'}-m^{''}-m}\int^{T_{R}}_{-\infty}dt \langle \dot{h}_{l^{'}m^{'}}\dot{h}^{*}_{l^{''}m^{''}}\rangle.
% \end{aligned}   
% \label{eq:hlm_mode_of_nonlinear_memory}
% \end{equation}
Here $\frac{dE_{\text{GW}}}{dt d\Omega}$ is the GW flux related to $h_{lm}$ modes, $\mathcal{G}_{l_{1}l_{2}l_{3}m_{1}m_{2}m_{3}}^{s_1 s_2 s_3}$ is an angular ($\Omega = (\Theta, \Phi)$) integral proportional to three spin-weighted spherical harmonics \cite{Favata:2008yd} and $\langle \cdot \rangle$ is the average over several wavelengths of GW which is required for construction of a well-defined GW stress tensor \cite{PhysRev.166.1272}.

Since the full $h_{lm}$ modes already include the nonlinear memory term from Eq. (\ref{eq:hlm_mode_in_terms_of_mass_quad_moments_and_memory}), expressing the nonlinear memory modes $h^{\textup{(mem)}}_{lm}$  in terms of the full $h_{lm}$ introduces a `nonlinear memory contribution to nonlinear memory'; however, this contribution is negligible \cite{Favata:2011qif}. Therefore, it is sufficient to substitute only the $\mathcal{I}_{lm}^{(l)}$ part of the $h_{lm}$ waveform in the nonlinear memory calculation. Also, we are only concerned with the leading-order nonlinear memory piece. Therefore, we only substitute the $l=2$ modes, i.e., $h_{2m}$ modes, on the right-hand side of Eq. (\ref{eq:hlm_mode_of_nonlinear_memory}).  When $h^{\textup{(mem)}}_{lm}$ is expressed in terms of $h_{2m}$ modes, one solves the angular integrals, implying certain selection rules (see Sec. III B of \cite{Favata:2008yd}) that will give a maximum value of `$l$' for which $h^{\textup{(mem)}}_{lm}$ are non-zero. So for the leading-order calculation only $h^{\textup{(mem)}}_{2m}$, $h^{\textup{(mem)}}_{3m}$ and $h^{\textup{(mem)}}_{4m}$ survive, {\color{black}and the rest of the} higher-$l$ nonlinear memory modes vanish. We can further simplify $h_{lm}$ by only considering the non-memory piece in the following way:

\begin{equation}
h^{\textup{N}}_{lm}\equiv \frac{\mathcal{I}^{(l)}_{lm}}{R\sqrt{2}}\,,
\label{eq:leading_order_hlm_modes}
\end{equation}
where $\textup{N}$ denotes Newtonian order and ${\color{black}\mathcal{I}_{lm}}\propto \mathcal{Y}^{*}_{lm}(\theta,\phi)$  ($\theta$ and $\phi$ are angles in the orbital frame), resulting in $\mathcal{I}^{*}_{lm}=(-1)^{m}\mathcal{I}_{l,-m}$, $h^{*}_{lm}=(-1)^{m}h_{l,-m}$. Assuming the equatorial orbital motion, i.e, orbit lies in the x-y plane ($\theta = \pi / 2$), then $h^{\textup{N}}_{2,\pm 1}\propto \mathcal{Y}^{*}_{2,\pm 1}\propto \sin 2\theta=0$, {\color{black}and then the $h^{\textup{N}}_{lm}$ modes simplify to:}
\begin{equation}
    h^{\textup{N}*}_{20} = h^{\textup{N}}_{20} \quad h^{\textup{N}*}_{2\pm 2} = h^{\textup{N}}_{2\mp 2} \quad h^{\textup{N}*}_{2\pm 1} = h^{\textup{N}}_{2\pm 1} = 0\,.
\label{eq:vanished_leading_order_hlm_modes}
\end{equation}
% Defining $h^{\textup{(mem)(1)}}_{lm} = \frac{d h^{\textup{(mem)}}_{lm}}{dT_R}$ where angular integrals are explicitly evaluated using $\dot{h}_{lm} \rightarrow \dot{h}^{\textup{N}}_{lm}$. Thus, after all simplifications, we obtain \cite{Favata:2011qif}
Following \cite{Favata:2011qif}, we can define $h^{\textup{(mem)(1)}}_{lm} \equiv \frac{d h^{\textup{(mem)}}_{lm}}{dT_R}$, then explicitly calculate the angular integrals in Eq. (\ref{eq:hlm_mode_of_nonlinear_memory}) and further replacing $\dot{h}_{lm}$  with $\dot{h}^{\textup{N}}_{lm}$ on the right-hand side gives

\begin{equation}
h^{\textup{(mem)(1)}}_{2\pm 1} = h^{\textup{(mem)(1)}}_{3m} = h^{\textup{(mem)(1)}}_{4\pm 1} = h^{\textup{(mem)(1)}}_{4\pm 3} = 0\,,
\label{eq:vanished_leading_order_hlm_modes_of_nonlinear_memory}
\end{equation}
\begin{equation}
  \begin{aligned}
        &h^{\textup{(mem)(1)}}_{20} = \frac{R}{42} \sqrt{\frac{15}{2\pi}} \Big{\langle}  2 |\dot{h}^{\textup{N}}_{22}|^2 - |\dot{h}^{\textup{N}}_{20}|^2   \Big{\rangle}\,,  \\& 
       h^{\textup{(mem)(1)}}_{2\pm 2} = \frac{R}{21} \sqrt{\frac{15}{2\pi}} \Big{\langle} \dot{h}^{\textup{N}}_{2\pm 2} \dot{h}^{\textup{N}}_{20}  \Big{\rangle}\,,  \\& 
       h^{\textup{(mem)(1)}}_{40} = \frac{R}{1260} \sqrt{\frac{5}{2\pi}} \Big{\langle} |\dot{h}^{\textup{N}}_{22}|^2 + 3 |\dot{h}^{\textup{N}}_{20}|^2   \Big{\rangle}\,,  \\&
       h^{\textup{(mem)(1)}}_{4\pm 2} = \frac{R}{252} \sqrt{\frac{3}{2\pi}} \Big{\langle} \dot{h}^{\textup{N}}_{2\pm 2} \dot{h}^{\textup{N}}_{20}  \Big{\rangle}\,,  \\&
       h^{\textup{(mem)(1)}}_{4\pm 4} = \frac{R}{504} \sqrt{\frac{14}{2\pi}} \Big{\langle} (\dot{h}^{\textup{N}}_{2\pm 2})^2  \Big{\rangle}\,.
  \end{aligned}
\label{eq:time_derivative_of_leading_order_nonlinear_memory_modes}
\end{equation}
As we know, Newtonian mass multipole moments are given as:
\begin{equation}
\begin{aligned}
    &\mathcal{I}^{\text{N}}_{20}=-4\sqrt{\frac{\pi}{15}}\eta M r(t)^{2}, \\&
    \mathcal{I}^{\text{N}}_{2,\pm 2}=2\sqrt{\frac{2\pi}{5}}\eta M r(t)^{2}e^{\mp 2i\phi(t)},
\end{aligned}
\label{eq:mass_quadrupole_moments}
\end{equation}
and taking their time derivatives, we can compute $h^{\textup{N}}_{lm}$ modes from Eq. (\ref{eq:leading_order_hlm_modes}) which further gives the time derivative of leading-order nonlinear memory modes $h^{\textup{(mem)(1)}}_{lm} $ from Eq. (\ref{eq:time_derivative_of_leading_order_nonlinear_memory_modes}).

{\color{black}Thus, Sec. (\ref{environment}) and (\ref{memory}) naturally motivate a systematic investigation of the conceptual connection between these two notions together: astrophysical environments and nonlinear memory. Building on this framework, we demonstrate in the subsequent sections how DM minispikes can modify the cumulative nonlinear memory amplitude relative to the vacuum case, given the spike exponent. This framework motivates a systematic study of how DM minispikes modify the cumulative nonlinear memory relative to the vacuum. For IMRIs observable by future space-based missions, it is then natural to ask whether these environmental modifications can produce memory signatures with astrophysically relevant mismatch in the low-frequency band, and whether the cumulative memory can encode information about the binary's evolution inside the ambient medium. This implies a nontrivial interplay between these two elusive components of the universe surrounding IMBHs. Next, we systematically examine this effect across distinct orbital configurations to assess its dependence on orbital dynamics and observables.}
%While the instantaneous growth of nonlinear memory is enhanced in the presence of spike profiles, the total accumulated memory is governed by the inspiral time impacted by environmental effects--dark matter gravity, dynamical friction, and accretion.
%%%%%%%%%%%%%%%%%%%%%%%%%%%%%%%%%%%%%%%%%%%%%%%%%%%%%%%%%%%%%%%%%%%
\section{Imprints of environment on nonlinear memory}\label{sec2}
In this section, we describe how we include the effects of a DM environment on top of the Newtonian equations of motion, focusing on how these effects alter the leading-order nonlinear memory term in GW signals relative to the vacuum case. We begin by formulating the perturbed Kepler problem, introducing the effects of DM, namely DM gravity, DF, and accretion, as independent perturbative forces. Each contribution is treated as a small correction to the standard Newtonian binary dynamics. \cite{Poisson_Will_2014}. Using all these contributions, we derive expressions for the ${h_{lm}}^{\text{mem(1)}}$ modes and incorporate these effects into the orbital parameters of the binary using the osculating orbit method. The osculating orbit method is used to solve the perturbed Kepler problem, which is a set of coupled differential equations that we numerically integrate to obtain the evolution of orbital parameters, which further helps in computing the leading-order nonlinear memory for distinct orbits: elliptical, hyperbolic, and quasi-circular. {\color{black} In this work, we consider a non-spinning binary system with component masses $m_1 = 10^3 M_{\odot}$ (central IMBH) and $m_2 = 10 M_{\odot}$, located at a luminosity distance of $R=10\,\textup{Mpc}$ from Earth throughout our calculations.}

\begin{figure}
%\centering
\includegraphics[width=1.1\linewidth]{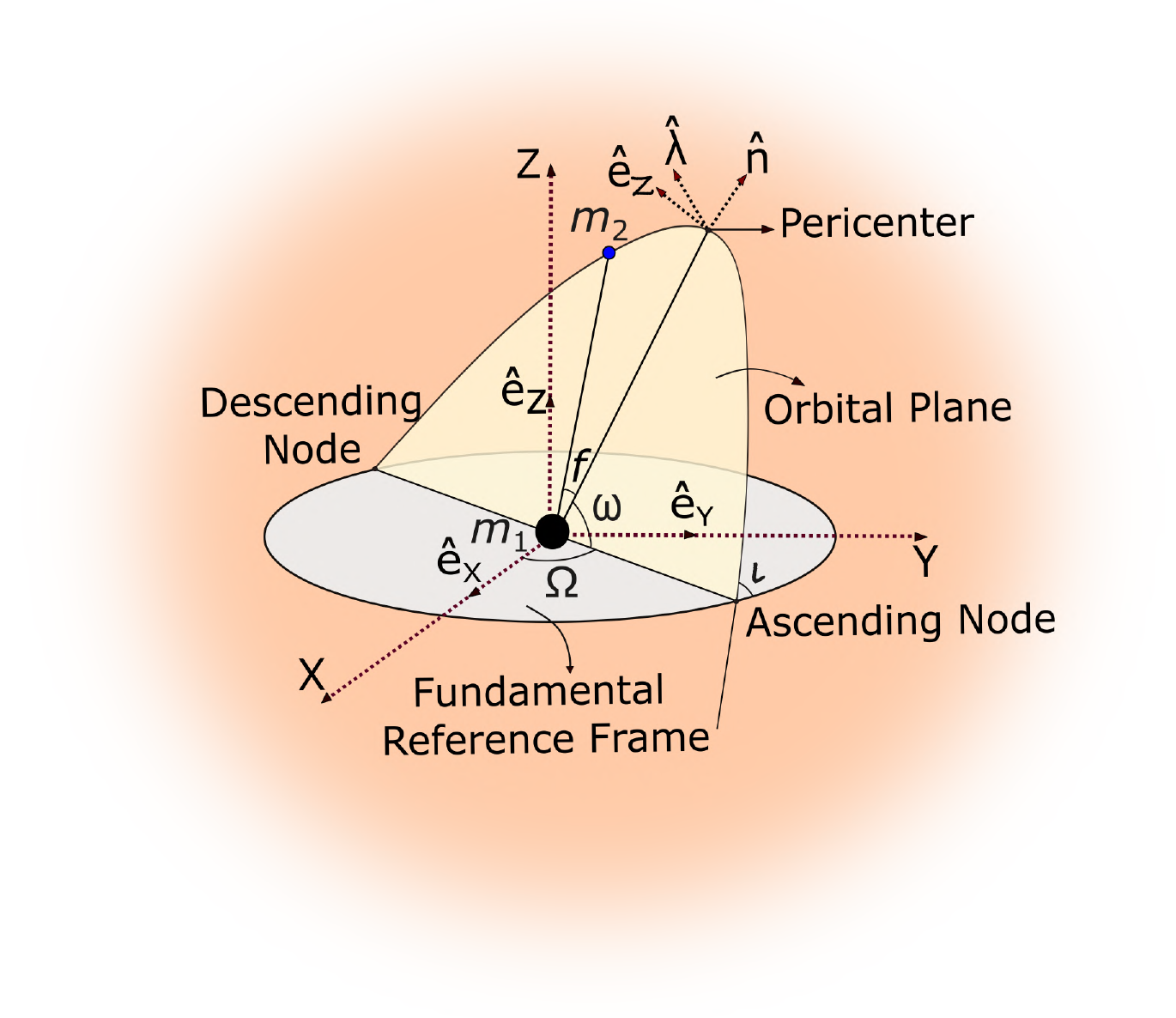}%{Memory_DM_V3NEW.pdf}
    \caption{Schematic of the orbital motion described in the fundamental reference frame in the presence of a surrounding DM distribution (in orange colour), representing a generic binary orbit in the fundamental reference frame $(X,Y,Z)$. Here, $m_1$ and $m_2$ denote the masses of the binary components; $\hat{e}_X, \hat{e}_Y, \hat{e}_Z$ are the basis vectors of the reference frame, while $\hat{e}_z, \hat{\lambda}, \hat{n}$ are the orthonormal basis vectors of the orbital plane. The orbital configuration is characterized by the inclination angle $\iota$, longitude of ascending node $\Omega$, argument of pericenter $\omega$, and true anomaly $f$.}
    \label{DM_Orbital_Motion}
\end{figure}

\subsection{Perturbed Kepler problem}\label{sec:perturbed_kep_prob}

The perturbation to the Kepler problem can be formulated as \cite{Poisson_Will_2014}:

\begin{equation}
\ddot{\vec{r}} = - \frac{M}{r^2} \hat{n} + \vec{f},
\label{eq:pertubed_kepler_eqn}
\end{equation}
where $\vec{r} = \vec{r}_1 - \vec{r}_2$, $r = |\vec{r}|$ is the distance between binaries and $M = m_1 + m_2$ is the total mass of the binaries. The first term in Eq. (\ref{eq:pertubed_kepler_eqn}) is the contribution of Newtonian gravity, while the second term is due to the perturbation forces (per unit mass) to the system. A general perturbation force can be modeled as \cite{Poisson_Will_2014}:
\begin{equation}
    \Vec{f}=\mathcal{R}\hat{n}+\mathcal{S}\hat{\lambda}+\mathcal{W}\hat{e_{z}},
\label{eq:general_perturbing_force}
\end{equation}
where $\mathcal{R}$, $\mathcal{S}$, and $\mathcal{W}$ are the components of the general perturbing force $\vec{f}$ along the radial, tangential, and perpendicular direction to the orbital plane, respectively. 

As can be intuitively understood, the presence of such perturbation forces leads to a finite change in the angular momentum vector $\Vec{h}$ (which otherwise remains conserved for a generic central force motion). The Runge-Lenz vector $\vec{e}_z$, too, is no longer conserved, and one can write down the time variation of these quantities in the following way \cite{Poisson_Will_2014}:
\begin{equation}
\begin{aligned}
    &\frac{d\Vec{h}}{dt}=\Vec{r}\times \vec{f}=-r\mathcal{W}\hat{\lambda}+r\mathcal{S}\hat{e_{z}}, \\& \frac{dh}{dt}\sim r\mathcal{S}, \, \,  h\frac{d\vec{e}_{z}}{dt}=-r\mathcal{W}\hat{\lambda}\,.
\end{aligned}
\label{eq:derivative_of_angular_momentum_and_runge_lenz_vector}
\end{equation}
The above relations show that the changes in orbital parameters and various other dynamical quantities depend heavily on how we define the perturbing force $\vec{f}$. A few models motivating our scenario are discussed below. 

%\subsubsection*{DM gravity}
%\label{sec:DM gravity}
%%%%%%%%%%%%%%%%%%%%%%%%%%%%%%%%%%%%%
\begingroup
\let\oldaddcontentsline\addcontentsline
\renewcommand{\addcontentsline}[3]{} % disables TOC entry
\subsubsection{DM gravity} \label{sec:DM gravity}
\endgroup
%%%%%%%%%%%%%%%%%%%%%%%%%%%%%%%%%%%%%%

Due to the gravitational potential of the DM density around IMBH, the stellar mass secondary BH will feel an additional force apart from the central IMBH gravity. This is termed as DM gravity and can be expressed as \cite{Eda:2013gg}:
% \begin{equation}
% {\vec{f}}_{DM} =
% \begin{cases}
%  -\frac{F}{r^{\alpha - 1}} \hat{n}, & \textup{Minispike}\\
%  -\frac{M_{\textrm{DM}}}{r^2} \hat{n}, & \textup{NFW}
% \label{eq:DM_gravity_force}
% \end{cases}
% \end{equation}
% where $F = \frac{4\pi \rho_{\textup{sp}} {r_{\textup{sp}}}^{\alpha} }{(3-\alpha)} $ and $M_{\textrm{DM}} (r) = 4 \pi  \rho_s {r_s}^3 \left[ -\frac{r}{r+r_s}+\log (r+r_s)+\frac{r_{\textup{min}}}{r_{\textup{min}}+r_s}-\log (r_{\textup{min}}+r_s) \right] $.
{\color{black}\begin{equation}
{\vec{f}}_{\textrm{DM}} = -\frac{M_{\textrm{DM}} (r)}{r^2} \hat{n},
\label{eq:DM_gravity_force}
\end{equation}
where
% \begin{equation}
% M_{\textrm{DM}}(r) =
% \begin{cases}
%  F \left[ r^{3-\alpha} - r_{\textrm{min}}^{3-\alpha} \right], & \textup{Minispike}\\
%  4 \pi  \rho_s {r_s}^3 \left[ -\frac{r}{r+r_s}+\log (r+r_s) \right. \\[4pt]
% \qquad \left. +\frac{r_{\textup{min}}}{r_{\textup{min}}+r_s}-\log (r_{\textup{min}}+r_s) \right], & \textup{NFW}
% \end{cases}
% \label{eq:DM_mass}
% \end{equation}
\begin{equation}
M_{\textrm{DM}}(r) = F \left[ r^{3-\alpha} - r_{\textrm{min}}^{3-\alpha} \right]
\label{eq:DM_mass}
\end{equation}
and $F = \frac{4\pi \rho_{\textup{sp}} {r_{\textup{sp}}}^{\alpha} }{(3-\alpha)} $.}
\vspace{0.3cm}
%\subsubsection*{Dynamical friction}
%%%%%%%%%%%%%%%%%%%%%%%%%%%%%%%%%%%%%
\begingroup
\let\oldaddcontentsline\addcontentsline
\renewcommand{\addcontentsline}[3]{} % disables TOC entry
\subsubsection{Dynamical Friction (DF)}
\endgroup
%%%%%%%%%%%%%%%%%%%%%%%%%%%%%%%%%%%%%%
A stellar-mass BH moving in a DM environment will experience gravitational drag from DM particles, perturbing its motion relative to the vacuum case. This collisionless gravitational drag is termed the Chandrasekhar DF force \cite{1943ApJ....97..255C}. This DF depends on the density of the environment $\rho_{\textup{DM}}$, the velocity of the stellar object $v$, and the speed of sound in the medium \cite{Ostriker_1999}. In the supersonic regime, the DF force per unit mass is given as \cite{PhysRevD.103.023015}:
\begin{equation}
{\vec{f}}_{\textrm{DF}} = -\frac{4\pi m_2 \rho_{\textup{DM}} I_v {\color{black} \xi(v)} }{ {v^3} } \vec{v},
\label{eq:dynamical_friction__force}
\end{equation}
where $\vec{v} = \dot{r} \hat{n} + r \dot{\phi} \hat{\lambda} $ and $ \xi(v)= (1-v^2)^{-1}(1+v^2)^2$ account for the relativistic effect \cite{10.1111/j.1365-2966.2007.12408.x} and $I_v$ is the Coulomb logarithm given as $ I_v = \ln \Lambda = \ln \sqrt{m_1 / m_2} \approx 3$ \cite{Eda:2014kra}.
\vspace{0.3cm}
%\subsubsection*{Accretion}
%%%%%%%%%%%%%%%%%%%%%%%%%%%%%%%%%%%%%
\begingroup
\let\oldaddcontentsline\addcontentsline
\renewcommand{\addcontentsline}[3]{} % disables TOC entry
\subsubsection{Accretion}
\endgroup
%%%%%%%%%%%%%%%%%%%%%%%%%%%%%%%%%%%%%%
Due to the gravity of a stellar BH, it will accrete DM matter around it. This leads to a change in the mass of a stellar object given as \cite{Macedo:2013qea, 10.1093/mnras/104.5.273}:
\begin{equation}
\dot{m_2} = \frac{4\pi m_{2}^2 \rho_{\textup{DM}} \lambda }{ {(v^2 + c_{s}^2)}^{3/2}}, 
\end{equation}
where $c_s$ is the speed of sound in the DM medium and $\lambda$ is a number that depends on the medium. In this paper, we assume $\lambda = 1$ following \cite{Macedo:2013qea}. Assuming that the accreted mass is much smaller than the mass of the stellar BH, we can model the accretion as a perturbing force per unit mass as \cite{Dai:2021olt}: 
\begin{equation}
{\vec{f}}_{acc} \simeq \frac{\dot{m_{2}} \vec{v}}{m_2} =  -\frac{4\pi m_2 \rho_{\textup{DM}} \lambda }{ {v^3} } \vec{v}\,.
\label{eq:accretion_force}
\end{equation}
\vspace{0.3cm}
%\subsubsection*{PN corrections}
%%%%%%%%%%%%%%%%%%%%%%%%%%%%%%%%%%%%%
\begingroup
\let\oldaddcontentsline\addcontentsline
\renewcommand{\addcontentsline}[3]{} % disables TOC entry
\subsubsection{PN Corrections}
\label{sec:PN_correcions}
\endgroup
%%%%%%%%%%%%%%%%%%%%%%%%%%%%%%%%%%%%%%
As the binaries evolve, they will lose energy and angular momentum in the form of GW. Due to this, the binary dynamics changes and can be modeled as a perturbation force to the Newtonian binary, following \cite{blanchet2024postnewtoniantheorygravitationalwaves}:
\begin{equation}
{\vec{f}}_{\textrm{GR}} = - \frac{M}{r^2} [ \mathbf{A} \hat{n} + \mathbf{B} \vec{v} ]\,,
\label{eq:GR_correction_force}
\end{equation}
where 
\begin{align}
A_{\textrm{1PN}} &= -\frac{ 3 {\dot{r}}^2 \eta}{2} + v^2 + 3 \eta v^2 - \frac{M}{r} (4 + 2 \eta)\,, \\[0.4em] A_{\textrm{2PN}} &= \frac{15 {\dot{r}}^4 \eta }{8} - \frac{45 {\dot{r}}^4 {\eta}^2 }{8} - \frac{9 {\dot{r}}^2 \eta v^2}{2} + 6{\dot{r}}^2 {\eta}^2 v^2 + 3\eta v^4 \notag\\
&\quad  - 4 {\eta}^2 v^4 + \frac{ M}{r} \Big[ - 2 {\dot{r}}^2 - 25 {\dot{r}}^2 {\eta} -2{\dot{r}}^2 {\eta}^2  - \frac{13 \eta v^2}{2} \notag\\
&\quad + 2 {\eta}^2 v^2 \Big] + \frac{M^2}{r^2} \Big[ 9 + \frac{87 \eta}{4} \Big]\,, \\[0.4em] A_{\textrm{2.5PN}} &= \frac{8 M \eta}{5 r} \dot{r}  \Big[ -\frac{17}{3} \frac{ M}{r} - 3 v^2 \Big]\,,  \\[0.4em] B_{\textrm{1PN}} &= - 4 \dot{r} + 2 \dot{r} \eta , \\[0.4em]  B_{\textrm{2PN}} &= \frac{9 {\dot{r}}^3 \eta}{2} + 3 {\dot{r}}^3 {\eta}^2 - \frac{15 \dot{r} \eta v^2}{2} - 2 \dot{r} {\eta}^2 v^2 \notag \\
&\quad + \frac{ M}{r} \Big[ 2 \dot{r} + \frac{41 \dot{r} \eta }{2} + 4 \dot{r} {\eta}^2 \Big]\,, \\[0.4em] B_{\textrm{2.5PN}} &= \frac{8 M \eta }{5 r } \Big[ 3 \frac{ M}{r } + v^2 \Big]\,,
\label{eq:PN_correction_terms}    
\end{align}
where $M = m_1 + m_2$, $\eta = \frac{m_1 m_2}{M^2}\,.$ {\color{black}Note that the effect of the GW backreaction to the binary equation of motion appears at 2.5PN order. However,  in this work, we have considered all the corrections up to 2.5PN while solving the orbital evolution of binaries for the hyperbolic orbit case and only the GW backreaction term at 2.5PN for elliptic orbits. This is because the leading-order nonlinear memory for hyperbolic orbits appears at 2.5PN in the waveform and 0PN for elliptic orbits \cite{Favata:2011qif} .}
 \vspace{0.3cm}
\subsection{Osculating orbit method}\label{sec:osculating_orbit_method}
The osculating orbit method \cite{Poisson_Will_2014} provides the evolution of orbital elements, such as eccentricity, semi-latus rectum, etc., over time. For a Keplerian orbit, these quantities are constants of motion, but for a perturbed scenario, these become functions of time. Hence, the orbit equation for a perturbed Kepler orbit is given as
\begin{equation}
    r(t)=\frac{p(t)}{1+e(t)\cos(\phi(t)-\omega)},
\label{eq:osculating_eqn}
\end{equation}
where $p$ is the semi-latus rectum, $e$ is the orbital eccentricity and $\omega$ is the angle of closest approach. Similarly, the rest of the orbital quantities are replaced by orbital parameters as a function of time.
\begin{equation}
    %\begin{aligned}
        \dot{r}=\Big(  \frac{M}{p(t)}\Big)^{1/2}e(t)\sin (\phi (t) - \omega) \hspace{2mm}; \hspace{2mm} \dot{\phi}=\frac{\sqrt{M p(t)}}{r^{2}}.
    \label{eq:radial_and_angular_velocity}
    %\end{aligned}
\end{equation}
These evolutions of orbital elements are governed by a set of differential equations with respect to time. However, for small perturbations, we can also write these differential equations in terms of true anomaly ($f = \phi - \omega$) as an independent variable. The instantaneous osculating equations for an elliptic orbit are given as \cite{Poisson_Will_2014, PhysRevD.109.124056}:
\begin{widetext}
\begin{equation}
\begin{aligned}
\frac{dp}{df}
&\simeq 2 \frac{p^3}{M}
   \frac{1}{(1 + e \cos f)^3} \, \mathcal{S}; \hspace{3mm}
\frac{de}{df}
\simeq \frac{p^2}{M}
   \Bigg[
   \frac{\sin f}{(1 + e \cos f)^2} \, \mathcal{R}
   +  \frac{2 \cos f + e (1 + \cos^2 f)}{(1 + e \cos f)^3}
     \, \mathcal{S}
   \Bigg],
\\[3pt]
\frac{d\iota}{df}
&\simeq \frac{p^2}{M}
   \frac{\cos (\omega + f)}{(1 + e \cos f)^3}
   \, \mathcal{W}; \hspace{3mm}
\sin\iota \, \frac{d\Omega}{df}
\simeq \frac{p^2}{M}
   \frac{\sin (\omega + f)}{(1 + e \cos f)^3}
   \, \mathcal{W},
\\[3pt]
\frac{d\omega}{df}
&\simeq \frac{1}{e} \frac{p^2}{M}
   \Bigg[
   - \frac{\cos f}{(1 + e \cos f)^2} \, \mathcal{R}
   + \frac{2 + e \cos f}{(1 + e \cos f)^3}
     \sin f \, \mathcal{S} - e \cot\iota
     \frac{\sin (\omega + f)}{(1 + e \cos f)^3}
     \, \mathcal{W}
   \Bigg],
\\[3pt]
\frac{dt}{df}
&\simeq \sqrt{\frac{p^3}{M}}
   \frac{1}{(1 + e \cos f)^2} \Bigg[
   1 - \frac{1}{e} \frac{p^2}{M}
   \Bigg(
   \frac{\cos f}{(1 + e \cos f)^2} \, \mathcal{R}
   - \frac{2 + e \cos f}{(1 + e \cos f)^3}
     \sin f \, \mathcal{S}
   \Bigg)
   \Bigg]\,.
\end{aligned}
\label{eq:elliptical_osculating_eqns}
\end{equation}  
\end{widetext}
% \begin{equation}
% \begin{aligned}
% & \frac{dp}{df} \simeq 2  \frac{p^3}{G M}   \frac{1}{(1 + e \cos (f))^3}  \mathcal{S}, 
% \\& \frac{de}{df} \simeq  \frac{p^2}{G M} \Big[ \frac{\sin f}{(1 + e \cos f)^2} \mathcal{R} + \frac{2 \cos f + e (1+ \cos^2 f)}{(1 + e \cos f)^3} \mathcal{S}  \Big]\,,
% \\& \frac{d \iota}{df} \simeq \frac{p^2}{G M} \frac{\cos (\omega + f ) }{(1 + e \cos f)^3} \mathcal{W}\,,
% \\& \sin \iota \frac{d \Omega}{df} \simeq \frac{p^2}{G M} \frac{\sin (\omega + f ) }{(1 + e \cos f)^3} \mathcal{W}\,,
% \\& \frac{d\omega}{df} \simeq \frac{1}{e} \frac{p^2}{G M}  \left[ -\frac{\cos f}{(1 + e \cos f)^2} \mathcal{R} +\frac{(2 + e \cos f)}{(1+e \cos f)^3} \sin f \, \mathcal{S} - e \cot \iota  \frac{ \sin (\omega +f)}{(1+e \cos f)^3} \mathcal{W} \right]\,, 
% \\& \frac{dt}{df} \simeq \sqrt{\frac{p^3}{G M}}  \frac{1}{(1 + e \cos f)^2} \Big( 1 - \frac{1}{e}  \frac{p^2}{G M} \Big[ \frac{\cos f}{(1 + e \cos f)^2} \mathcal{R} - \frac{(2 + e \cos f)}{(1 + e \cos f)^3}  \sin f \, \mathcal{S} \Big] \Big)\,.   
% \label{eq:elliptical_osculating_eqns}
% \end{aligned}
% \end{equation}
{\color{black} For the hyperbolic orbit $e = - \frac{1}{\cos (\omega)}$, this gives the instantaneous osculating equation for a hyperbolic orbit \cite{PhysRevD.109.124056, Poisson_Will_2014}:} 
\begin{equation}
\begin{aligned}
& \frac{de}{df} =   - \frac{\sin \omega}{\cos^2 \omega} \frac{d\omega}{df},
% \\&  \frac{dp}{df} \simeq 2  \frac{p^3}{G M}   \frac{1}{(1 + e \cos f)^3} \mathcal{S}\,,
% \\& \frac{d\omega}{df} \simeq \frac{1}{e} \frac{p^2}{G M}  \left[ -\frac{\cos f}{(1 + e \cos f)^2} \mathcal{R}  +\frac{(2 + e \cos f)}{(1+e \cos f)^3} \sin f \, \mathcal{S}  - e \cot \iota \frac{ \sin (\omega +f)}{(1+e \cos f)^3} \mathcal{W}  \right]\,, 
% \\&  \frac{dt}{df} \simeq \sqrt{\frac{p^3}{G M}} \frac{1}{(1 + e \cos f)^2} \left( 1 - \frac{1}{e} \frac{p^2}{G M} \Big[ \frac{\cos f}{(1 + e \cos f)^2} \mathcal{R}  - \frac{2 + e \cos f}{(1 + e \cos f)^3} \sin f \, \mathcal{S}  \Big] \right)\,.   
\label{eq:hyperbolic_osculating_eqns}
\end{aligned}
\end{equation}
{\color{black} and $\frac{dp}{df}, \frac{d\omega}{df}$, $\frac{dt}{df}$ are the same as mentioned above in Eq. (\ref{eq:elliptical_osculating_eqns}). Here, $\mathcal{R}$, $\mathcal{S}$, and $\mathcal{W}$ represent the components of the perturbing forces in a Newtonian binary system. For the type of perturbations considered in this work, the out-of-orbital-plane component $\mathcal{W}$ vanishes (see Eq.~(\ref{eq:general_perturbing_force})). Consequently, the orbital inclination $\iota$ and the longitude of the ascending node $\Omega$ remain constants of motion, simplifying the subsequent analysis. This can be understood from Fig. (\ref{DM_Orbital_Motion}).} 
% Also $\mathcal{R}$, $\mathcal{S}$, and $\mathcal{W}$ represent the components of the perturbing forces in a Newtonian binary system. Since the perturbations considered in this study do not have a $\mathcal{W}$ component (as in Eq.~(\ref{eq:general_perturbing_force})), the equations simplify in further calculations leaving $\iota $ and $\Omega$ constants of motion.

\subsection{Elliptical orbits}\label{sec3.3}

Having laid out the theoretical structure of our approach in this manuscript, we can now focus on specific orbital motions. In this section, we derive the leading-order nonlinear memory arising from binary interactions in elliptical orbits in the presence of non-trivial environmental contributions. {\color{black} We first derive $h^{\textup{(mem)}(1)}_{lm}$ using the perturbed Newtonian binary equation of motion, considering DM environment effects defined in section (\ref{sec:perturbed_kep_prob}) and GW radiation reaction (2.5PN correction) because nonlinear memory appears at ($0$PN) in waveform \cite{Favata:2011qif} for elliptic orbits and also assuming contribution of other PN corrections from GR is very less compared to DM effects in elliptic orbits.} Using this, we can calculate $h^{\textup{(mem)}}_{lm}$ by integrating with respect to eccentricity, which requires expressing other orbital parameters as functions of eccentricity. These parameters are obtained by solving the averaged osculating equations.

% \subsubsection{\textup{DM minispike profile} \& memory modes}\label{subsec:hlm_memo_dot_minispike}
% \stoptoc
% \subsubsection[Deriving $h_{lm}^{\textup{(mem)(1)}}$ for perturbed Newtonian binaries]%
% {Deriving $h_{lm}^{\textup{(mem)(1)}}$ for perturbed Newtonian binaries}
%\subsubsection*{Deriving $h_{lm}^{\textup{(mem)(1)}}$ for perturbed Newtonian binaries}
%%%%%%%%%%%%%%%%%%%%%%%%%%%%%%%%%%%%%
% \begingroup
% \let\oldaddcontentsline\addcontentsline
% \renewcommand{\addcontentsline}[3]{} % disables TOC entry
% \subsubsection{Deriving $h_{lm}^{\textup{(mem)(1)}}$ for perturbed Newtonian binaries}
% %\label{subsec:hlm_memo_dot_minispike}
% \endgroup
%%%%%%%%%%%%%%%%%%%%%%%%%%%%%%%%%%%%%%
{\it \textbf{Deriving $h_{lm}^{\textup{(mem)(1)}}$ for perturbed Newtonian binaries}.} The perturbed Newtonian binary equation, considering the effects of DM environment, including DM minispike gravity, DF, and accretion as perturbing forces from Eq.(\ref{eq:DM_gravity_force}-\ref{eq:accretion_force}), takes the following form:
\begin{equation}
\ddot{\vec{r}} = - \frac{M}{r^2} \hat{n} + \vec{f}_{\textrm{DM}}+ \vec{f}_{\textrm{DF}} + \vec{f}_{\textrm{acc}} +{\color{black}\vec{f}_{\textrm{2.5PN}}}. 
% = - \frac{M_{\textrm{eff}}}{r^2} \hat{n} - \frac{F}{r^{\alpha -1}} \hat{n}+ \vec{f}_{\textrm{DF}} + \vec{f}_{\textrm{acc}} +\vec{f}_{\textrm{2.5PN}}.
\label{eq:minispike_perturbed_binary_eom}
\end{equation}
{\color{black}This equation can be resolved into components for $r$ and $\phi$ using Eq. (\ref{eq:DM_gravity_force}), (\ref{eq:DM_mass}), (\ref{eq:dynamical_friction__force}), (\ref{eq:accretion_force}) and 2.5PN correction term from sec. (\ref{sec:PN_correcions}) }:
%\begin{widetext}
\begin{equation}
\begin{aligned}
    \ddot{r} =\;& r\dot{\phi}^{2}-\frac{M_{\textrm{eff}}}{r^{2}}-\frac{F}{r^{\alpha -1}} - \frac{4\pi m_2 \rho_{\textrm{DM}} (I_v {\color{black} \xi(v)} + \lambda) \dot{r}}{v^3} \\
    & {\color{black}- \frac{M}{r^2} \Big[ A_{2.5\text{PN}} 
    + \dot{r} B_{2.5\text{PN}} \Big]}, \\  
    \ddot{\phi} =& -\frac{2\dot{r}\dot{\phi}}{r} - \frac{4\pi m_2 \rho_{\textrm{DM}} (I_v {\color{black} \xi(v)} + \lambda) \dot{\phi}}{v^3} {\color{black}-\frac{M}{r} \Big[ B_{2.5\text{PN}} \Big] \dot{\phi}},
\label{eq:minispike_perturbed_radial_and_angular_eom}
\end{aligned}
\end{equation}   
%\end{widetext}
{\color{black}where $M_{\textrm{eff}}$ arises from the contribution of DM mass for the minispike profile mentioned in Eq. (\ref{eq:DM_mass}) to the total binary mass and given as}
\begin{equation}
M_{\textrm{eff}} (r) = 
\begin{cases} 
M - \frac{4 \pi\rho_{\textup{sp}} r_{\textup{sp}}^{\alpha} r_{\textup{min}}^{{\color{black}3-\alpha}} }{(3-\alpha)} & ,\,\, r_{\textup{min}} \leq r \leq r_{\textup{sp}} \\
M & , \,\, r \leq r_{\textup{min}}
\label{eq:minispike_effective_mass}
\end{cases}
\end{equation}  
\begin{equation}
F =
\begin{cases} 
\frac{4 \pi\rho_{\textup{sp}} r_{\textup{sp}}^{\alpha} }{(3-\alpha)} &  , \,\,r_{\textup{min}} \leq r \leq r_{\textup{sp}} \\
0 &  ,\,\,r \leq r_{\textup{min}}
\label{eq:minispike_F}
\end{cases}
\end{equation}  
\begin{equation}
\begin{aligned}
& \dot{r} = \sqrt{\frac{M_{\textrm{eff}}}{p(t)}} e(t) \sin{f(t)}, \\& \dot{\phi} = \sqrt{\frac{M_{\textrm{eff}}}{p^3 (t)}}(1 + e(t) \cos{f(t)} )^2, \\& v = \sqrt{ {\dot{r}}^2 + r^2 {\dot{\phi}}^2},
\label{eq:minispike_perturbed_radial_and_angular_velocity}
\end{aligned}
\end{equation}
and $\rho_{\textrm{DM}}$ is the minispike density defined in Eq. (\ref{eq:minispike_density}).

Using Eq. (\ref{eq:leading_order_hlm_modes}) and (\ref{eq:mass_quadrupole_moments}), we can write expressions for $h^{\textup{N}}_{2m}$. Then replacing $\dot{r}$, $\dot{\phi} $, $\ddot{r}$, and $\ddot{\phi}$ from Eq. (\ref{eq:minispike_perturbed_radial_and_angular_eom}) and (\ref{eq:minispike_perturbed_radial_and_angular_velocity}), we obtain a lengthy and complicated expression for $h^{\textup{N}}_{2m}$ as a function of orbital parameters ($e$, $p$, $\omega$, $f$). Further, taking one more time derivative of $h^{\textup{N}}_{2m}$ and a similar replacement from Eq. (\ref{eq:minispike_perturbed_radial_and_angular_eom}) and (\ref{eq:minispike_perturbed_radial_and_angular_velocity}), we get $\dot{h}^{N}_{2m}$. We replace this long complicated\footnote{For all the lengthy calculations, we have used the $\textsf{Python}$ symbolic calculation package $\textsf{SymPy}$ to reduce the labor of writing big expressions used in further calculations. Also, we don't show those expressions explicitly in the manuscript as they are not very illuminating.} expression by $G_{2m}$, where $G_{2m}$ is a function of  ($e,p,f,\omega$):
\begin{equation}
\dot{h}^{\text{N}}_{2m}\equiv \frac{\dddot{\mathcal{I}}^{\text{N}}_{2m}}{R\sqrt{2}} = G_{2m} (e,p,f, \omega)\,.
\label{eq:h2m_dot_expression}
\end{equation}
Finally, the above equation, along with Eq. (\ref{eq:time_derivative_of_leading_order_nonlinear_memory_modes}), gives us the time derivative of the leading-order nonlinear memory modes $h_{lm}^{\textup{(mem)(1)}}$:
\begin{widetext}
\begin{equation}
  \begin{aligned}
        &h^{\textup{(mem)(1)}}_{20} = \frac{R}{42} \sqrt{\frac{15}{2\pi}} \Big{\langle}  2 |G_{22} (e,p,f, \omega)|^2 - |G_{20} (e,p,f, \omega)|^2   \Big{\rangle}\,,  \\& 
       h^{\textup{(mem)(1)}}_{2 \pm 2} = \frac{R}{21} \sqrt{\frac{15}{2\pi}} \Big{\langle} G_{2\pm 2} (e,p,f, \omega) G_{20} (e,p,f, \omega)  \Big{\rangle}\,,  \\& 
       h^{\textup{(mem)(1)}}_{40} = \frac{R}{1260} \sqrt{\frac{5}{2\pi}} \Big{\langle} |G_{22} (e,p,f, \omega)|^2 + 3 |G_{20} (e,p,f, \omega)|^2   \Big{\rangle}\,,  \\&
       h^{\textup{(mem)(1)}}_{4\pm 2} = \frac{R}{252} \sqrt{\frac{3}{2\pi}} \Big{\langle} G_{2\pm 2} (e,p,f, \omega) G_{20} (e,p,f, \omega)  \Big{\rangle}\,,  \\&
       h^{\textup{(mem)(1)}}_{4\pm 4} = \frac{R}{504} \sqrt{\frac{14}{2\pi}} \Big{\langle} (G_{2\pm 2} (e,p,f, \omega))^2  \Big{\rangle}\,.
\label{eq:elliptical_nonlinear_memory_dot_expression}
  \end{aligned}
\end{equation}    
\end{widetext}

Note that after simplifying the above expression, we observe that only $\{2,\pm 2\}$, $\{4,\pm 2\}$, and $\{4,\pm 4\}$ are functions of $e$, $p$, $f$, and $\omega$. {\color{black}The $\{2, 0\}$ and $\{4, 0\}$} do not explicitly depend on $\omega$, and are just functions of $e$, $p$, and $f$. {\color{black}This will further help our calculation of dominant leading-order nonlinear memory modes, i.e, $\{2,0\}$ and $\{4,0\}$.}

%\subsubsection*{Calculating nonlinear memory}

{\it \textbf{Calculating nonlinear memory}.} To obtain $h_{lm}^{\textup{(mem)}}$ from Eq. (\ref{eq:elliptical_nonlinear_memory_dot_expression}) for an elliptical orbit, we resort to averaging over the orbital period of the binary $T_{\textrm{orb}}$. The orbital-averaging formula for any function $F(t)$ is given as:
\begin{equation}
\begin{aligned}
\langle F(t) \rangle &= \frac{1}{T_{\textrm{orb}}} \int_{0}^{T_{\textrm{orb}}} F(t) dt\,, \\& = \frac{(1 - e^2)^{3/2}}{2 \pi} \int_{0}^{2 \pi} df \, \frac{F(f)}{(1 + e \cos f)^2}\,,
\label{eq:orbital_averaging_formaula}
\end{aligned}
\end{equation}
where $t$ is the time, $f$ is the true anomaly and $e$ is the orbital eccentricity. Owing to the complicated expression of $h_{lm}^{\textup{(mem)(1)}}$ in Eq. (\ref{eq:elliptical_nonlinear_memory_dot_expression}), we resort to computing the average numerically and define the numerical function to be $H$, which depends on ($e$, $p$, $\omega$).
\begin{equation}
h_{lm}^{\textup{(mem)(1)}} = H(e, p, \omega)\,.
\label{eq:hlm_dot_elliptical}
\end{equation}
This averaging technique eliminates the high-frequency component from the memory waveform \cite{Favata:2011qif} and also implies an adiabatic approximation, i.e, the time scale on which orbital parameters change is much larger than the orbital period \cite{PhysRevD.66.044002, Kumar_2024}. Such approximations are valid in our cases since the perturbations to the Newtonian binary are very small.

{\color{black} Given the numerical solution of the above Eq. (\ref{eq:hlm_dot_elliptical}), we use it to compute a time integral to get $h^{\textup{(mem)}}_{lm}$. Since orbital parameters $e$, $p$, and $\omega$ are functions of true anomaly due to the perturbing forces, their instantaneous evolution is governed by osculating equations Eq. (\ref{eq:elliptical_osculating_eqns}). Using adiabatic approximation, we can thus write the averaged osculating equations and solve this set of coupled differential equations given as: 
\begin{equation}
\begin{aligned}
& \Big\langle \frac{dp}{df} \Big\rangle = \Big\langle \frac{dp}{df} \Big\rangle_{\textrm{DM}} + \Big\langle \frac{dp}{df} \Big\rangle_{\textrm{DF}} + \Big\langle \frac{dp}{df} \Big\rangle_{\textrm{acc}} + \Big\langle \frac{dp}{df} \Big\rangle_{\textrm{GW}}, \\& \Big\langle \frac{de}{df} \Big\rangle = \Big\langle \frac{de}{df} \Big\rangle_{\textrm{DM}} + \Big\langle \frac{de}{df} \Big\rangle_{\textrm{DF}} + \Big\langle \frac{de}{df} \Big\rangle_{\textrm{acc}} + \Big\langle \frac{de}{df} \Big\rangle_{\textrm{GW}}\,, \\& \Big\langle \frac{d \omega}{df} \Big\rangle = \Big\langle \frac{d \omega}{df} \Big\rangle_{\textrm{DM}} + \Big\langle \frac{d \omega}{df} \Big\rangle_{\textrm{DF}} + \Big\langle \frac{d \omega}{df} \Big\rangle_{\textrm{acc}} + \Big\langle \frac{d \omega}{df} \Big\rangle_{\textrm{GW}}\,, \\& \Big\langle \frac{dt}{df} \Big\rangle = \Big\langle \frac{dt}{df} \Big\rangle_{\textrm{DM}} + \Big\langle \frac{dt}{df} \Big\rangle_{\textrm{DF}} + \Big\langle \frac{dt}{df} \Big\rangle_{\textrm{acc}} + \Big\langle \frac{dt}{df} \Big\rangle_{\textrm{GW}}\,, 
\end{aligned}
\label{eq:elliptical_avg_osculating_eqn}
\end{equation}
where $\langle \rangle_{\textrm{DM}}$, $\langle \rangle_{\textrm{DF}}$, $\langle \rangle_{\textrm{acc}}$, and $\langle \rangle_{\textrm{GW}}$ are averaged osculating equations due to DM gravity, DF, accretion, and GW backreaction perturbation, respectively. We numerically solve these coupled differential equations starting from an initial eccentricity $e_0$ and semi-latus rectum $p_0$ till the secondary reaches the last stable orbit (LSO).

Using the solution of Eq.~(\ref{eq:elliptical_avg_osculating_eqn}), we numerically integrate the nonlinear memory modes as a function of eccentricity using the Euler method:
\begin{equation}
h_{lm}^{\textup{(mem)}} = \int_{-\infty}^{T_R} h_{lm}^{\textup{(mem)(1)}} \, dt \, = \int_{e_0}^{e(t)} h_{lm}^{\textup{(mem)(1)}} \frac{\langle \frac{dt}{df} \rangle}{\langle \frac{de}{df} \rangle} \, de\,.
\label{eq:elliptical_nonlinear_memo}
\end{equation}}

\resumetoc

\subsection{Hyperbolic orbits}

In this section, we derive the leading-order nonlinear memory arising from binaries in scattering/hyperbolic orbits in a DM environment. Similar to the elliptical orbit case, we first derive ${h_{lm}}^{\textup{(mem)(1)}}$ using the perturbed Newtonian binary equation of motion, but here we consider both the DM environment as well as PN corrections to orbital equations up to $2.5$PN order defined in section (\ref{sec:perturbed_kep_prob}) because nonlinear memory contribution appears at $2.5$PN in waveforms for hyperbolic orbits. Thus, we cannot neglect the PN corrections to the orbit equation up to $2.5$PN \cite{Favata:2011qif}. We calculate ${h_{lm}}^{(\text{mem})}$ by integrating with respect to the true anomaly variable, which requires expressing other orbital parameters as functions of true anomaly which can again be obtained by solving the instantaneous osculating equations for the hyperbolic case following Eq.~(\ref{eq:hyperbolic_osculating_eqns}).

% \subsubsection{\textup{DM minispike profile: memory modes and ``jump''}}

% \stoptoc

% \subsubsection*[Deriving $h_{lm}^{\textup{(mem)}(1)}$ for perturbed Newtonian binaries]%
% {Deriving $h_{lm}^{\textup{(mem)}(1)}$ for perturbed Newtonian binaries}

{\it\textbf{Deriving $h_{lm}^{\textup{(mem)}(1)}$ for perturbed Newtonian binaries}.} For the hyperbolic case, in the perturbed Newtonian binary equations that we are setting up, we  include the effects of DM environment and PN corrections up to 2.5PN terms as perturbing forces (\ref{eq:DM_gravity_force}-\ref{eq:PN_correction_terms}):
\begin{equation}
\ddot{\vec{r}} = - \frac{M}{r^2} \hat{n} + \vec{f}_{\textrm{DM}}+ \vec{f}_{\textrm{DF}} + \vec{f}_{\textrm{acc}} + \vec{f}_{\textrm{GR}}\,. 
% = - \frac{M_{\textrm{eff}}}{r^2} \hat{n} - \frac{F}{r^{\alpha -1}} \hat{n}+ \vec{f}_{\textrm{DF}} + \vec{f}_{\textrm{acc}} - \frac{M}{r^2} [ \mathbf{A} \hat{n} + \mathbf{B} \vec{v} ].
\label{eq:hyperbolic_minispike_eom}
\end{equation}
Resolving these equations into components, we get, 
\begin{widetext}
\begin{equation}
\begin{aligned}
    & \ddot{r}= r\dot{\phi}^{2} - \frac{M_{\text{eff}}}{r^{2}} - \frac{F}{r^{\alpha - 1}}
    - \frac{4\pi m_2 \rho_{\text{DM}} (I_v {\color{black} \xi(v)} + \lambda) \dot{r}}{v^3} - \frac{M}{r^2} \Big[ (A_{1\text{PN}} + A_{2\text{PN}} + A_{2.5\text{PN}}) 
    + \dot{r} (B_{1\text{PN}} + B_{2\text{PN}} + B_{2.5\text{PN}}) \Big]\,, \\
    & \ddot{\phi} = - \frac{2\dot{r}\dot{\phi}}{r}
    - \frac{4\pi m_2 \rho_{\text{DM}} (I_v {\color{black} \xi(v)} + \lambda) \dot{\phi}}{v^3} - \frac{M}{r} \Big[ B_{1\text{PN}} + B_{2\text{PN}} + B_{2.5\text{PN}} \Big] \dot{\phi}\,.
\end{aligned}
\label{eq:hyperbolic_radial_and_angular_eom}
\end{equation} 
\end{widetext}
Following similar steps described in section (\ref{sec3.3}), we can write the time derivative of the leading-order nonlinear memory modes ${h_{lm}}^{\textup{(mem)(1)}}$ as a function $K_{2m}$ (owing to the complicated expression) depending on $e$, $p$, $f$ and $\omega$ \cite{Favata:2011qif}:
\begin{widetext}
\begin{equation}
\begin{aligned}
    &h^{\textup{(mem)(1)}}_{20} = \frac{R}{42} \sqrt{\frac{15}{2\pi}} \Big{\langle}  2 |K_{22} (e,p,f, \omega)|^2 - |K_{20} (e,p,f, \omega)|^2   \Big{\rangle}\,,  \\& 
   h^{\textup{(mem)(1)}}_{2 \pm 2} = \frac{R}{21} \sqrt{\frac{15}{2\pi}} \Big{\langle} K_{2\pm 2} (e,p,f, \omega) K_{20} (e,p,f, \omega)  \Big{\rangle}\,,  \\& 
   h^{\textup{(mem)(1)}}_{40} = \frac{R}{1260} \sqrt{\frac{5}{2\pi}} \Big{\langle} |K_{22} (e,p,f, \omega)|^2 + 3 |K_{20} (e,p,f, \omega)|^2   \Big{\rangle}\,,  \\&
   h^{\textup{(mem)(1)}}_{4\pm 2} = \frac{R}{252} \sqrt{\frac{3}{2\pi}} \Big{\langle} K_{2\pm 2} (e,p,f, \omega) K_{20} (e,p,f, \omega)  \Big{\rangle}\,, \\&
   h^{\textup{(mem)(1)}}_{4\pm 4} = \frac{R}{504} \sqrt{\frac{14}{2\pi}} \Big{\langle} (K_{2\pm 2} (e,p,f, \omega))^2  \Big{\rangle}\,.
\end{aligned}
\label{eq:hyperbolic_nonlinear_memo_dot}
\end{equation}
\end{widetext}

%\subsubsection*{Nonlinear memory and the ``jump"}
{\it\textbf{Nonlinear memory and the ``jump"}.}
While integrating the above expressions for $h^{\textup{(mem)}}_{lm}$ as a function of true anomaly $f$, we can also compute an overall ``nonlinear memory jump'' $\Delta h^{\textup{(mem)}}_{lm}$ owing to the periastron passage which happens very quickly for hyperbolic encounter. This nonlinear memory jump can be obtained as a function of different initial eccentricities $e_0$'s. Consequently, different values of $e_0$ correspond to different values of the overall nonlinear memory jump for the whole orbit.

Since an unbound orbit is not periodic, an averaging procedure of the type used for elliptical orbits is not applicable. We can only integrate expressions in Eq. (\ref{eq:hyperbolic_nonlinear_memo_dot}) with respect to time to obtain $h_{lm}^{\textup{(mem)}}$. While doing so, we need to change variables from time $t$ to true anomaly $f$ using the instantaneous osculating equation for hyperbolic orbits in Eq. (\ref{eq:hyperbolic_osculating_eqns}) and replace $e$, $p$, and $\omega$ as functions of $f$, which comes after numerically solving the instantaneous osculating equations. Therefore, integrating expressions in Eq.~(\ref{eq:hyperbolic_nonlinear_memo_dot}) with respect to $f$ from initial angle $f_{\textrm{in}} = \cos^{-1} (-\frac{1}{e_0})$ (correspond to a situation when the binaries are very far apart, ideally $-\infty$) to a value $f(t)$ at a later time. We can then write $h_{lm}^{(mem)}$ as \cite{Favata:2011qif}:

\begin{equation}
h_{lm}^{\textup{(mem)}} = \int_{-\infty}^{T_R} h_{lm}^{\text{mem}(1)} \, dt \, = \int_{f_{in}}^{f(t)} h_{lm}^{\text{mem}(1)} \frac{dt}{df}  \, df\,.
\label{eq:hyperbolic_nonlinear_memo}
\end{equation}
We can then define an overall nonlinear memory ``jump'' as an integration from $f_{\textrm{in}}$ to $f_{\text{final}}$, which corresponds to binaries that start from very far away (ideally $-\infty$) distance will go to far away (ideally $\infty$) after the whole hyperbolic encounter \cite{Favata:2011qif}:

\begin{equation}
\Delta h_{lm}^{\textup{(mem)}} = \int_{f_{in}}^{f_{\text{final}}} h_{lm}^{\text{mem}(1)} \frac{dt}{df}  \, df\,.
\label{eq:hyperbolic_nonlinear_memo_jump}
\end{equation}

In order to compute this jump, we consider the instantaneous osculating equations for the hyperbolic case, Eq. (\ref{eq:hyperbolic_osculating_eqns}), considering all DM environment effects and PN corrections up to 2.5PN (which can be written using the perturbing force expressions from Eq. (\ref{eq:DM_gravity_force}-\ref{eq:PN_correction_terms})) and solve this set of differential equations numerically \cite{our_github_repo_nonlinear_memo}\footnote{The Github repository of our detailed code can be found here \cite{our_github_repo_nonlinear_memo}.}. This gives the orbital quantities ($e,p,\omega$) as a function of $f$, and when one puts these quantities back into the above Eq. (\ref{eq:hyperbolic_nonlinear_memo}-\ref{eq:hyperbolic_nonlinear_memo_jump}), it gives us the nonlinear memory and overall memory jump for the hyperbolic orbit case. 
% {\color{black} Again, using the same methodology, we compute the nonlinear memory modes for the NFW profile, employing the corresponding equations of motion given in the Appendix~(\ref{append}). In the next section, we will conduct a comparative study of the nonlinear memory embedded in these two different static DM environments, i.e., the DM spike and NFW models for hyperbolic orbits. }
\resumetoc
% \subsubsection{DM NFW profile}

% The approach for studying the effects of an NFW profile in GW memory relies on using the equation of motion of binaries defined in Eq. (\ref{eq:hyperbolic_minispike_eom}) (we have to use Eq.~(\ref{eq:NFW_density}) for the NFW density and the corresponding expression for NFW in (\ref{eq:DM_gravity_force})) and following the similar steps described above for the hyperbolic orbit in the DM spike profile. Following these steps, one can calculate the nonlinear memory and the overall memory jump for the NFW profile.

\subsection{Quasi-circular orbits}

In the previous sections, we modeled the binary evolution assuming a static DM density profile around the IMBH. However, a more realistic description should account for the fact that the DM distribution itself may respond to the evolution of the binary. As the compact object inspirals, the surrounding DM environment is continuously perturbed by energy and angular momentum exchange. This modifies the local DM distribution and, in turn, feeds back into the orbital evolution.

For quasi-circular binaries, such coupled evolution has been discussed in Ref. \cite{PhysRevD.105.043009}, where the dominant mechanisms are GW backreaction and DF exerted by the DM overdensity surrounding the secondary object. Motivated by the qualitative behaviour identified there (see Sec. IV of Ref. \cite{PhysRevD.105.043009}), we incorporate an empirical effective density profile (EDP) to approximately capture the time-dependent response of the DM minispike during inspiral.

Physically, the evolution of the DM profile can be understood as follows. At large orbital separations, the secondary black hole moves slowly enough that DF continuously redistributes nearby DM particles, leading to a gradual depletion and flattening of the initially steep minispike profile. As the binary shrinks and GW emission becomes increasingly efficient, the inspiral timescale becomes shorter than the environmental response timescale. Consequently, the DM distribution has less time to rearrange, and the density profile approaches its original minispike form near the final stages of inspiral. The EDP introduced below is intended to reproduce this qualitative behaviour rather than provide a self-consistent dynamical evolution.

Accordingly, we consider a quasi-circular binary evolving inside an initially static DM minispike and allow the profile to evolve empirically during the inspiral. The binary reaches a characteristic radius $r_b$, associated with the dephasing break frequency $f_b$ introduced in Ref. \cite{PhysRevD.105.043009}. We identify this radius with the critical radius $p_c$ discussed in Ref. \cite{Dai:2021olt}, where the eccentricity growth induced by DF balances the circularizing effect of GW radiation reaction.

Motivated by the behaviour reported in Ref. \cite{PhysRevD.105.043009}, we assume that at separations larger than the break scale ($r\geq r_b$), the effective DM profile becomes steeper and follows
\begin{equation}
    \rho_{\textup{DM}} (r) \propto r^{-(\alpha + \gamma)},
\end{equation}
where $\gamma$ depends only weakly on the initial DM parameters. In contrast, at sufficiently small separations ($r \ll r_b$), the inspiral becomes dominated by GW backreaction and the effective profile gradually returns toward the initial minispike scaling,
\begin{equation}
    \rho_{\textup{DM}} (r, t=0) \propto r^{- \alpha }.
\end{equation}
To phenomenologically model this transition, we adopt the following effective density profile \cite{PhysRevD.105.043009}:
\begin{equation}
\rho_{\textrm{DM}} (r) = \rho_{\textup{sp}} \Big( \frac{r_{\textup{sp}}}{r} \Big)^{\alpha + \gamma (\kappa)}, \, \quad r_{\textrm{ISCO}} < r < r_{\textup{sp}}\, ,
\label{eq:evolving_density}
\end{equation}
where the correction exponent $\gamma$ evolves through the parameter $\kappa$.
We choose $\gamma (\kappa )$ so that it vanishes near the beginning of the inspiral, reaches its maximum close to $r=r_b$, and decreases again at small separations. This behaviour is modeled empirically as
\begin{equation}
\gamma ( \kappa ) = (2.5 - \alpha) \, e^{-( \kappa (r) - \kappa_{\textrm{max}} )^2},
\label{eq:gamma_kappa}
\end{equation}
with
\begin{equation}
\kappa ( r ) = \kappa_f \frac{ ( r - r_{\textrm{init}} )}{(r_{\textrm{ISCO}} -r_{\textrm{init}})}\,,
\label{eq:kappa}
\end{equation}\\
where $\kappa_{\textup{max}}$ corresponds to the value of $\kappa$ at $r=r_b$. The prefactor $(2.5-\alpha)$ ensures that the effective slope remains within the expected minispike range $2.25 \leq \alpha \leq 2.5$ \cite{Gondolo:1999ef}. We set $\kappa_f = 10$, although any sufficiently large positive value produces qualitatively similar behaviour and ensures recovery of the original minispike profile at small separations.

Since we are focusing on quasi-circular binaries, i.e., binaries with eccentricity $e\sim 0$, we can have an estimate by analytically expanding our previously obtained complicated expressions in a small eccentricity expansion (hence, the word ``quasi-circular''). With this model of evolving DM density, which depends on binary separation, the perturbed binary equations for a quasi-circular orbit look like:
\begin{equation}
\ddot{\vec{r}} = - \frac{M}{r^2} \hat{n} + \vec{f}_{\textrm{DF}} + \vec{f}_{\textrm{2.5PN}}\,.
\label{eq:quasicircular_eom}
\end{equation}

This equation can again be resolved into $r$ and $\phi$ motions:
\begin{equation}
\begin{aligned}
    & \ddot{r}=r\dot{\phi}^{2}-\frac{M}{r^{2}} - \frac{4\pi m_2 \rho_{\textrm{DM}} I_v {\color{black} \xi(v)} \dot{r}}{v^3} - \frac{M}{r^2} \Big[ A_{2.5\text{PN}} 
    + \dot{r} B_{2.5\text{PN}} \Big]\,, \\ & 
    \ddot{\phi}=-\frac{2\dot{r}\dot{\phi}}{r} - \frac{4\pi m_2 \rho_{\textrm{DM}} I_v {\color{black} \xi(v)} \dot{\phi}}{v^3} - \frac{M}{r} \Big[ B_{2.5\text{PN}} \Big] \dot{\phi}\,.
    \label{eq:quasicircular_radial_and_angular_eom}
\end{aligned}
\end{equation}

Using the above equation of motion and following a similar procedure as in the elliptical case, we can express the first time derivative of the nonlinear memory mode $h_{lm}^{\textup{(mem)(1)}}$. To compute the nonlinear memory, we integrate over eccentricity as given in Eq. (\ref{eq:elliptical_nonlinear_memo}) which requires the evolution of the orbital parameters ($e(f), p(f), \omega (f), t(f)$) (we obtain these by solving the averaged osculating equations for a quasi-circular binary). While writing the instantaneous osculating equations we perform a small eccentricity expansion keeping terms $\sim \mathcal{O}(e^4)$ followed by an orbital averaging over the whole time period of motion of the binary to obtain the averaged osculating equation\footnote{For $\langle \tfrac{de}{df} \rangle_{\text{quasi}}$, the next higher-order term appears at $e^5$.}:
\begin{widetext}
\begin{equation}
\begin{aligned}
% \Big\langle \frac{de}{df} \Big\rangle_{\textup{quasi}}
% = {} &
% - \frac{\eta}{15} \left(\frac{M}{p}\right)^{5/2}
% e \left(304 + 121 e^2\right) + \frac{\pi I_v m_2 \rho_{sp}}{2 M^2}
% \left(\frac{r_{sp}}{p}\right)^{\alpha}
% \Bigg\{
% 4 e
% \Big[
% (\alpha + 3) M^3
% + (\alpha + 1) M^2 p
% - (\alpha - 1) M p^2 \\ &\quad
% - (\alpha - 3) p^3
% \Big] - e^3
% \Big[
% (\alpha + 3)(\alpha^2 + \alpha + 3) M^3 + (\alpha + 1)(\alpha^2 - 3\alpha + 3) M^2 p - (\alpha - 1)(\alpha^2 - 7\alpha + 11) M p^2 \\ &\quad
%  - (\alpha - 3)(\alpha^2 - 11\alpha + 27) p^3
% \Big]
% \Bigg\},
\Big\langle \frac{de}{df} \Big\rangle_{\textup{quasi}}
= {} &
- \frac{\eta}{15} \left(\frac{M}{p}\right)^{5/2}
e \left(304 + 121 e^2\right) + \frac{\pi I_v m_2 \rho_{sp} p^2}{48 M^2 (M-p)^6}
\left(\frac{r_{sp}}{p}\right)^{\alpha}
\Bigg\{
e \mathcal{B}_1(M,p, \alpha) + e^3 \mathcal{B}_2(M,p, \alpha)
\Bigg\},
\\[1em]
\Big\langle \frac{dp}{df} \Big\rangle_{\textup{quasi}}
= {} &
- \frac{8\eta}{5}
\left(\frac{M}{p}\right)^{5/2}
p \left(8 + 7 e^2\right) + \frac{\pi I_v m_2 p \rho_{sp} p^3}{8 M^2 (M-p)^5}
\left(\frac{r_{sp}}{p}\right)^{\alpha}
\Bigg\{
\mathcal{A}_1(M,p)
+ e^2 \,
\mathcal{A}_2(M,p,\alpha) - e^4 \,
\mathcal{A}_3(M,p,\alpha)
\Bigg\},
\\[0.6em]
% \Big\langle \frac{dp}{df} \Big\rangle_{\textup{quasi}}
% = {} &
% - \frac{8\eta}{5}
% \left(\frac{M}{p}\right)^{5/2}
% p \left(8 + 7 e^2\right) + \frac{\pi I_v m_2 p \rho_{sp}}{8 M^2}
% \left(\frac{r_{sp}}{p}\right)^{\alpha}
% \Bigg\{
% \mathcal{A}_1(M,p)
% + e^2 \,
% \mathcal{A}_2(M,p,\alpha) - e^4 \,
% \mathcal{A}_3(M,p,\alpha)
% \Bigg\},
% \\[0.6em]
%
\Big\langle \frac{d\omega}{df} \Big\rangle_{\textup{quasi}}
= {} & 0,
\\[0.6em]
\Big\langle \frac{dt}{df} \Big\rangle_{\textup{quasi}}
= {} &
\left(\frac{p}{M}\right)^{1/2}
\frac{p}{4}
\left(8 + 12 e^2 + 15 e^4\right),
\label{eq:quasicircular_avg_osculating_eqns}
\end{aligned}
\end{equation}
where 
% \begin{align}
% \mathcal{A}_3 &=4 M^3 p^3 \left(-1359 + \alpha \left(344 + (\alpha - 17)(\alpha - 5)\alpha \right)\right)
% - M^4 p^2 \left(765 + \alpha \left(-2544 + \alpha \left(821 + (\alpha - 74)\alpha \right)\right)\right) \notag\\ &\quad
% - 2 M p^5 \left(3009 + \alpha \left(-1960 + \alpha \left(437 + (\alpha - 38)\alpha \right)\right)\right)
% + p^6 \left(1389 + \alpha \left(-976 + \alpha \left(245 + (\alpha - 26)\alpha \right)\right)\right) \notag\\ &\quad
% + M^6 \left(309 + \alpha \left(-320 + \alpha \left(117 + (\alpha - 18)\alpha \right)\right)\right)
% - 2 M^5 p \left(-1311 + \alpha \left(669 + \alpha \left(-75 + (\alpha - 6)\alpha \right)\right)\right) \notag\\ &\quad
% - M^2 p^4 \left(-9435 + \alpha \left(5216 + \alpha \left(-843 + \alpha (30 + \alpha)\right)\right)\right)
% \end{align}
\begin{align}
\mathcal{B}_1
&= -192 M^2 p^5 \left(\alpha - 23\right)
   + 576 M p^6 \left(\alpha - 5\right)
   - 192 p^7 \left(\alpha - 3\right)
   + 192 M^7 \left(\alpha - 1\right)
   - 576 M^6 p \left(\alpha + 1\right) \notag\\
&\quad
   + 192 M^5 p^2 \left(\alpha + 19\right)
   + 192 M^4 p^3 \left(5\alpha - 23\right)
   - 192 M^3 p^4 \left(5\alpha + 3\right),
\\[0.4em]
\mathcal{B}_2
&= -24 p^7 \left(\alpha - 3\right)
      \left(27 - 11\alpha + \alpha^2\right)
   + 24 M^7 \left(\alpha - 1\right)
      \left(11 - 7\alpha + \alpha^2\right) 
   - 24 M^2 p^5
      \left(-1005 + 576\alpha - 74\alpha^2 + \alpha^3\right) \notag \\
&\quad 
    + 72 M p^6 \left(-151 + 102\alpha - 20\alpha^2 + \alpha^3\right) - 72 M^6 p
      \left(59 - 24\alpha - 2\alpha^2 + \alpha^3\right) + 24 M^5 p^2 \left(273 - 306\alpha + 52\alpha^2 + \alpha^3\right) \notag\\
&\quad
   + 24 M^4 p^3
      \left(371 + 180\alpha - 94\alpha^2 + 5\alpha^3\right)
   - 24 M^3 p^4
      \left(1089 - 360\alpha - 16\alpha^2 + 5\alpha^3\right),
\\[0.4em]
\mathcal{A}_1 &=64 M^6 - 128 M^5 p - 64 M^4 p^2 + 256 M^3 p^3 - 64 M^2 p^4 - 128 M p^5 + 64 p^6,
\\[0.4em]
\mathcal{A}_2
&= -16 M^4 p^2
   \left(91 - 35\alpha + \alpha^2\right)
   - 32 M p^5
   \left(51 - 17\alpha + \alpha^2\right)
   + 16 p^6   \left(27 - 11\alpha + \alpha^2\right) + 64 M^3 p^3
   \left(3 - 9\alpha + \alpha^2\right) \notag\\
&\quad
   + 16 M^6
   \left(11 - 7\alpha + \alpha^2\right)
   - 32 M^5 p
   \left(-13 - \alpha + \alpha^2\right)
   - 16 M^2 p^4
   \left(-117 + 17\alpha + \alpha^2\right),
\\[0.4em]
\mathcal{A}_3
&= 4 M^3 p^3
    \left[-1359+ \alpha \left\{344 + (\alpha - 17)(\alpha - 5)\alpha \right\} \right] - M^4 p^2 \left[765 + \alpha \left\{-2544 + \alpha \left[ 821 + (\alpha - 74)\alpha \right] \right\} \right] \notag\\
&\quad
    - 2 M p^5 \left[3009 + \alpha \left\{-1960 + \alpha \left[ 437 + (\alpha - 38)\alpha \right] \right\} \right] + p^6 \left[ 1389+ \alpha \left\{ -976
    + \alpha \left[ 245 + (\alpha - 26)\alpha \right] \right\} \right] \notag\\
&\quad
    + M^6 \left[309 + \alpha \left\{ -320
    + \alpha \left[ 117 + (\alpha - 18)\alpha \right] \right\} \right] - 2 M^5 p \left[-1311 + \alpha \left\{ 669 + \alpha \left[-75 + (\alpha - 6)\alpha \right]\right\}\right] \notag\\
&\quad
    - M^2 p^4 \left[-9435 + \alpha \left\{
    5216 + \alpha \left[-843 + \alpha(30 + \alpha)\right]\right\}\right].
\end{align}
\end{widetext}

Solving this set of differential equations numerically gives the evolution of orbital parameters, which we use to solve the integral in Eq. (\ref{eq:elliptical_nonlinear_memo}) to obtain nonlinear memory for a quasi-circular orbit in an evolving DM environment. \textcolor{black}{In the next section, we will conduct a comparative study of the nonlinear memory embedded in this evolving DM minispike and static profiles with vacuum for quasi-circular orbits.}

%%%%%%%%%%%%%%%%%%%%%%%%%%%%%%%%%%%%%%%%%%%%%%%%%%%%%%%%%%%%%%%%%%

\section{Numerical results and Detection prospects}\label{sec3}

In this section, we present our results and estimate the prospects for the detection of nonlinear GW memory {\color{black}using the future proposed space-based detector LISA}. We plot the nonlinear memory as a function of eccentricity for an elliptical orbit, first considering only the GW backreaction (only 2.5PN term) for the vacuum case (without DM) and then the net effect (with DM), including DM gravity, DF, accretion, and GW backreaction. We also plot the nonlinear memory as a function of true anomaly for a hyperbolic orbit, considering PN correction terms up to 2.5PN (without DM) and the net effect. Both plots highlight the hereditary nature of nonlinear memory build-up. {\color{black}Note that we consider a binary located at a distance of $R=10$ Mpc.}

Further, to assess detectability, we numerically computed the mismatch for LISA detector and found that the signals are detectable within the present approximations for elliptical binaries. However, distinguishing whether a detected signal originates from a binary evolving in vacuum or within a DM environment is expected to be highly challenging. Also, the waveform mismatch quantifies the impact of memory on signals from binaries embedded in a DM environment around an IMBH for different values of DM spike index ($\alpha$). This analysis was carried out using the stationary phase approximation (SPA) for waveforms from low-eccentricity elliptical binaries \cite{Cutler:1994ys, Droz:1999qx, Damour:2000gg}.

\subsection{Elliptical orbits}

{ \color{black}

Before presenting the results for elliptical orbits, we emphasize that two distinct mechanisms influence the buildup of nonlinear memory. First, it is directly affected by additional terms arising from DM environmental effects in the nonlinear memory expression [Eq. (\ref{eq:hlm_dot_elliptical})]. Second, it is indirectly affected through the total inspiral time of the binary system. Both mechanisms act to enhance the nonlinear memory. However, increasing the DM parameter $\alpha$ simultaneously shortens the inspiral time as shown in the Table~(\ref{table:binary_inspiral_time}) while amplifying the instantaneous memory contribution. As a result, the nonlinear memory increases only up to certain values of $\alpha$ and the initial binary parameters. Beyond this regime, even though a larger $\alpha$ tends to strengthen the memory contribution, the reduced inspiral time prevents the total accumulated memory from exceeding that of the vacuum case. Note that, for most of the inspiral cases considered in this work, we focus on elliptical binaries whose orbital evolution requires more than one year to reach the LSO, considering the long observation period of LISA.

\begin{table}
\centering
\setlength{\tabcolsep}{6pt}
\renewcommand{\arraystretch}{1.3}

\caption{Total inspiral time (in years) for different values of initial eccentricities $e_0$ and semi-latus rectum $p_0$ in vacuum and DM spike ($\alpha$) environments. All distances are expressed in units of the ISCO scale of the IMBH.}
\label{tab:inspiral_time}

%==================== (a) Vacuum ====================

% \centering
% \begin{tabular}{|c|c|c|c|c|c|}
% \multicolumn{6}{c}{Vacuum} \\ \hline
% $e_0 \backslash p_0$ & 50 & 70 & 100 & 120 & 150 \\ \hline
% 0.1 & 1.2 & 4.7 & 19.7 & 40.9 & 99.7 \\ \hline
% 0.5 & 1.4 & 5.3 & 22.2 & 46.0 & 112.4 \\ \hline
% 0.9 & 3.2 & 12.3 & 51.3 & 106.4 & 259.8 \\ \hline 
% \noalign{\vskip 6pt}
% \multicolumn{6}{c}{DM ($\alpha = 2.25$)} \\ \hline
% $e_0 \backslash p_0$ & 50 & 70 & 100 & 120 & 150 \\ \hline
% 0.1 & 3.8 & 10.6 & 24.8 & 35.3 & 51.1 \\ \hline
% 0.5 & 4.2 & 11.2 & 25.1 & 35.0 & 49.8 \\ \hline
% 0.9 & 6.9 & 14.6 & 26.3 & 33.4 & 43.3 \\ \hline
% \noalign{\vskip 6pt}
% \multicolumn{6}{c}{DM ($\alpha = 2.35$)} \\ \hline
% $e_0 \backslash p_0$ & 50 & 70 & 100 & 120 & 150 \\ \hline
% 0.1 & 2.0 & 3.9 & 7.0 & 9.1 & 12.1 \\ \hline
% 0.5 & 2.1 & 4.0 & 7.0 & 9.0 & 11.9 \\ \hline
% 0.9 & 2.6 & 4.4 & 7.0 & 8.6 & 10.8 \\ \hline
% \noalign{\vskip 6pt}
% \multicolumn{6}{c}{DM ($\alpha = 2.5$)} \\ \hline
% $e_0 \backslash p_0$ & 50 & 70 & 100 & 120 & 150 \\ \hline
% 0.1 & 0.3 & 0.4 & 0.7 & 0.8 & 1.1 \\ \hline
% 0.5 & 0.3 & 0.4 & 0.7 & 0.8 & 1.1 \\ \hline
% 0.9 & 0.3 & 0.5 & 0.8 & 1.0 & 1.2 \\ \hline
% \end{tabular}
% \label{table:binary_inspiral_time}
% \end{table}
\centering
\begin{tabular}{|c|c|c|c|c|c|}
\multicolumn{6}{c}{Vacuum} \\ \hline
$e_0 \backslash p_0$ & 50 & 70 & 100 & 120 & 150 \\ \hline
0.1 &2.4 & 9.4 & 39.2  & 81.3 & 198.5 \\ \hline
0.5 &2.8 & 10.6 & 44.2 & 91.6 & 223.7 \\ \hline
0.9 &6.4 & 24.5 & 102.1 & 211.3 & 517.1 \\ \hline 
\noalign{\vskip 6pt}
\multicolumn{6}{c}{DM ($\alpha = 2.25$)} \\ \hline
$e_0 \backslash p_0$ & 50 & 70 & 100 & 120 & 150 \\ \hline
0.1 & 6.7 & 17.0 & 36.0 & 49.2 & 68.9 \\ \hline
0.5 & 7.2 & 17.6 & 35.9 & 48.3 & 66.3 \\ \hline
0.9 & 10.7 & 20.6 & 34.6 & 43.0 & 54.7 \\ \hline
\noalign{\vskip 6pt}
\multicolumn{6}{c}{DM ($\alpha = 2.35$)} \\ \hline
$e_0 \backslash p_0$ & 50 & 70 & 100 & 120 & 150 \\ \hline
0.1 & 3.0 & 5.7 & 9.8 & 12.5 & 16.4 \\ \hline
0.5 & 3.2 & 5.8 & 9.7 & 12.3 & 16.1 \\ \hline
0.9 & 3.8 & 6.1 & 9.3 & 11.3 & 14.2 \\ \hline
\noalign{\vskip 6pt}
\multicolumn{6}{c}{DM ($\alpha = 2.5$)} \\ \hline
$e_0 \backslash p_0$ & 50 & 70 & 100 & 120 & 150 \\ \hline
0.1 & 0.4 & 0.6 & 1.0 & 1.2 & 1.5 \\ \hline
0.5 & 0.4 & 0.6 & 1.0 & 1.2 & 1.6 \\ \hline
0.9 & 0.5 & 0.7 & 1.1 & 1.4 & 1.7 \\ \hline
\end{tabular}
\label{table:binary_inspiral_time}
\end{table}

\begin{figure}
    \centering
    \begin{subfigure}{\linewidth}
        \includegraphics[width=\linewidth]{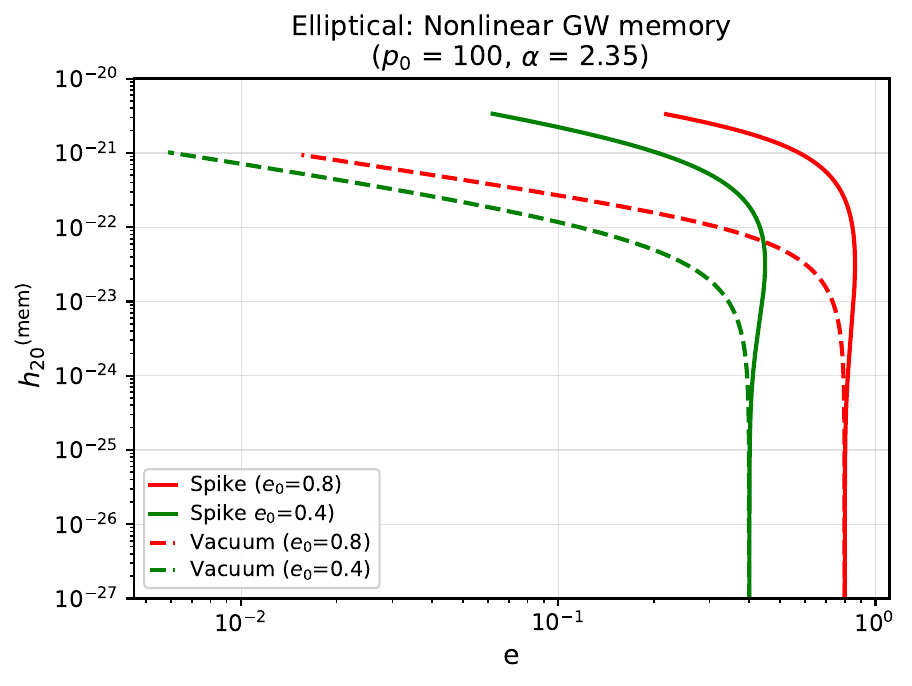}
        \subcaption{}
        \label{fig:elliptical_spike_plot_3}
    \end{subfigure}

    \begin{subfigure}{\linewidth}
        \includegraphics[width=\linewidth]{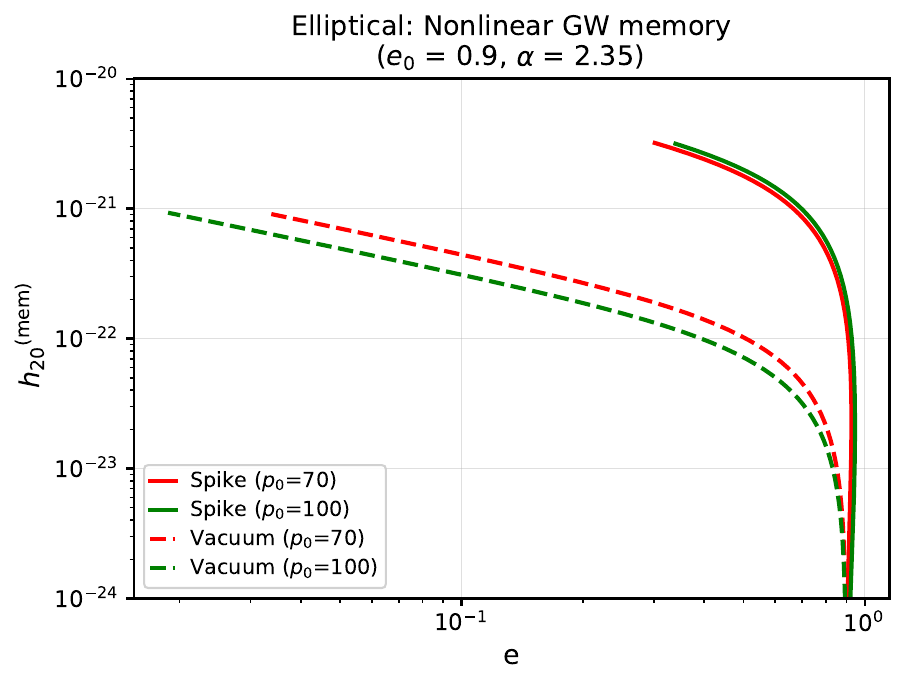}
        \subcaption{}
        \label{fig:elliptical_spike_plot_5}
    \end{subfigure}

    \begin{subfigure}{\linewidth}
        \includegraphics[width=\linewidth]{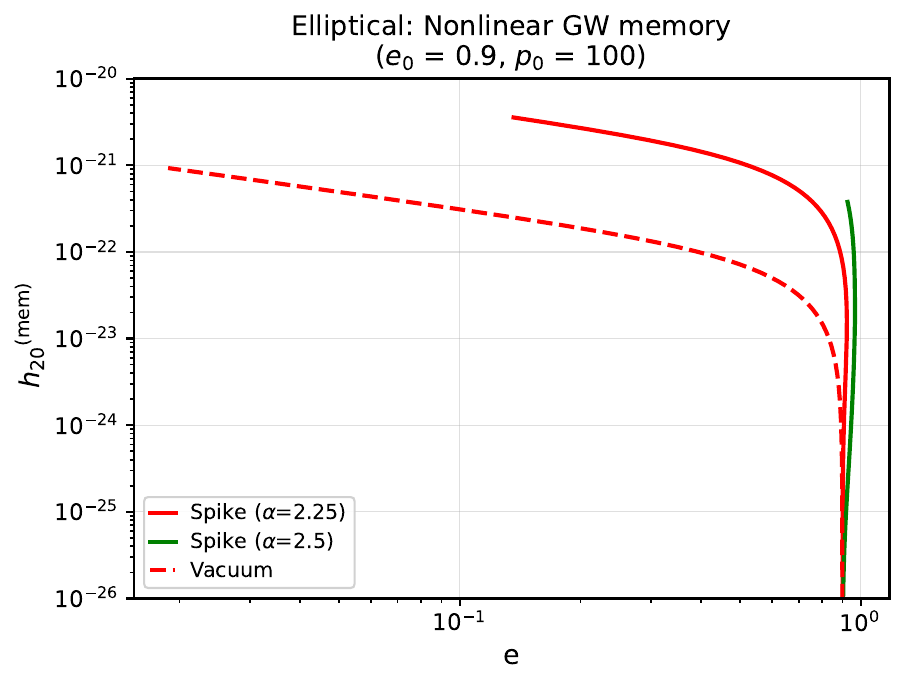}
        \subcaption{}
        \label{fig:elliptical_spike_plot_7}
    \end{subfigure}

    \caption{The above plots show the $h_{20}^{\textup{(mem)}}$ as functions of \textcolor{black}{eccentricity} with different values of initial eccentricities $e_0$, initial semi-latus rectum $p_0$ and DM spike exponent $\alpha$ in panels (a), (b), and (c), respectively, for elliptical binaries. We have considered binaries of masses $m_1 = 10^3 M_{\odot}, \, m_2 = 10 M_{\odot}$, which are at a distance $R=10$ Mpc away from observation point.}
    \label{fig:elliptical_spike_plot}
\end{figure}

\subsubsection{Memory evolution vs eccentricity}

For elliptical orbits, we compute the dominant nonlinear memory modes \( h_{20}^{\textup{(mem)}} \) and \( h_{40}^{\textup{(mem)}} \) as functions of the orbital eccentricity. The results for $h_{20}^{\textup{(mem)}}$ mode are shown in Figs.~(\ref{fig:elliptical_spike_plot_3}), (\ref{fig:elliptical_spike_plot_5}), and (\ref{fig:elliptical_spike_plot_7}), corresponding to different  values of initial eccentricity \(e_0\), the initial semi-latus rectum \(p_0\), and the DM spike exponent \(\alpha\), respectively. The $h_{40}^{\textup{(mem)}}$ mode shows a similar dependence as the 20 mode; however, its overall amplitude is additionally suppressed by roughly $\mathcal{O}(10^{-2})$ compared to 20 mode. As illustrated in Fig.~(\ref{fig:elliptical_spike_plot_3}), a larger initial eccentricity leads to a slightly enhanced nonlinear memory, consistent with the hereditary nature of the memory effect. Fig.~(\ref{fig:elliptical_spike_plot_5}) demonstrates that increasing the initial semi-latus rectum \(p_0\) results in a marginally larger nonlinear memory amplitude and a longer inspiral duration prior to reaching the LSO. Finally, Fig.~(\ref{fig:elliptical_spike_plot_7}) shows the dependence of nonlinear memory on the DM spike exponent \(\alpha\): a larger value, \(\alpha = 2.5\), leads to more rapid orbital evolution and reduced time for memory accumulation, causing earlier truncation of the memory signal, whereas a smaller value, \(\alpha = 2.25\), allows for a slower inspiral and consequently a greater overall buildup of nonlinear memory.

\subsubsection{Mismatch Computation: Nonlinear memory and DM around IMBH}

To observe the connection between detection of nonlinear memory and DM medium around an IMBH, we compute the mismatch between waveforms generated in a DM spike environment with and without memory. The analysis is done in two ways: (i) the mismatch accumulated month-by-month during the final year prior to the LSO, shown in Fig.~(\ref{fig:elliptical_mismatch_plot_1}), and (ii) the mismatch of the full accumulated signal as a function of the DM spike exponent $\alpha$, shown in Fig.~(\ref{fig:elliptical_mismatch_plot_2}). To find mismatch between signals $h_1$ and $h_2$, we use:
\begin{equation}
    \mathcal{M} = 1 - \mathcal{O} (h_1, h_2)\,,
\end{equation}
where $\mathcal{O} (h_1, h_2)$ is the overlap between the signals $h_1$ and $h_2$, written as:
\begin{equation}
    \mathcal{O} (h_1, h_2) = \frac{(h_1, h_2)}{\sqrt{(h_1, h_1) (h_2, h_2)}}\,,
\end{equation}
and
\begin{equation}
    (h_1, h_2) =  4 \mathcal{R} \text{e} \int_{{\nu}_{\text{initial}}}^{{\nu}_{\text{final}}} \frac{\tilde{h}_1({\nu}) \tilde{h}_2^{*}({\nu})  }{S_n ({\nu})} \, d{\nu}\,,
\end{equation}
where $\tilde{h}_1({\nu})$ and $\tilde{h}_2({\nu})$ are signals in frequency domain which are obtained using the SPA \cite{Cutler:1994ys, Droz:1999qx, Damour:2000gg} for low eccentricity elliptical orbit, $\nu$ is the instantaneous GW frequency, and $S_n({\nu})$ is the LISA noise PSD \cite{PhysRevD.84.064023},
\begin{equation}
\begin{aligned}
     S_{n} (\nu) = \text{min} \big[ & \frac{S_n^{\text{inst}} (\nu)}{\text{exp}(-\beta T^{-1}_{\text{mission}} dN/d\nu)} + S_n^{\text{exgal}} (\nu)\,, \\& S_n^{\text{inst}} (\nu) + S_n^{\text{gal}} (\nu) + S_n^{\text{exgal}} (\nu) \big]\,, 
     \label{eq:lisa_noise_rewritten}
\end{aligned}
\end{equation}
where
\begin{equation}
    \begin{aligned}
    S_n^{\text{inst}} (\nu) &= 9.18 \times 10^{-52} \nu^{-4} + 1.59 \times 10^{-42} + \notag\\
    &\quad 9.18 \times 10^{-38} \nu^2 \, \text{Hz}^{-1}, \\
    S_n^{\text{gal}} (\nu) &= 2.1 \times 10^{-45} \big( \frac{\nu}{1\textup{Hz}} \big)^{-7/3} \,  \text{Hz}^{-1},  \\
    S_n^{\text{exgal}} (\nu) &= 4.2 \times 10^{-47} \big( \frac{\nu}{1\textup{Hz}}  \big)^{-7/3} \,  \text{Hz}^{-1},
    \end{aligned}
\end{equation}
and $ \frac{dN}{d\nu} = 2 \times 10^{-3} \big( \frac{1 \text{Hz}}{\nu} \big)^{11/3} $ and $ \beta T^{-1}_{\text{mission}} = 1.5 / \text{yr}$ are taken from \cite{PhysRevD.84.064023}.

\begin{figure}
    \centering
    \begin{subfigure}{\linewidth}
        \includegraphics[width=\linewidth]{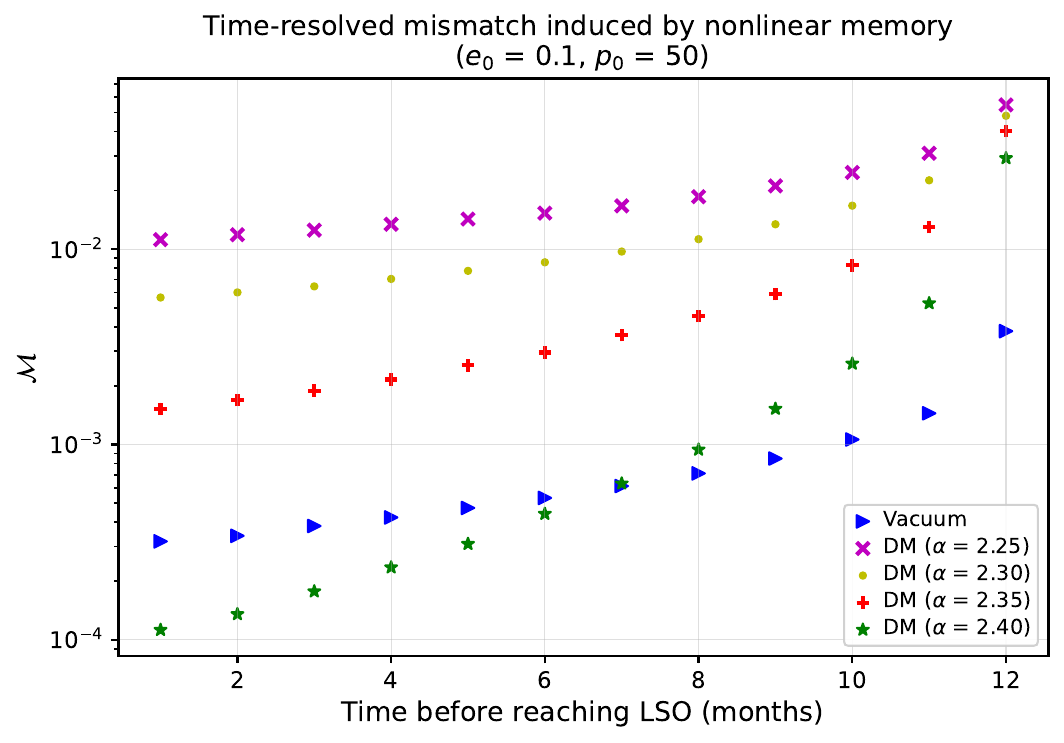}
        \subcaption{}
        \label{fig:elliptical_mismatch_plot_1}
    \end{subfigure}

    \begin{subfigure}{\linewidth}
        \includegraphics[width=\linewidth]{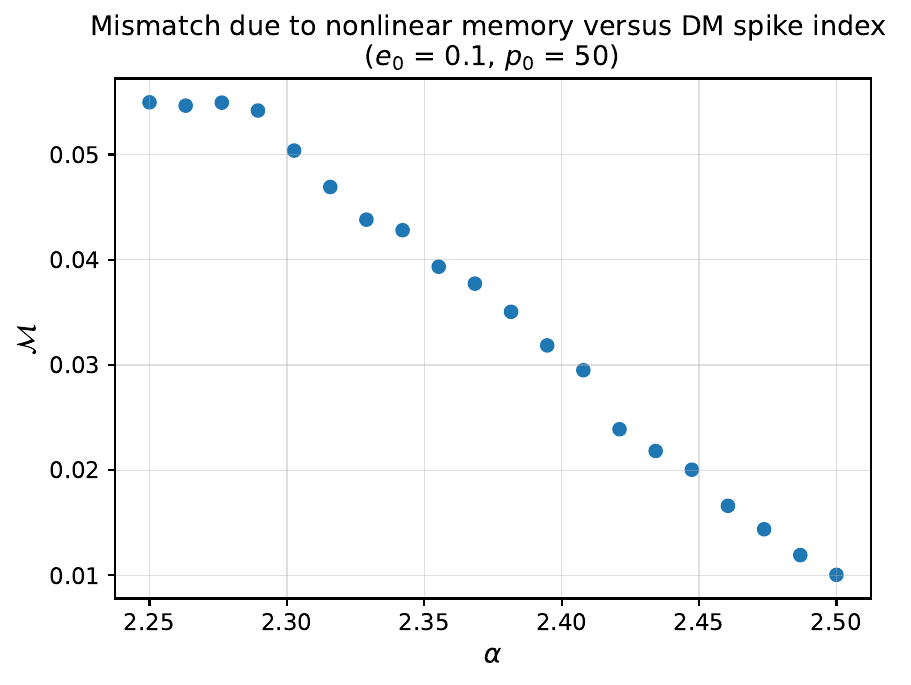}
        \subcaption{}
        \label{fig:elliptical_mismatch_plot_2}
    \end{subfigure}
    \caption{Mismatch induced by nonlinear memory in elliptical binaries evolving in vacuum and in a DM environment. In panel (a), we show the time-resolved mismatch between waveforms computed with and without nonlinear memory, evaluated over successive monthly segments during the final year of evolution prior to the LSO. In panel (b), we show the mismatch computed using the full accumulated signal, as a function of the DM spike exponent $\alpha$. In all cases, the waveforms are constructed using the SPA for low-eccentricity orbits, and the comparison isolates the impact of nonlinear memory by toggling its contribution while keeping all other parameters fixed. Results are shown for binaries with $e_0=0.1$ and $p_0=50$.}
    \label{fig:elliptical_mismatch_plot}
\end{figure}

{\color{black} To obtain signals, we used the GW polarization mode decomposition of Eq.~(\ref{eq:gw_polarization_decomposition}), truncated at $\ell=2$:
\begin{equation}
\begin{aligned}
h_{+} - i h_{\times}
&\approx \sum_{m=-2}^{2} H_{2m}\, Y_{2m}(\Theta,\Phi)\,,  \\
&= H_{20} Y_{20} + H_{2+2} Y_{2+2} + H_{2-2} Y_{2-2}\,,
\end{aligned}
\label{eq:hplus_hcross_rewritten}
\end{equation}
where $H_{lm}=h^{\mathrm{N}}_{lm}$ (excluding memory) or $H_{lm}=h_{lm}$ (including memory), as defined in Eq.~(\ref{eq:hlm_mode_in_terms_of_mass_quad_moments_and_memory}), $\{2\pm 1\}$ modes satisfy $h^{\textup{N}}_{2\pm 1} = 0 \implies H_{2\pm 1}=0$ from Eq.~(\ref{eq:vanished_leading_order_hlm_modes}) and $Y_{lm}$ denote spin-weighted spherical harmonics, $\Theta=\pi/3$ is the inclination angle (defined as the angle between the orbital angular momentum (the z-axis of our coordinate system) and the line of sight to the observer) taken, and $\Phi$ is the coalescence phase. Using these GW polarizations and SPA analysis, we construct our signal in frequency domain for low eccentric orbit for binaries embedded in vacuum and DM.

{\color{black}We also calculate the minimum mismatch required for the distinguishability \cite{PhysRevD.78.124020} of eccentric binary signals in the different scenario mentioned in Table~(\ref{tab:snr_mismatch}). For this, we first calculate signal-to-noise ($\rho$) using \cite{PhysRevD.84.064023}:
\begin{equation}
\rho = \sqrt{ 2 \sum_{\alpha = I, II} \int_{t_{\text{init}}}^{t_{\text{LSO}}} \frac{h_{\alpha}^2 (t)}{S_{n} (\nu(t))}\, dt}
\end{equation}
where $\alpha = I, II$ are two different LISA interferometers and $h_{\alpha} (t) $ are LISA response given as \cite{PhysRevD.84.064023}:
\begin{equation}
h_{\alpha}(t)= h_{+} (t) F_{\alpha}^{+} (t) + h_{\times}(t) F_{\alpha}^{\times} (t)
\end{equation}
where $h_{+, \times} (t)$ and $F_{\alpha}^{+, \times} (t)$ are time-domain GW polarizations and LISA antenna-pattern functions \cite{PhysRevD.84.064023} respectively.

To assess whether the memory-induced mismatches shown in Fig.~(\ref{fig:elliptical_mismatch_plot_1}) are astrophysically meaningful, we compare them against the standard distinguishability criterion for waveform mismatches \cite{PhysRevD.78.124020}.
Two waveforms differing by a mismatch $\mathcal{M} > \mathcal{M}_{\text{min}}$ are, in this sense, statistically distinguishable given the measurement precision afforded by that SNR. Across the full range of DM spike indices considered, $\mathcal{M}_{\text{min}}$ ranges from $\sim \, 6.40 \times 10^{-7}$
(vacuum, memory-ON) to $\sim \, 3.62 \times 10^{-5}$  (DM, $\alpha = 2.35$, memory-OFF). The memory-induced mismatches computed in Fig.~(\ref{fig:elliptical_mismatch_plot_1}) are reaching
$\mathcal{O}(10^{-2})$, exceeding $\mathcal{M}_{\text{min}}$ by 2–3 orders of magnitude in every case. This confirms that the inclusion of nonlinear memory leads to a substantial increase in the mismatch for binaries embedded in a DM spike, reaching values of order $\mathcal{O}(10^{-2})$ (see Fig.~(\ref{fig:elliptical_mismatch_plot_1})).}
The impact of nonlinear memory is more pronounced for smaller values of $\alpha$. This is because a smaller $\alpha$ results in a longer inspiral evolution before reaching the LSO, allowing greater accumulation of memory and hence a larger mismatch. In contrast, a larger $\alpha$ shortens the duration of the inspiral, reducing the buildup of memory and consequently lowering the mismatch.

% We find that the inclusion of nonlinear memory leads to a substantial increase in the mismatch for binaries embedded in a DM spike, reaching values of order $\mathcal{O}(10^{-2})$ (see Fig.~\ref{fig:elliptical_mismatch_plot_1}). The impact of nonlinear memory is more pronounced for smaller values of $\alpha$. This is because a smaller $\alpha$ results in a longer inspiral evolution before reaching the LSO, allowing greater accumulation of memory and hence a larger mismatch. In contrast, a larger $\alpha$ shortens the duration of the inspiral, reducing the buildup of memory and consequently lowering the mismatch.

Nevertheless, for a wide range of $\alpha$, the mismatch between signals with and without memory in a DM spike remains large enough to distinguish the memory effect. As shown in Fig.~(\ref{fig:elliptical_mismatch_plot_2}), the full-signal mismatch in the DM case decreases with increasing $\alpha$. This trend indicates that the contribution of nonlinear memory becomes weaker as $\alpha$ increases, or equivalently, that the influence of the DM environment on memory buildup diminishes for larger $\alpha$.}

% {\color{red}We note that $\mathcal{M}_{\text{min}}$ scales as $\sim 1/\rho^2 \sim R^2$ that is for sources at larger distances (lower SNR), a correspondingly larger memory-induced mismatch would be required for distinguishability. Given the several-order-of-magnitude margin found here, our qualitative conclusions should remain robust for IMRIs somewhat farther than our case R = 10 Mpc source, though a dedicated study of the detectability horizon is left to future work.}

{\color{black}We note that the $\mathcal{M}_{\text{min}}$, scales as $ \propto \rho^{-2} \propto R^{2}$ \cite{PhysRevD.78.124020, Cutler:1994ys}. Consequently, for sources located at larger distances (and hence with lower SNR), a correspondingly larger memory-induced mismatch is required for distinguishability. Nevertheless, given the several-orders-of-magnitude separation between the predicted mismatch and $\mathcal{M}_{\text{min}}$ in our analysis, we expect our qualitative conclusions to remain valid for IMRIs at distances somewhat beyond the fiducial case of $R = 10 \, \text{Mpc}$. A detailed investigation of the corresponding detectability horizon is deferred to future work.}

\begin{table}[ht]
\centering
\setlength{\tabcolsep}{6pt}
\renewcommand{\arraystretch}{1.3}

\caption{Minimum mismatch ($\mathcal{M}_{\text{min}}$) required for distinguishability of signal from eccentric binary ($e_0 = 0.1$, $p_0=50$ and R = 10 Mpc) in vacuum and different DM environments with and without memory.}
\label{tab:snr_mismatch}
\begin{tabular}{|c|c|c|c|} \hline
% \toprule
\textbf{Environment} & \textbf{Memory} & \textbf{SNR}($\rho$) & $\mathcal{M}_{\min}\ \sim \, \frac{1}{2 \rho^2}$ \\ \hline
% \midrule

\multirow{2}{*}{Vacuum}
& OFF & 881.01 & $6.44 \times 10^{-7}$ \\
& ON  & 884.20 & $6.40 \times 10^{-7}$ \\ \hline
% \midrule

\multirow{2}{*}{DM ($\alpha = 2.25$)}
& OFF & 249.31 & $8.04 \times 10^{-6}$ \\
& ON  & 283.00 & $6.24 \times 10^{-6}$ \\ \hline
% \midrule

\multirow{2}{*}{DM ($\alpha = 2.30$)}
& OFF & 155.80 & $2.06 \times 10^{-5}$ \\
& ON  & 174.99 & $1.63 \times 10^{-5}$ \\ \hline
% \midrule

\multirow{2}{*}{DM ($\alpha = 2.35$)}
& OFF & 117.48 & $3.62 \times 10^{-5}$ \\
& ON  & 121.83 & $3.37 \times 10^{-5}$ \\ \hline
% \midrule

\multirow{2}{*}{DM ($\alpha = 2.40$)}
& OFF & 277.92 & $6.47 \times 10^{-6}$ \\
& ON  & 303.60 & $5.42 \times 10^{-6}$ \\ \hline
% \bottomrule

\end{tabular}
\end{table}

\subsection{Quasi-circular orbits}

{\color{black}In Fig.~(\ref{fig:elliptical_spike_plot_9}), we show the dominant nonlinear memory mode $h_{20}^{\textup{(mem)}}$ from a quasi-circular binary as a function of eccentricity in the presence of static and dynamic DM spike environments.} In both cases, the memory amplitude is larger than in vacuum, with the dynamic DM case yielding slightly higher values than the static case. Moreover, the memory buildup in the dynamic case is truncated earlier, reflecting the faster orbital decay driven by DF in the evolving DM environment in addition to GW backreaction. The $h_{40}^{\textup{(mem)}}$ mode has a similar evolution with amplitude $\mathcal{O}(10^{-2})$ less than 20 modes.

One should note that for proper self-consistent treatment of the coupled evolution between the binary and the surrounding DM environment would require evolving the DM spike phase-space distribution function as the binary inspirals \cite{PhysRevD.105.043009}, rather than relying on an effective phenomenological description. Such an analysis involves significantly more sophisticated modeling and numerical treatment and therefore lies beyond the scope of the present work. In this paper, we therefore adopt an empirical prescription to capture the qualitative effect of DM density evolution and use it to study its impact on nonlinear memory.

{\color{black} 

\begin{figure}
    \includegraphics[width=\linewidth]{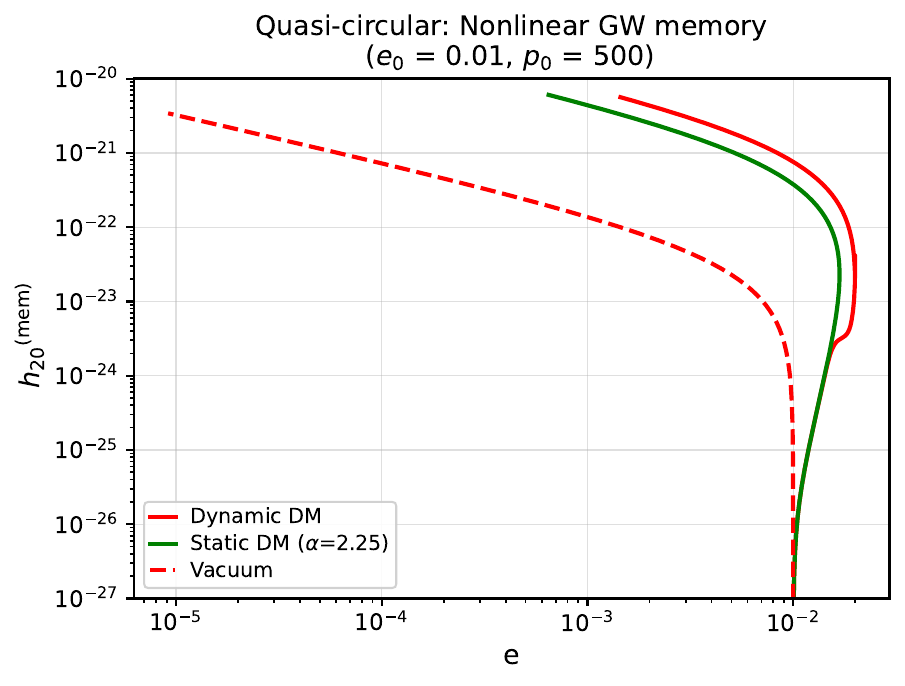}
    \caption{The above plot shows $h_{20}^{\textup{(mem)}}$ nonlinear memory modes as functions of eccentricity due to quasi-circular binaries. The dashed line represents the memory in vacuum, accounting only for GW backreaction. The solid red line includes the effects of a dynamically evolving DM spike through DF apart from GW backreaction, while the solid green line corresponds to a static DM spike profile.}
    \label{fig:elliptical_spike_plot_9}
\end{figure}

}

\subsection{Hyperbolic orbits}

{\color{black}

% In the section we present the results of nonlinear memory build up for unbounded (hyperbolic) binaries embedded in DM environment. First, we present the memory dependency for hyperbolic binaries on initial orbital parameters ($e_0, \, p_0$) and the DM spike exponent ($\alpha$) in Fig. \ref{fig:hyperbolic_spike_plot}. Then in Fig. \ref{fig:hyperbolic_compare_plot}, we showed the comparison of the effect of DM spike and NFW profile on nonlinear memory with respect to vacuum. In all cases, nonlinear memory is significantly enhanced due to DM environment compare to vacuum case however memory amplitude is quite low up to the order of $\mathcal{O} (10^{-26})$ and might be extremely challenging to detect even by future proposed space based detectors such as LISA and GWSat.

In this section, we present results for the nonlinear memory for unbounded (hyperbolic) binaries embedded in a DM environment. We examine the dependence of the memory amplitude on the initial orbital parameters ($e_0$, $p_0$) and the DM spike exponent $\alpha$, as shown in Fig.~(\ref{fig:hyperbolic_spike_plot1}) and (\ref{fig:hyperbolic_spike_plot2}). In all scenarios considered, the presence of a DM environment enhances the nonlinear memory compared to the vacuum. However, the memory amplitude remains small, at $\mathcal{O}(10^{-26})$, making its detection extremely challenging even for future space-based detectors such as LISA.

% We then compare the effects of a DM spike and an NFW profile relative to the vacuum case in Fig.~(\ref{fig:hyperbolic_compare_plot}). 
%\newpage
\subsubsection{Dependency on orbital parameters}

{\color{black}Fig.~(\ref{fig:hyperbolic_spike_plot1}) presents the dominant nonlinear memory mode $h_{20}^{\textup{(mem)}}$ as functions of the true anomaly for hyperbolic binaries evolving in a DM spike environment and in vacuum. We find that the nonlinear memory generated during hyperbolic encounters is strongly suppressed, reaching amplitudes of order $\mathcal{O} (10^{-24})$. The corresponding $h_{40}^{\textup{(mem)}}$ mode exhibits a similar growth behaviour as the 20 mode, although its amplitude remains suppressed by approximately $\mathcal{O}(10^{-2})$ relative to the 20 mode.} Such small amplitudes would be challenging to detect even with the proposed future GW detectors. Nevertheless, in all cases considered, the nonlinear memory in the presence of a DM spike remains larger than that obtained in the vacuum case.

Here, the \textit{vacuum case} accounts only for the PN correction terms, whereas the \textit{DM case} includes the environmental contributions from DF, accretion, and the gravitational potential of the DM distribution, together with PN corrections up to 2.5PN order.

We further observe a clear dependence on the initial orbital parameters, including the eccentricity $e_0$ and the semi-latus rectum $p_0$, which is related to the impact parameter of the hyperbolic orbit. As shown in Figs.~(\ref{fig:hyperbolic_spike_plot_6}) and (\ref{fig:hyperbolic_spike_plot_8}), larger $e_0$ leads to slightly enhanced nonlinear memory. In contrast, smaller values of $p_0$ produce a significantly larger memory signal compared to the corresponding vacuum case with identical orbital parameters.

\begin{figure}
    \centering
    \begin{subfigure}{\linewidth}
        \includegraphics[width=\linewidth]{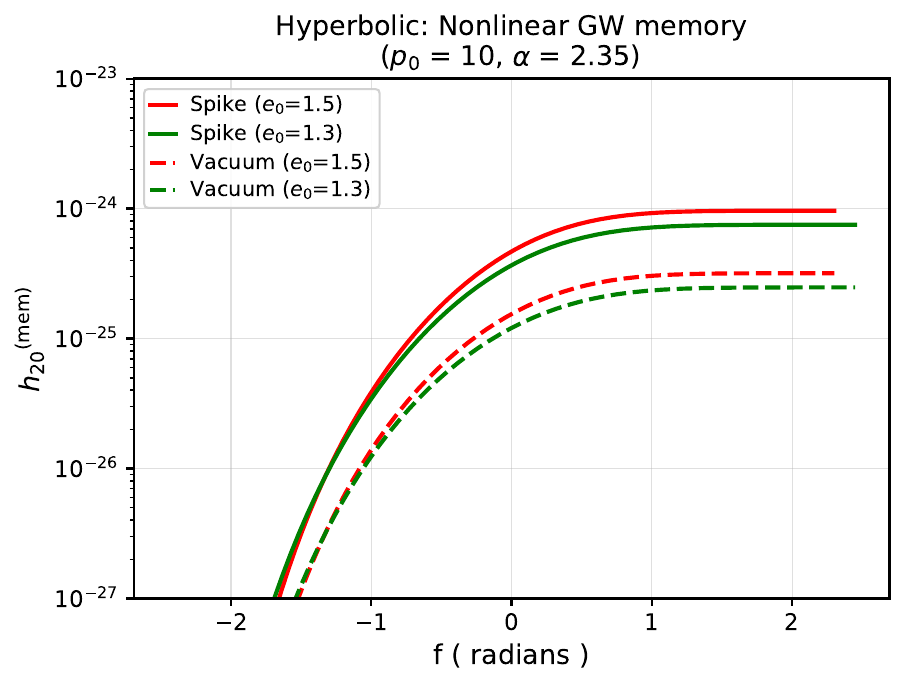}
        \subcaption{}
        \label{fig:hyperbolic_spike_plot_6}
    \end{subfigure}

    \begin{subfigure}{\linewidth}
        \includegraphics[width=\linewidth]{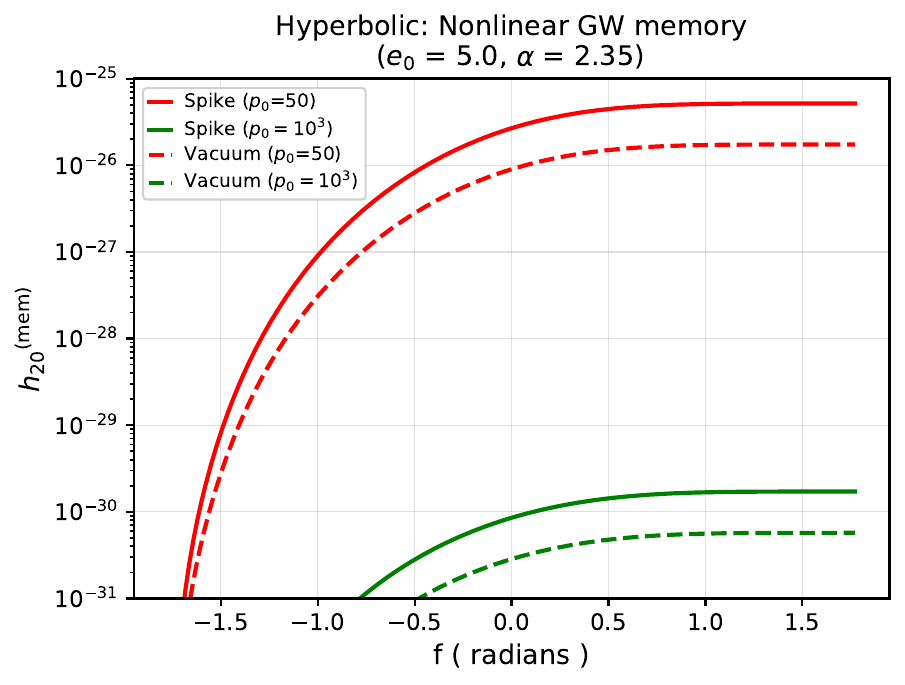}
        \subcaption{}
        \label{fig:hyperbolic_spike_plot_8}
    \end{subfigure}

    % \begin{subfigure}{\linewidth}
    %     \includegraphics[width=\linewidth]{V3_PRD_figures/hyperbolic_spike_plot_10.pdf}
    %     \subcaption{}
    %     \label{fig:hyperbolic_spike_plot_10}
    % \end{subfigure}

    \caption{The above plots show the $h_{20}^{\textup{(mem)}}$ as functions of true anomaly with different values of initial eccentricities $e_0$ and initial semi-latus rectum $p_0$ in panels (a) and (b) respectively for hyperbolic binaries.}
    \label{fig:hyperbolic_spike_plot1}
\end{figure}

\subsubsection{Dependence on the DM spike exponent}

{\color{black}Fig.~(\ref{fig:hyperbolic_spike_plot_10}) shows the dependence of the nonlinear memory on the DM density power-law exponent $\alpha$ for a hyperbolic binary while keeping $e_0$ and $p_0$ fixed. For the range of $\alpha$ values considered, the nonlinear memory exhibits only a weak sensitivity to the DM spike exponent. In particular, the difference between the memory amplitudes for $\alpha=2.25$ and $\alpha=2.5$ remains at the level of $\mathcal{O}(10^{-4})$, as highlighted in the inset of the figure. This suggests that varying the steepness of the DM profile has only a marginal impact on the accumulated nonlinear memory for hyperbolic encounters.

This weak dependence can be understood physically from the short interaction timescale of hyperbolic encounters. Unlike inspiralling binaries, where the secondary object remains embedded in the DM environment for an extended period and accumulates environmental effects over many orbital cycles, hyperbolic binaries interact with the surrounding medium only briefly during the close encounter. As a result, the influence of DM induced effects on the orbital evolution and consequently on the nonlinear memory remains comparatively small.}
% {\color{red}A similar weak dependence on $\alpha$ is also observed in the overall nonlinear memory jump shown in Fig.~(\ref{fig:hyperbolic_spike_plot_5}).
{\color{black}Furthermore, Fig.~(\ref{fig:hyperbolic_spike_plot_5}) shows the overall nonlinear memory jump for hyperbolic binaries, indicating that the presence of a DM spike enhances the memory jump relative to the vacuum case. {\color{black}Here the choppiness at the higher values of $e_0$ is due to the numerical resolution error, which does not affect the qualitative behaviour or overall trend of the curve. }}

% \subsubsection{Dependency on DM spike exponent}

% Fig.~(\ref{fig:hyperbolic_spike_plot_10}) illustrates the effect of the DM density power-law exponent $\alpha$ on the nonlinear memory from a hyperbolic binary, with $e_0$ and $p_0$ held fixed. {\color{red}For the range of $\alpha$ values considered, the nonlinear memory curves shows difference of order $\mathcal{O}(10^{-4})$ between $\alpha = 2.25$ and $\alpha=2.5$ as shown in inset plot indicating very weak dependence on $\alpha$. This might be due to the lesser interaction time between secondary BH with environment as compare to inspiral case, reflecting weaker DM effects. A similar trend was also observed in the nonlinear jump as shown in Fig.~(\ref{fig:hyperbolic_spike_plot_5})}. 

% At early stages of the encounter, the nonlinear memory in the presence of DM is slightly smaller than that in vacuum; however, as the orbit evolves, it becomes marginally larger than the vacuum result.

\begin{figure}

    \begin{subfigure}{\linewidth}
        \includegraphics[width=\linewidth]{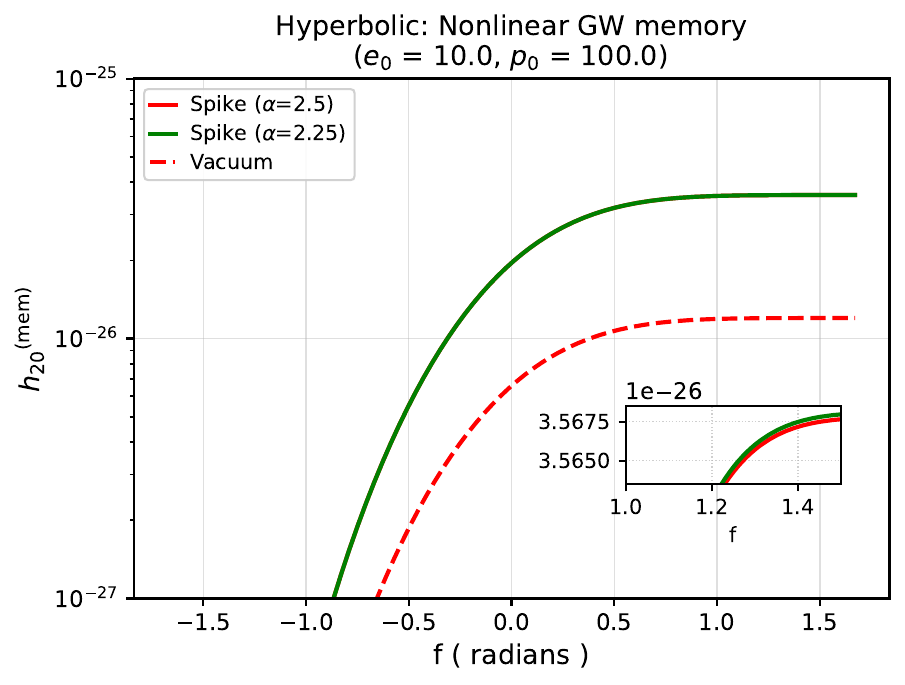}
        \subcaption{}
        \label{fig:hyperbolic_spike_plot_10}
    \end{subfigure}

    \begin{subfigure}{\linewidth}
        \includegraphics[width=\linewidth]{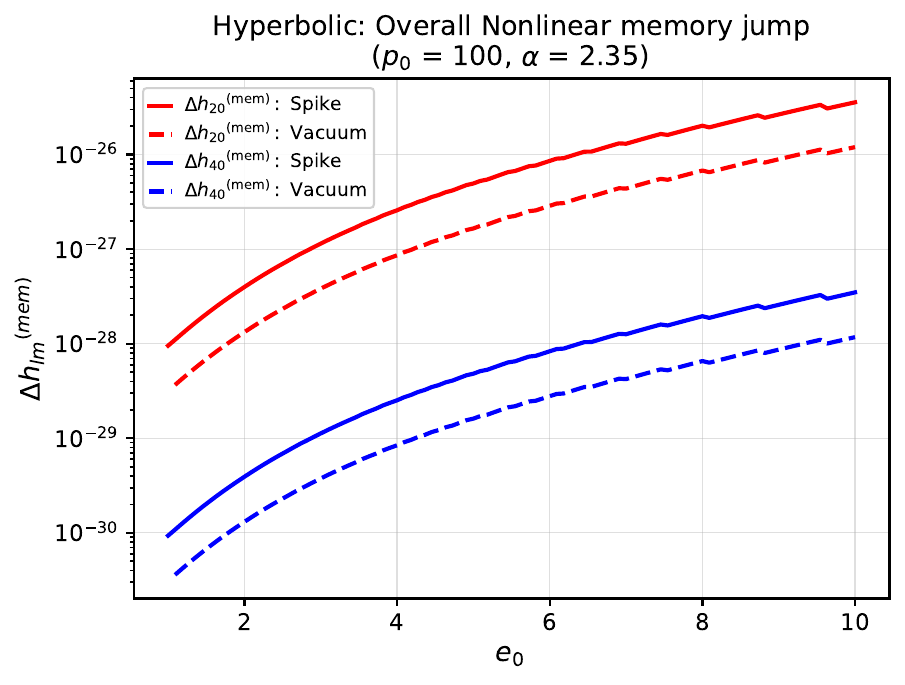}
        \subcaption{}
        \label{fig:hyperbolic_spike_plot_5}
    \end{subfigure}

    \caption{The above plots shows dependency of nonlinear memory on DM spike index $\alpha$ in hyperbolic orbit. Panel (a) shows the $h_{20}^{\textup{(mem)}}$ as functions of true anomaly with different values of $\alpha$. Panel (b) shows the nonlinear memory jump as function of initial eccentricities ($e_0$) for hyperbolic binaries.}
    \label{fig:hyperbolic_spike_plot2}
\end{figure}
}

\section{Discussion}\label{dscn}
{\color{black}Nonlinear GW memory is a hereditary effect; it depends on the cumulative radiated energy and encodes the past evolution of the source rather than only its instantaneous configuration. This makes it a natural observable to probe long-lived environmental effects. Motivated by this feature, in this work we studied how DM environments surrounding IMBHs modify the leading-order nonlinear memory generated by non-spinning IMR binaries. Our analysis is focused on elliptical, hyperbolic, and quasi-circular configurations, with the orbital dynamics evolved using the osculating-orbit method in the presence of DM gravity, DF, accretion, and GW backreaction. Our main result is that DM environments can reshape the buildup of nonlinear memory by altering the orbital evolution of the binary. In this sense, the environment affects the memory both directly, through the modified source dynamics entering the memory integrand, and indirectly, through changes in the duration of the orbital evolution over which the hereditary contribution accumulates. The relative importance of these effects depends strongly on the orbital class and on the properties of the DM profile.} We now briefly summarize the findings of our analysis for the various orbital classes:{\color{black}
\begin{itemize}
    \item For \textbf{elliptical binaries}, we find that the memory grows monotonically as the orbit circularizes, but the cumulative amplitude at the end of the inspiral is modified in a nontrivial way by the surrounding DM distribution. In a minispike environment, steeper spikes increase the strength of environmental forces and can enhance the instantaneous memory contribution. At the same time, they also accelerate the inspiral and reduce the accumulation time available before the LSO. The resulting competition implies that the total memory does not vary monotonically with the spike exponent $\alpha$. 
    
    To isolate the impact of nonlinear gravitational memory across distinct values of DM spike exponent $\alpha$, we evaluate the memory-induced waveform mismatch using the SPA for low-eccentricity elliptical binaries ($e_0 = 0.1$, $p_0 = 50$). As illustrated by the time-resolved analysis in Figure (\ref{fig:elliptical_mismatch_plot_1}), the inclusion of nonlinear memory leads to a substantial ({\color{black}as quantified in Table~(\ref{tab:snr_mismatch}), corresponds to mismatches well above the SNR-dependent distinguishability threshold for our fiducial IMRI)} observational changes during the last one year of evolution prior to the LSO, yielding notable mismatches on the order of $\mathcal{O}(10^{-2})$ and higher. This memory-induced mismatch is more pronounced for smaller $\alpha$ because larger spikes drive a faster orbital evolution, thereby shortening the inspiral duration and restricting the time available for the hereditary buildup of the memory signal. Conversely, as confirmed by the full accumulated signal analysis in Figure (\ref{fig:elliptical_mismatch_plot_2}), the total mismatch monotonically decreases as a function of increasing $\alpha$. This downward trend reflects a distinct physical competition where steeper DM environments shorten the overall inspiral window, truncating the signal earlier and restricting the net accumulation of memory. Nonetheless, across the astrophysically relevant range of $\alpha$, the resulting mismatch remains large enough to distinguish a waveform that includes memory from one that does not. Note that the mismatch calculation is intended as representative order-of-magnitude estimates within an approximate low-eccentricity framework, rather than as a full data-analysis forecast.
    %{\color{red}Instead, for fixed orbital parameters, the memory, the memory-only SNR, and the memory-induced mismatch can all show parameter-dependent local extrema. These features should therefore be interpreted as reflecting a competition between stronger instantaneous environmental forcing and shorter inspiral duration, rather than as universal signatures of the model.

   % The dependence on the initial orbital parameters follows the same logic. Larger initial eccentricities and larger initial semi-latus recta generally allow more time for hereditary buildup and hence tend to increase the accumulated memory. However, this tendency can be partially counteracted by environmental effects, especially for steeper minispikes that shorten the inspiral substantially. Our comparison between minispike, NFW, and vacuum cases further shows that a spike can produce a larger instantaneous memory than either vacuum or NFW, while the final accumulated memory may remain comparable because the stronger environment also drives a faster inspiral.}

    \item For \textbf{hyperbolic binaries}, the situation is qualitatively different. Although the DM environment enhances the nonlinear memory relative to the vacuum case, the overall amplitude remains very small, typically of order $\mathcal{O}(10^{-26}-10^{-24})$ in the cases considered here. This makes direct detection highly challenging even for future space-based detectors. {\color{black}We also find that the overall memory jump is modified by the presence of the environment, but within the present setup for spike profile produce quantitatively similar effects.} Thus, while hyperbolic encounters remain useful for illustrating how environments modify the nonlinear memory jump, they appear much less promising observationally than the elliptical systems studied here.%Fig (7-8)
    %The effect depends on both the spike exponent and the initial orbital parameters.

    \item For \textbf{quasi-circular binaries}, we considered both a static minispike and an empirical prescription for an evolving effective density profile. In both cases, the nonlinear memory is enhanced relative to vacuum. The evolving profile leads to a modestly larger memory amplitude than the static case, but the buildup terminates earlier because the inspiral proceeds faster under the combined action of GW backreaction and DF. Compared to the elliptical case, the quasi-circular setup provides a cleaner secular picture, though the quantitative results remain model-dependent because of the empirical nature of the evolving-density prescription. %Fig 9
    %For quasi-circular binaries, a dark matter spike enhances the nonlinear memory relative to vacuum. A dynamic spike yields a slightly larger amplitude, but the memory growth ends earlier due to faster orbital decay driven by environmental effects in addition to GW backreaction.
    
    %\item {\color{red}We also used the computed waveforms to estimate representative SNRs and mismatches in the context of space-based detectors such as LISA and GWSat. These estimates suggest that, for low-eccentricity elliptical binaries, nonlinear memory can leave a noticeable imprint on the waveform in the parameter ranges considered here. At the same time, the present detection-related results should be interpreted with appropriate caution. The SNR and mismatch calculations are intended as representative, order-of-magnitude estimates within an approximate low-eccentricity framework, rather than as a full data-analysis forecast. In particular, the detailed dependence of the memory-only SNR and the mismatch on $\alpha$ is sensitive to the choice of initial orbital parameters, and localized enhancements in these curves should be viewed as parameter-specific crossover features rather than generic predictions. }  
\end{itemize}
The broader implication of our analysis is therefore not that nonlinear memory provides the dominant observational probe of DM environments, but rather that it offers a complementary one. Environmental effects are generally expected to be more prominent in the oscillatory sector of the waveform, especially through their imprint on the orbital phase evolution. However, because nonlinear memory is hereditary, it responds to the integrated history of that evolution and can therefore encode environmental information in a qualitatively distinct manner. From this perspective, the memory sector may offer an additional handle on the long-timescale influence of DM surrounding IMBHs.
 
To our knowledge, this work provides the first systematic study of how DM environments modify leading-order nonlinear GW memory from IMR binaries. The main lesson is that the surrounding medium can imprint itself on the hereditary sector of the waveform in ways that are subtle, orbit-dependent, and often non-monotonic. While a full assessment of observational prospects will require more refined waveform modeling and parameter-estimation tools, the present results provide an order-of-magnitude foundation for exploring nonlinear memory as a complementary probe of astrophysical environments around IMBHs.}

Further, our analysis incorporates several simplifying assumptions: leading-order nonlinear memory corrections to the waveform, small perturbative effects from the environment on the binary dynamics, a static DM profile for elliptical and hyperbolic orbits, and non-spinning binary components. These assumptions define the scope of the current study and highlight several directions for future work. Relaxing each of these assumptions can give us a more detailed study of this interplay, specifically including non-trivial effects of disks \cite{Jiang:2025jbd}, where motions can also have intricate effects on their detection prospects \cite{Pan:2021oob}. It would be interesting to incorporate such features into our future analysis to better model the waveforms. Further, one can also perform the Fisher analysis in our context to infer the degeneracy or correlation among binary black hole parameters, including environmental effects on nonlinear memory \cite{Gasparotto:2023fcg}. Also, studying the effects of the medium in relativistic binaries embedded in a dynamically evolving DM environment through $N$-body simulations \cite{karydas2024sharpeningdarkmattersignature} will be a great way to push this line of study. Such systems may yield a substantially larger memory contribution to the waveform and offer improved detection prospects, especially with next-generation space-based detectors. Additionally, such analyses can be extended to spinning binaries with generic orbits \cite{Zi:2025lio} and to compact binaries exhibiting non-equatorial eccentric motion (inclined orbits). These scenarios are relevant and also realistic for more robust waveform modeling in the context of future detectors. We hope to address some of these aspects in our future studies.

%\newpage
%%%%%%%%%%%%%%%%%%%%%%%%%%%%%%%%%%%%%%%%%%%%%%%%%%%%%%%%%%%%%%%%%%%%%%%%%%%%%%%%%%%%%%%%%%%%%%%%%%%
\stoptoc
\section*{Acknowledgements} 
The research of S.K. is funded by National Post-Doctoral Fellowship (Grant No. PDF/2023/000369) from the ANRF (formerly SERB), Department of Science and Technology (DST), Government of India. AC is funded by Beijing Natural Science Foundation (BNSF)
International Scientists Project (Grant No. IS25033) and also from a postdoctoral fellowship (Grant No. 202504) through
the Department of Astronomy, Tsinghua University. A.C. would also like to acknowledge the support of European Union's Horizon Europe research and innovation programme under grant agreement No 101131928, project ACME. A.B. is supported by the Core Research Grant (CRG/2023/005112) by the Department of Science and Technology Science and Anusandhan National Research Foundation (formerly SERB), Government of India. A.B thank the speakers of ``Testing Aspects General Relativity-IV" for the illuminating discussion.  A.B. also acknowledges the associateship program of the Indian Academy of Science, Bengaluru. R.K.S. acknowledges the support in part by the International Center for Theoretical Sciences (ICTS) for participating in the program Beyond the Horizon: Testing the black hole paradigm (code: ICTS/BTH2025/03). Authors also acknowledge Srijit Bhattacharjee, Abhishek Sharma and Divya Tahelyani for useful discussions.\\
\resumetoc
%%%%%%%%%%%%%%%%%%%%%%%%%%%%%%%%%%%%%%%%%%%%%%%%%%%%%%%%%%%%%%%%%%%%%%%%%%%%%%%%%%%%%%%%%%%%%%%%%%
% \appendix 
% \section{Equations for the case of NFW}
% \label{append}
% {\color{black}The perturbed Newtonian binary in case of NFW profile is given as:
% \begin{equation}
% \ddot{\vec{r}} = - \frac{M}{r^2} \hat{n} + \vec{f}_{\textrm{DM}}+ \vec{f}_{\textrm{DF}} + \vec{f}_{acc} + \vec{f}_{\textrm{2.5PN}}.
% \label{eq:elliptical_NFW_eom}
% \end{equation}
% This equation can be resolved into the equations for $r$ and $\phi$ as:
% \begin{widetext}
% \begin{equation}
% \begin{aligned}
% & \ddot{r} = r \dot{\phi}^2 - \frac{M}{r^2} - \frac{M_{\textrm{DM} (r)}}{r^{2}} - \frac{4\pi m_2 \rho_{\textrm{DM}} (I_v {\color{black} \xi(v)} + \lambda) \dot{r}}{v^3} -\frac{M}{r^2} \Big[ A_{2.5\text{PN}} + \dot{r} B_{2.5\text{PN}} \Big], \\& \ddot{\phi} = - \frac{2 \dot{r} \dot{\phi}}{r} - \frac{4\pi m_2 \rho_{\textrm{DM}} ({\color{black}I_v \xi (v)} + \lambda) \dot{\phi}}{v^3} - \frac{M}{r} \Big[ B_{2.5\text{PN}} \Big] \dot{\phi},
% \end{aligned}
% \label{eq:elliptical_NFW_radial_and_angular_eom}
% \end{equation}    
% \end{widetext}
% where, $ M_{\textrm{DM}} (r)$ is the mass of NFW DM (given in section~(\ref{sec:DM gravity}) ) as function of distance $r$ and NFW profile $\rho_{DM}$ mentioned in Eq. (\ref{eq:NFW_density}).}

%\bibliographystyle{apsrev4-2f}
%\bibliographystyle{unsrt.bst}
\bibliographystyle{apsrev}
\bibliography{JN1}

@article{Pan:2021oob,
    author = "Pan, Zhen and Lyu, Zhenwei and Yang, Huan",
    title = "{Wet extreme mass ratio inspirals may be more common for spaceborne gravitational wave detection}",
    eprint = "2104.01208",
    archivePrefix = "arXiv",
    primaryClass = "astro-ph.HE",
    doi = "10.1103/PhysRevD.104.063007",
    journal = "Phys. Rev. D",
    volume = "104",
    number = "6",
    pages = "063007",
    year = "2021"
}

@article{Cutler:1994ys,
    author = "Cutler, Curt and Flanagan, Eanna E.",
    title = "{Gravitational waves from merging compact binaries: How accurately can one extract the binary's parameters from the inspiral wave form?}",
    eprint = "gr-qc/9402014",
    archivePrefix = "arXiv",
    reportNumber = "GRP-369",
    doi = "10.1103/PhysRevD.49.2658",
    journal = "Phys. Rev. D",
    volume = "49",
    pages = "2658--2697",
    year = "1994"
}

@article{Droz:1999qx,
    author = "Droz, Serge and Knapp, Daniel J. and Poisson, Eric and Owen, Benjamin J.",
    title = "{Gravitational waves from inspiraling compact binaries: Validity of the stationary phase approximation to the Fourier transform}",
    eprint = "gr-qc/9901076",
    archivePrefix = "arXiv",
    doi = "10.1103/PhysRevD.59.124016",
    journal = "Phys. Rev. D",
    volume = "59",
    pages = "124016",
    year = "1999"
}

@article{Damour:2000gg,
    author = "Damour, Thibault and Iyer, Bala R. and Sathyaprakash, B. S.",
    title = "{Frequency domain P approximant filters for time truncated inspiral gravitational wave signals from compact binaries}",
    eprint = "gr-qc/0001023",
    archivePrefix = "arXiv",
    doi = "10.1103/PhysRevD.62.084036",
    journal = "Phys. Rev. D",
    volume = "62",
    pages = "084036",
    year = "2000"
}

@article{Porto:2024cwd,
    author = "Porto, Rafael A. and Riva, Massimiliano M. and Yang, Zixin",
    title = "{Nonlinear gravitational radiation reaction: failed tail, memories {\&} squares}",
    eprint = "2409.05860",
    archivePrefix = "arXiv",
    primaryClass = "gr-qc",
    reportNumber = "DESY 24-133",
    doi = "10.1007/JHEP04(2025)050",
    journal = "JHEP",
    volume = "04",
    pages = "050",
    year = "2025"
}

@article{Pollney:2010hs,
    author = "Pollney, Denis and Reisswig, Christian",
    title = "{Gravitational memory in binary black hole mergers}",
    eprint = "1004.4209",
    archivePrefix = "arXiv",
    primaryClass = "gr-qc",
    doi = "10.1088/2041-8205/732/1/L13",
    journal = "Astrophys. J. Lett.",
    volume = "732",
    pages = "L13",
    year = "2011"
}

@article{Jiang:2025jbd,
    author = "Jiang, Ning and Pan, Zhen",
    title = "{Embers of Active Galactic Nuclei: Tidal Disruption Events and Quasiperiodic Eruptions}",
    eprint = "2503.17609",
    archivePrefix = "arXiv",
    primaryClass = "astro-ph.HE",
    doi = "10.3847/2041-8213/adc456",
    journal = "Astrophys. J. Lett.",
    volume = "983",
    number = "1",
    pages = "L18",
    year = "2025"
}

@article{LIGOScientific:2016aoc,
    author = "Abbott, B. P. and others",
    collaboration = "LIGO Scientific, Virgo",
    title = "{Observation of Gravitational Waves from a Binary Black Hole Merger}",
    eprint = "1602.03837",
    archivePrefix = "arXiv",
    primaryClass = "gr-qc",
    reportNumber = "LIGO-P150914",
    doi = "10.1103/PhysRevLett.116.061102",
    journal = "Phys. Rev. Lett.",
    volume = "116",
    number = "6",
    pages = "061102",
    year = "2016"
}

@article{LIGOScientific:2016vlm,
    author = "Abbott, B. P. and others",
    collaboration = "LIGO Scientific, Virgo",
    title = "{Properties of the Binary Black Hole Merger GW150914}",
    eprint = "1602.03840",
    archivePrefix = "arXiv",
    primaryClass = "gr-qc",
    reportNumber = "LIGO-P1500218",
    doi = "10.1103/PhysRevLett.116.241102",
    journal = "Phys. Rev. Lett.",
    volume = "116",
    number = "24",
    pages = "241102",
    year = "2016"
}

@article{LIGOScientific:2016sjg,
    author = "Abbott, B. P. and others",
    collaboration = "LIGO Scientific, Virgo",
    title = "{GW151226: Observation of Gravitational Waves from a 22-Solar-Mass Binary Black Hole Coalescence}",
    eprint = "1606.04855",
    archivePrefix = "arXiv",
    primaryClass = "gr-qc",
    reportNumber = "LIGO-P151226",
    doi = "10.1103/PhysRevLett.116.241103",
    journal = "Phys. Rev. Lett.",
    volume = "116",
    number = "24",
    pages = "241103",
    year = "2016"
}

@article{LIGOScientific:2017bnn,
    author = "Abbott, Benjamin P. and others",
    collaboration = "LIGO Scientific, VIRGO",
    title = "{GW170104: Observation of a 50-Solar-Mass Binary Black Hole Coalescence at Redshift 0.2}",
    eprint = "1706.01812",
    archivePrefix = "arXiv",
    primaryClass = "gr-qc",
    reportNumber = "LIGO-P170104",
    doi = "10.1103/PhysRevLett.118.221101",
    journal = "Phys. Rev. Lett.",
    volume = "118",
    number = "22",
    pages = "221101",
    year = "2017",
    note = "[Erratum: Phys.Rev.Lett. 121, 129901 (2018)]"
}

@article{LIGOScientific:2018dkp,
    author = "Abbott, B. P. and others",
    collaboration = "LIGO Scientific, Virgo",
    title = "{Tests of General Relativity with GW170817}",
    eprint = "1811.00364",
    archivePrefix = "arXiv",
    primaryClass = "gr-qc",
    reportNumber = "LIGO-P1800059",
    doi = "10.1103/PhysRevLett.123.011102",
    journal = "Phys. Rev. Lett.",
    volume = "123",
    number = "1",
    pages = "011102",
    year = "2019"
}

@article{Lasky:2016knh,
    author = "Lasky, Paul D. and Thrane, Eric and Levin, Yuri and Blackman, Jonathan and Chen, Yanbei",
    title = "{Detecting gravitational-wave memory with LIGO: implications of GW150914}",
    eprint = "1605.01415",
    archivePrefix = "arXiv",
    primaryClass = "astro-ph.HE",
    doi = "10.1103/PhysRevLett.117.061102",
    journal = "Phys. Rev. Lett.",
    volume = "117",
    number = "6",
    pages = "061102",
    year = "2016"
}

@article{Islo:2019qht,
    author = "Islo, Kristina and Simon, Joseph and Burke-Spolaor, Sarah and Siemens, Xavier",
    title = "{Prospects for Memory Detection with Low-Frequency Gravitational Wave Detectors}",
    eprint = "1906.11936",
    archivePrefix = "arXiv",
    primaryClass = "astro-ph.HE",
    month = "6",
    year = "2019"
}

@article{Boersma:2020gxx,
    author = "Boersma, Oliver M. and Nichols, David A. and Schmidt, Patricia",
    title = "{Forecasts for detecting the gravitational-wave memory effect with Advanced LIGO and Virgo}",
    eprint = "2002.01821",
    archivePrefix = "arXiv",
    primaryClass = "astro-ph.HE",
    doi = "10.1103/PhysRevD.101.083026",
    journal = "Phys. Rev. D",
    volume = "101",
    number = "8",
    pages = "083026",
    year = "2020"
}

@article{Grant:2022bla,
    author = "Grant, Alexander M. and Nichols, David A.",
    title = "{Outlook for detecting the gravitational-wave displacement and spin memory effects with current and future gravitational-wave detectors}",
    eprint = "2210.16266",
    archivePrefix = "arXiv",
    primaryClass = "gr-qc",
    doi = "10.1103/PhysRevD.107.064056",
    journal = "Phys. Rev. D",
    volume = "107",
    number = "6",
    pages = "064056",
    year = "2023",
    note = "[Erratum: Phys.Rev.D 108, 029901 (2023)]"
}

@article{LISAConsortiumWaveformWorkingGroup:2023arg,
    author = "Afshordi, Niayesh and others",
    collaboration = "LISA Consortium Waveform Working Group",
    title = "{Waveform Modelling for the Laser Interferometer Space Antenna}",
    eprint = "2311.01300",
    archivePrefix = "arXiv",
    primaryClass = "gr-qc",
    month = "11",
    year = "2023"
}

@article{Goncharov:2023woe,
    author = "Goncharov, Boris and Donnay, Laura and Harms, Jan",
    title = "{Inferring Fundamental Spacetime Symmetries with Gravitational-Wave Memory: From LISA to the Einstein Telescope}",
    eprint = "2310.10718",
    archivePrefix = "arXiv",
    primaryClass = "gr-qc",
    doi = "10.1103/PhysRevLett.132.241401",
    journal = "Phys. Rev. Lett.",
    volume = "132",
    number = "24",
    pages = "241401",
    year = "2024"
}

@article{Zeldovich:1974gvh,
    author = "Zel'dovich, Y. B. and Polnarev, A. G.",
    title = "{Radiation of gravitational waves by a cluster of superdense stars}",
    journal = "Sov. Astron.",
    volume = "18",
    pages = "17",
    year = "1974"
}

@article{Braginsky:1985vlg,
    author = "Braginsky, V. B. and Grishchuk, L. P.",
    title = "{Kinematic Resonance and Memory Effect in Free Mass Gravitational Antennas}",
    journal = "Sov. Phys. JETP",
    volume = "62",
    pages = "427--430",
    year = "1985"
}

@ARTICLE{1987Natur.327..123B,
       author = {{Braginsky}, V.~B. and {Thorne}, Kip S.},
        title = "{Gravitational-wave bursts with memory and experimental prospects}",
      journal = {Nature},
         year = 1987,
        month = may,
       volume = {327},
       number = {6118},
        pages = {123-125},
          doi = {10.1038/327123a0},
       adsurl = {https://ui.adsabs.harvard.edu/abs/1987Natur.327..123B}
}

@misc{strominger2018lecturesinfraredstructuregravity,
      title={Lectures on the Infrared Structure of Gravity and Gauge Theory}, 
      author={Andrew Strominger},
      year={2018},
      eprint={1703.05448},
      archivePrefix={arXiv},
      primaryClass={hep-th},
      url={https://arxiv.org/abs/1703.05448}, 
}

@article{Christodoulou:1991cr,
    author = "Christodoulou, D.",
    title = "{Nonlinear nature of gravitation and gravitational wave experiments}",
    doi = "10.1103/PhysRevLett.67.1486",
    journal = "Phys. Rev. Lett.",
    volume = "67",
    pages = "1486--1489",
    year = "1991"
}

@article{Zhang:2017rno,
    author = "Zhang, P. -M and Duval, C. and Gibbons, G. W. and Horvathy, P. A.",
    title = "{The Memory Effect for Plane Gravitational Waves}",
    eprint = "1704.05997",
    archivePrefix = "arXiv",
    primaryClass = "gr-qc",
    doi = "10.1016/j.physletb.2017.07.050",
    journal = "Phys. Lett. B",
    volume = "772",
    pages = "743--746",
    year = "2017"
}

@article{Zhang:2017geq,
    author = "Zhang, P. -M. and Duval, C. and Gibbons, G. W. and Horvathy, P. A.",
    title = "{Soft gravitons and the memory effect for plane gravitational waves}",
    eprint = "1705.01378",
    archivePrefix = "arXiv",
    primaryClass = "gr-qc",
    doi = "10.1103/PhysRevD.96.064013",
    journal = "Phys. Rev. D",
    volume = "96",
    number = "6",
    pages = "064013",
    year = "2017"
}

@article{Pasterski:2015tva,
    author = "Pasterski, Sabrina and Strominger, Andrew and Zhiboedov, Alexander",
    title = "{New Gravitational Memories}",
    eprint = "1502.06120",
    archivePrefix = "arXiv",
    primaryClass = "hep-th",
    doi = "10.1007/JHEP12(2016)053",
    journal = "JHEP",
    volume = "12",
    pages = "053",
    year = "2016"
}

@article{Mao:2018xcw,
    author = "Mao, Pujian and Wu, Xiaoning",
    title = "{More on gravitational memory}",
    eprint = "1812.07168",
    archivePrefix = "arXiv",
    primaryClass = "gr-qc",
    reportNumber = "CJQS-2019-014",
    doi = "10.1007/JHEP05(2019)058",
    journal = "JHEP",
    volume = "05",
    pages = "058",
    year = "2019"
}

@article{Tahura:2020vsa,
    author = "Tahura, Shammi and Nichols, David A. and Saffer, Alexander and Stein, Leo C. and Yagi, Kent",
    title = "{Brans-Dicke theory in Bondi-Sachs form: Asymptotically flat solutions, asymptotic symmetries and gravitational-wave memory effects}",
    eprint = "2007.13799",
    archivePrefix = "arXiv",
    primaryClass = "gr-qc",
    doi = "10.1103/PhysRevD.103.104026",
    journal = "Phys. Rev. D",
    volume = "103",
    number = "10",
    pages = "104026",
    year = "2021"
}

@article{Hou:2020tnd,
    author = "Hou, Shaoqi and Zhu, Zong-Hong",
    title = "{Gravitational memory effects and Bondi-Metzner-Sachs symmetries in scalar-tensor theories}",
    eprint = "2005.01310",
    archivePrefix = "arXiv",
    primaryClass = "gr-qc",
    doi = "10.1007/JHEP01(2021)083",
    journal = "JHEP",
    volume = "01",
    pages = "083",
    year = "2021"
}

@article{Seraj:2021qja,
    author = "Seraj, Ali",
    title = "{Gravitational breathing memory and dual symmetries}",
    eprint = "2103.12185",
    archivePrefix = "arXiv",
    primaryClass = "hep-th",
    doi = "10.1007/JHEP05(2021)283",
    journal = "JHEP",
    volume = "05",
    pages = "283",
    year = "2021"
}

@article{Tahura:2021hbk,
    author = "Tahura, Shammi and Nichols, David A. and Yagi, Kent",
    title = "{Gravitational-wave memory effects in Brans-Dicke theory: Waveforms and effects in the post-Newtonian approximation}",
    eprint = "2107.02208",
    archivePrefix = "arXiv",
    primaryClass = "gr-qc",
    doi = "10.1103/PhysRevD.104.104010",
    journal = "Phys. Rev. D",
    volume = "104",
    number = "10",
    pages = "104010",
    year = "2021"
}

@article{Heisenberg:2023prj,
    author = "Heisenberg, Lavinia and Yunes, Nicol\'as and Zosso, Jann",
    title = "{Gravitational wave memory beyond general relativity}",
    eprint = "2303.02021",
    archivePrefix = "arXiv",
    primaryClass = "gr-qc",
    doi = "10.1103/PhysRevD.108.024010",
    journal = "Phys. Rev. D",
    volume = "108",
    number = "2",
    pages = "024010",
    year = "2023"
}

@article{Hawking:2016sgy,
    author = "Hawking, Stephen W. and Perry, Malcolm J. and Strominger, Andrew",
    title = "{Superrotation Charge and Supertranslation Hair on Black Holes}",
    eprint = "1611.09175",
    archivePrefix = "arXiv",
    primaryClass = "hep-th",
    doi = "10.1007/JHEP05(2017)161",
    journal = "JHEP",
    volume = "05",
    pages = "161",
    year = "2017"
}

@article{Hawking:2016msc,
    author = "Hawking, Stephen W. and Perry, Malcolm J. and Strominger, Andrew",
    title = "{Soft Hair on Black Holes}",
    eprint = "1601.00921",
    archivePrefix = "arXiv",
    primaryClass = "hep-th",
    doi = "10.1103/PhysRevLett.116.231301",
    journal = "Phys. Rev. Lett.",
    volume = "116",
    number = "23",
    pages = "231301",
    year = "2016"
}

@phdthesis{Kumar:2021qrg,
    author = "Kumar, Shailesh",
    title = "{Gravitational Memory Effect for Near-Horizon Asymptotic Symmetries}",
    eprint = "2203.08983",
    archivePrefix = "arXiv",
    primaryClass = "gr-qc",
    school = "IIIT, Allahabad",
    month = "10",
    year = "2021"
}

@article{Cunningham:2024dog,
    author = "Cunningham, Kevin and Kavanagh, Chris and Pound, Adam and Trestini, David and Warburton, Niels and Neef, Jakob",
    title = "{Gravitational memory: new results from post-Newtonian and self-force theory}",
    eprint = "2410.23950",
    archivePrefix = "arXiv",
    primaryClass = "gr-qc",
    month = "10",
    year = "2024"
}

@article{Blanchet2014,
author={Blanchet, Luc},
title={Gravitational Radiation from Post-Newtonian Sources and Inspiralling Compact Binaries},
journal={Living Reviews in Relativity},
year={2014},
month={Feb},
day={13},
volume={17},
number={1},
pages={2},
issn={1433-8351},
doi={10.12942/lrr-2014-2},
url={https://doi.org/10.12942/lrr-2014-2}
}

@article{Mitman:2020pbt,
    author = "Mitman, Keefe and Moxon, Jordan and Scheel, Mark A. and Teukolsky, Saul A. and Boyle, Michael and Deppe, Nils and Kidder, Lawrence E. and Throwe, William",
    title = "{Computation of displacement and spin gravitational memory in numerical relativity}",
    eprint = "2007.11562",
    archivePrefix = "arXiv",
    primaryClass = "gr-qc",
    doi = "10.1103/PhysRevD.102.104007",
    journal = "Phys. Rev. D",
    volume = "102",
    number = "10",
    pages = "104007",
    year = "2020"
}

@article{Mitman:2021xkq,
    author = "Mitman, Keefe and others",
    title = "{Fixing the BMS frame of numerical relativity waveforms}",
    eprint = "2105.02300",
    archivePrefix = "arXiv",
    primaryClass = "gr-qc",
    doi = "10.1103/PhysRevD.104.024051",
    journal = "Phys. Rev. D",
    volume = "104",
    number = "2",
    pages = "024051",
    year = "2021"
}

@article{Mitman:2022kwt,
    author = "Mitman, Keefe and others",
    title = "{Fixing the BMS frame of numerical relativity waveforms with BMS charges}",
    eprint = "2208.04356",
    archivePrefix = "arXiv",
    primaryClass = "gr-qc",
    doi = "10.1103/PhysRevD.106.084029",
    journal = "Phys. Rev. D",
    volume = "106",
    number = "8",
    pages = "084029",
    year = "2022"
}

@misc{blanchet2024postnewtoniantheorygravitationalwaves,
      title={Post-Newtonian Theory for Gravitational Waves}, 
      author={Luc Blanchet},
      year={2024},
      eprint={1310.1528},
      archivePrefix={arXiv},
      primaryClass={gr-qc},
      url={https://arxiv.org/abs/1310.1528}, 
}

@article{Mitman:2024uss,
    author = "Mitman, Keefe and others",
    title = "{A review of gravitational memory and BMS frame fixing in numerical relativity}",
    eprint = "2405.08868",
    archivePrefix = "arXiv",
    primaryClass = "gr-qc",
    doi = "10.1088/1361-6382/ad83c2",
    journal = "Class. Quant. Grav.",
    volume = "41",
    number = "22",
    pages = "223001",
    year = "2024"
}

@article{Bhat:2024cyq,
    author = "Bhat, Sajad A. and Bhattacharjee, Srijit and Kapadia, Shasvath J.",
    title = "{Can the Near-Horizon Black Hole Memory be detected through Binary Inspirals?}",
    eprint = "2406.15604",
    archivePrefix = "arXiv",
    primaryClass = "gr-qc",
    month = "6",
    year = "2024"
}

@article{Sarkar:2021djs,
    author = "Sarkar, Subhodeep and Kumar, Shailesh and Bhattacharjee, Srijit",
    title = "{Can we detect a supertranslated black hole?}",
    eprint = "2110.03547",
    archivePrefix = "arXiv",
    primaryClass = "gr-qc",
    doi = "10.1103/PhysRevD.105.084001",
    journal = "Phys. Rev. D",
    volume = "105",
    number = "8",
    pages = "084001",
    year = "2022"
}

@article{Bhattacharjee:2020vfb,
    author = "Bhattacharjee, Srijit and Kumar, Shailesh and Bhattacharyya, Arpan",
    title = "{Displacement memory effect near the horizon of black holes}",
    eprint = "2010.16086",
    archivePrefix = "arXiv",
    primaryClass = "gr-qc",
    doi = "10.1007/JHEP03(2021)134",
    journal = "JHEP",
    volume = "03",
    pages = "134",
    year = "2021"
}

@article{Bhattacharjee:2020lgt,
    author = "Bhattacharjee, Srijit and Kumar, Shailesh",
    title = "{Memory effect and BMS symmetries for extreme black holes}",
    eprint = "2003.09334",
    archivePrefix = "arXiv",
    primaryClass = "hep-th",
    doi = "10.1103/PhysRevD.102.044041",
    journal = "Phys. Rev. D",
    volume = "102",
    number = "4",
    pages = "044041",
    year = "2020"
}

@article{Bhattacharjee:2019jaf,
    author = "Bhattacharjee, Srijit and Kumar, Shailesh and Bhattacharyya, Arpan",
    title = "{Memory Effect and BMS-like Symmetries for Impulsive Gravitational Waves}",
    eprint = "1905.12905",
    archivePrefix = "arXiv",
    primaryClass = "hep-th",
    reportNumber = "YITP-19-43",
    doi = "10.1103/PhysRevD.100.084010",
    journal = "Phys. Rev. D",
    volume = "100",
    number = "8",
    pages = "084010",
    year = "2019"
}

@article{Donnay:2018ckb,
    author = "Donnay, Laura and Giribet, Gaston and Gonz\'alez, Hern\'an A. and Puhm, Andrea",
    title = "{Black hole memory effect}",
    eprint = "1809.07266",
    archivePrefix = "arXiv",
    primaryClass = "hep-th",
    reportNumber = "CPHT-RR021.042018",
    doi = "10.1103/PhysRevD.98.124016",
    journal = "Phys. Rev. D",
    volume = "98",
    number = "12",
    pages = "124016",
    year = "2018"
}

@article{Donnay:2015abr,
    author = "Donnay, Laura and Giribet, Gaston and Gonzalez, Hernan A. and Pino, Miguel",
    title = "{Supertranslations and Superrotations at the Black Hole Horizon}",
    eprint = "1511.08687",
    archivePrefix = "arXiv",
    primaryClass = "hep-th",
    doi = "10.1103/PhysRevLett.116.091101",
    journal = "Phys. Rev. Lett.",
    volume = "116",
    number = "9",
    pages = "091101",
    year = "2016"
}

@article{Donnay:2016ejv,
    author = "Donnay, Laura and Giribet, Gaston and Gonz\'alez, Hern\'an A. and Pino, Miguel",
    title = "{Extended Symmetries at the Black Hole Horizon}",
    eprint = "1607.05703",
    archivePrefix = "arXiv",
    primaryClass = "hep-th",
    doi = "10.1007/JHEP09(2016)100",
    journal = "JHEP",
    volume = "09",
    pages = "100",
    year = "2016"
}

@article{Rahman:2019bmk,
    author = "Rahman, Adel A. and Wald, Robert M.",
    title = "{Black Hole Memory}",
    eprint = "1912.12806",
    archivePrefix = "arXiv",
    primaryClass = "gr-qc",
    doi = "10.1103/PhysRevD.101.124010",
    journal = "Phys. Rev. D",
    volume = "101",
    number = "12",
    pages = "124010",
    year = "2020"
}

@article{Blau:2015nee,
    author = "Blau, Matthias and O'Loughlin, Martin",
    title = "{Horizon Shells and BMS-like Soldering Transformations}",
    eprint = "1512.02858",
    archivePrefix = "arXiv",
    primaryClass = "hep-th",
    doi = "10.1007/JHEP03(2016)029",
    journal = "JHEP",
    volume = "03",
    pages = "029",
    year = "2016"
}

@article{Favata:2010zu,
    author = "Favata, Marc",
    editor = "Marka, Zsuzsa and Marka, Szabolcs",
    title = "{The gravitational-wave memory effect}",
    eprint = "1003.3486",
    archivePrefix = "arXiv",
    primaryClass = "gr-qc",
    doi = "10.1088/0264-9381/27/8/084036",
    journal = "Class. Quant. Grav.",
    volume = "27",
    pages = "084036",
    year = "2010"
}

@article{Favata:2009ii,
    author = "Favata, Marc",
    title = "{Nonlinear gravitational-wave memory from binary black hole mergers}",
    eprint = "0902.3660",
    archivePrefix = "arXiv",
    primaryClass = "astro-ph.SR",
    doi = "10.1088/0004-637X/696/2/L159",
    journal = "Astrophys. J. Lett.",
    volume = "696",
    pages = "L159--L162",
    year = "2009"
}

@article{Alnasheet:2025tpd,
    author = "Alnasheet, Qassim and Cardoso, Vitor and Duque, Francisco and Panosso Macedo, Rodrigo",
    title = "{Gravitational-wave tails and memory effect for mergers in astrophysical environments}",
    doi = "10.1103/yyv5-3y1c",
    journal = "Phys. Rev. D",
    volume = "112",
    number = "4",
    pages = "044066",
    year = "2025"
}

@article{Favata:2008yd,
    author = "Favata, Marc",
    title = "{Post-Newtonian corrections to the gravitational-wave memory for quasi-circular, inspiralling compact binaries}",
    eprint = "0812.0069",
    archivePrefix = "arXiv",
    primaryClass = "gr-qc",
    doi = "10.1103/PhysRevD.80.024002",
    journal = "Phys. Rev. D",
    volume = "80",
    pages = "024002",
    year = "2009"
}

@article{Favata_2009,
   title={Gravitational-wave memory revisited: Memory from the merger and recoil of binary black holes},
   volume={154},
   ISSN={1742-6596},
   url={http://dx.doi.org/10.1088/1742-6596/154/1/012043},
   DOI={10.1088/1742-6596/154/1/012043},
   journal={Journal of Physics: Conference Series},
   publisher={IOP Publishing},
   author={Favata, Marc},
   year={2009},
   month=mar, pages={012043} }

@article{Favata:2011qif,
    author = "Favata, Marc",
    title = "{The Gravitational-wave memory from eccentric binaries}",
    eprint = "1108.3121",
    archivePrefix = "arXiv",
    primaryClass = "gr-qc",
    doi = "10.1103/PhysRevD.84.124013",
    journal = "Phys. Rev. D",
    volume = "84",
    pages = "124013",
    year = "2011"
}

@article{Solanki:2023wmv,
    author = "Solanki, Divyesh N. and Bhattacharjee, Srijit",
    title = "{Soft theorems and memory effects at finite temperatures}",
    eprint = "2308.02445",
    archivePrefix = "arXiv",
    primaryClass = "hep-th",
    doi = "10.1140/epjc/s10052-023-12335-8",
    journal = "Eur. Phys. J. C",
    volume = "83",
    number = "12",
    pages = "1167",
    year = "2023"
}

@article{Solanki:2024oci,
    author = "Solanki, Divyesh N. and Bhattacharjee, Srijit",
    title = "{Supertranslations at Spatial and Timelike Infinities in the First-Order Formalism}",
    eprint = "2409.01783",
    archivePrefix = "arXiv",
    primaryClass = "gr-qc",
    month = "9",
    year = "2024"
}

@article{Chakraborty:2019yxn,
    author = "Chakraborty, Indranil and Kar, Sayan",
    title = "{Geodesic congruences in exact plane wave spacetimes and the memory effect}",
    eprint = "1901.11236",
    archivePrefix = "arXiv",
    primaryClass = "gr-qc",
    doi = "10.1103/PhysRevD.101.064022",
    journal = "Phys. Rev. D",
    volume = "101",
    number = "6",
    pages = "064022",
    year = "2020"
}

@article{Talbot:2018sgr,
    author = "Talbot, Colm and Thrane, Eric and Lasky, Paul D. and Lin, Fuhui",
    title = "{Gravitational-wave memory: waveforms and phenomenology}",
    eprint = "1807.00990",
    archivePrefix = "arXiv",
    primaryClass = "astro-ph.HE",
    doi = "10.1103/PhysRevD.98.064031",
    journal = "Phys. Rev. D",
    volume = "98",
    number = "6",
    pages = "064031",
    year = "2018"
}

@article{Gasparotto:2023fcg,
    author = "Gasparotto, Silvia and Vicente, Rodrigo and Blas, Diego and Jenkins, Alexander C. and Barausse, Enrico",
    title = "{Can gravitational-wave memory help constrain binary black-hole parameters? A LISA case study}",
    eprint = "2301.13228",
    archivePrefix = "arXiv",
    primaryClass = "gr-qc",
    doi = "10.1103/PhysRevD.107.124033",
    journal = "Phys. Rev. D",
    volume = "107",
    number = "12",
    pages = "124033",
    year = "2023"
}

@article{Blanchet:2023pce,
    author = "Blanchet, Luc and Comp{\`e}re, Geoffrey and Faye, Guillaume and Oliveri, Roberto and Seraj, Ali",
    title = "{Multipole expansion of gravitational waves: memory effects and Bondi aspects}",
    eprint = "2303.07732",
    archivePrefix = "arXiv",
    primaryClass = "gr-qc",
    doi = "10.1007/JHEP07(2023)123",
    journal = "JHEP",
    volume = "07",
    pages = "123",
    year = "2023"
}

@article{PhysRevD.89.084039,
  title = {Perturbative and gauge invariant treatment of gravitational wave memory},
  author = {Bieri, Lydia and Garfinkle, David},
  journal = {Phys. Rev. D},
  volume = {89},
  issue = {8},
  pages = {084039},
  numpages = {9},
  year = {2014},
  month = {Apr},
  publisher = {American Physical Society},
  doi = {10.1103/PhysRevD.89.084039},
  url = {https://link.aps.org/doi/10.1103/PhysRevD.89.084039}
}

@article{Tolish:2014bka,
    author = "Tolish, Alexander and Wald, Robert M.",
    title = "{Retarded Fields of Null Particles and the Memory Effect}",
    eprint = "1401.5831",
    archivePrefix = "arXiv",
    primaryClass = "gr-qc",
    doi = "10.1103/PhysRevD.89.064008",
    journal = "Phys. Rev. D",
    volume = "89",
    number = "6",
    pages = "064008",
    year = "2014"
}

@article{PhysRevD.45.520,
  title = {Gravitational-wave bursts with memory: The Christodoulou effect},
  author = {Thorne, Kip S.},
  journal = {Phys. Rev. D},
  volume = {45},
  issue = {2},
  pages = {520--524},
  numpages = {0},
  year = {1992},
  month = {Jan},
  publisher = {American Physical Society},
  doi = {10.1103/PhysRevD.45.520},
  url = {https://link.aps.org/doi/10.1103/PhysRevD.45.520}
}

@article{Strominger:2014pwa,
    author = "Strominger, Andrew and Zhiboedov, Alexander",
    title = "{Gravitational Memory, BMS Supertranslations and Soft Theorems}",
    eprint = "1411.5745",
    archivePrefix = "arXiv",
    primaryClass = "hep-th",
    doi = "10.1007/JHEP01(2016)086",
    journal = "JHEP",
    volume = "01",
    pages = "086",
    year = "2016"
}

@article{Sun:2022pvh,
    author = "Sun, Shuo and Shi, Changfu and Zhang, Jian-dong and Mei, Jianwei",
    title = "{Detecting the gravitational wave memory effect with TianQin}",
    eprint = "2207.13009",
    archivePrefix = "arXiv",
    primaryClass = "gr-qc",
    doi = "10.1103/PhysRevD.107.044023",
    journal = "Phys. Rev. D",
    volume = "107",
    number = "4",
    pages = "044023",
    year = "2023"
}

@article{BenAchour:2024ucn,
    author = "Ben Achour, Jibril and Uzan, Jean-Philippe",
    title = "{Displacement versus velocity memory effects from a gravitational plane wave}",
    eprint = "2406.07106",
    archivePrefix = "arXiv",
    primaryClass = "gr-qc",
    doi = "10.1088/1475-7516/2024/08/004",
    journal = "JCAP",
    volume = "08",
    pages = "004",
    year = "2024"
}

@article{Gaur:2024oms,
    author = "Gaur, Rudeep",
    title = "{The Kerr Memory Effect at Null Infinity}",
    eprint = "2403.07302",
    archivePrefix = "arXiv",
    primaryClass = "gr-qc",
    month = "3",
    year = "2024"
}

@article{Galoppo:2024vww,
    author = "Galoppo, Marco and Gaur, Rudeep and Harvey-Hawes, Christopher",
    title = "{Kerr--Newman Memory Effect}",
    eprint = "2407.15289",
    archivePrefix = "arXiv",
    primaryClass = "gr-qc",
    month = "7",
    year = "2024"
}

@ARTICLE{1990ApJ...356..359H,
       author = {{Hernquist}, Lars},
        title = "{An Analytical Model for Spherical Galaxies and Bulges}",
      journal = {apj},
         year = 1990,
        month = jun,
       volume = {356},
        pages = {359},
          doi = {10.1086/168845},
       adsurl = {https://ui.adsabs.harvard.edu/abs/1990ApJ...356..359H}
}

@article{Bertone:2004pz,
    author = "Bertone, Gianfranco and Hooper, Dan and Silk, Joseph",
    title = "{Particle dark matter: Evidence, candidates and constraints}",
    eprint = "hep-ph/0404175",
    archivePrefix = "arXiv",
    reportNumber = "FERMILAB-PUB-04-047-A",
    doi = "10.1016/j.physrep.2004.08.031",
    journal = "Phys. Rept.",
    volume = "405",
    pages = "279--390",
    year = "2005"
}

@article{Navarro:1995iw,
    author = "Navarro, Julio F. and Frenk, Carlos S. and White, Simon D. M.",
    title = "{The Structure of cold dark matter halos}",
    eprint = "astro-ph/9508025",
    archivePrefix = "arXiv",
    doi = "10.1086/177173",
    journal = "Astrophys. J.",
    volume = "462",
    pages = "563--575",
    year = "1996"
}

@article{Clowe:2006eq,
    author = "Clowe, Douglas and Bradac, Marusa and Gonzalez, Anthony H. and Markevitch, Maxim and Randall, Scott W. and Jones, Christine and Zaritsky, Dennis",
    title = "{A direct empirical proof of the existence of dark matter}",
    eprint = "astro-ph/0608407",
    archivePrefix = "arXiv",
    reportNumber = "SLAC-PUB-12078",
    doi = "10.1086/508162",
    journal = "Astrophys. J. Lett.",
    volume = "648",
    pages = "L109--L113",
    year = "2006"
}

@article{Corbelli:1999af,
    author = "Corbelli, Edvige and Salucci, Paolo",
    title = "{The Extended Rotation Curve and the Dark Matter Halo of M33}",
    eprint = "astro-ph/9909252",
    archivePrefix = "arXiv",
    doi = "10.1046/j.1365-8711.2000.03075.x",
    journal = "Mon. Not. Roy. Astron. Soc.",
    volume = "311",
    pages = "441--447",
    year = "2000"
}

@article{Navarro:1996gj,
    author = "Navarro, Julio F. and Frenk, Carlos S. and White, Simon D. M.",
    title = "{A Universal density profile from hierarchical clustering}",
    eprint = "astro-ph/9611107",
    archivePrefix = "arXiv",
    doi = "10.1086/304888",
    journal = "Astrophys. J.",
    volume = "490",
    pages = "493--508",
    year = "1997"
}

@article{Gondolo:1999ef,
    author = "Gondolo, Paolo and Silk, Joseph",
    title = "{Dark matter annihilation at the galactic center}",
    eprint = "astro-ph/9906391",
    archivePrefix = "arXiv",
    reportNumber = "MPI-PHT-99-10, OUAST-99-9",
    doi = "10.1103/PhysRevLett.83.1719",
    journal = "Phys. Rev. Lett.",
    volume = "83",
    pages = "1719--1722",
    year = "1999"
}

@article{Merritt:2002vj,
    author = "Merritt, David and Milosavljevic, Milos and Verde, Licia and Jimenez, Raul",
    title = "{Dark matter spikes and annihilation radiation from the galactic center}",
    eprint = "astro-ph/0201376",
    archivePrefix = "arXiv",
    reportNumber = "RUTGERS-AP-337",
    doi = "10.1103/PhysRevLett.88.191301",
    journal = "Phys. Rev. Lett.",
    volume = "88",
    pages = "191301",
    year = "2002"
}

@article{Ullio:2001fb,
    author = "Ullio, Piero and Zhao, HongSheng and Kamionkowski, Marc",
    title = "{A Dark matter spike at the galactic center?}",
    eprint = "astro-ph/0101481",
    archivePrefix = "arXiv",
    doi = "10.1103/PhysRevD.64.043504",
    journal = "Phys. Rev. D",
    volume = "64",
    pages = "043504",
    year = "2001"
}

@article{Bertone:2005xv,
    author = "Bertone, Gianfranco and Merritt, David",
    title = "{Dark matter dynamics and indirect detection}",
    eprint = "astro-ph/0504422",
    archivePrefix = "arXiv",
    reportNumber = "FERMILAB-PUB-05-051-A",
    doi = "10.1142/S0217732305017391",
    journal = "Mod. Phys. Lett. A",
    volume = "20",
    pages = "1021",
    year = "2005"
}

@article{Cardoso:2021wlq,
    author = "Cardoso, Vitor and Destounis, Kyriakos and Duque, Francisco and Macedo, Rodrigo Panosso and Maselli, Andrea",
    title = "{Black holes in galaxies: Environmental impact on gravitational-wave generation and propagation}",
    eprint = "2109.00005",
    archivePrefix = "arXiv",
    primaryClass = "gr-qc",
    doi = "10.1103/PhysRevD.105.L061501",
    journal = "Phys. Rev. D",
    volume = "105",
    number = "6",
    pages = "L061501",
    year = "2022"
}

@article{PhysRevD.103.023015,
  title = {Eccentricity evolution of compact binaries and applications to gravitational-wave physics},
  author = {Cardoso, Vitor and Macedo, Caio F. B. and Vicente, Rodrigo},
  journal = {Phys. Rev. D},
  volume = {103},
  issue = {2},
  pages = {023015},
  numpages = {13},
  year = {2021},
  month = {Jan},
  publisher = {American Physical Society},
  doi = {10.1103/PhysRevD.103.023015},
  url = {https://link.aps.org/doi/10.1103/PhysRevD.103.023015}
}

@article{Destounis:2022obl,
    author = "Destounis, Kyriakos and Kulathingal, Arun and Kokkotas, Kostas D. and Papadopoulos, Georgios O.",
    title = "{Gravitational-wave imprints of compact and galactic-scale environments in extreme-mass-ratio binaries}",
    eprint = "2210.09357",
    archivePrefix = "arXiv",
    primaryClass = "gr-qc",
    doi = "10.1103/PhysRevD.107.084027",
    journal = "Phys. Rev. D",
    volume = "107",
    number = "8",
    pages = "084027",
    year = "2023"
}

@article{Cardoso:2022whc,
    author = "Cardoso, Vitor and Destounis, Kyriakos and Duque, Francisco and Panosso Macedo, Rodrigo and Maselli, Andrea",
    title = "{Gravitational Waves from Extreme-Mass-Ratio Systems in Astrophysical Environments}",
    eprint = "2210.01133",
    archivePrefix = "arXiv",
    primaryClass = "gr-qc",
    doi = "10.1103/PhysRevLett.129.241103",
    journal = "Phys. Rev. Lett.",
    volume = "129",
    number = "24",
    pages = "241103",
    year = "2022"
}

@article{Shen:2023erj,
    author = "Shen, Zibo and Wang, Anzhong and Gong, Yungui and Yin, Shaoyu",
    title = "{Analytical models of supermassive black holes in galaxies surrounded by dark matter halos}",
    eprint = "2311.12259",
    archivePrefix = "arXiv",
    primaryClass = "gr-qc",
    doi = "10.1016/j.physletb.2024.138797",
    journal = "Phys. Lett. B",
    volume = "855",
    pages = "138797",
    year = "2024"
}

@article{Duque:2023seg,
    author = "Duque, Francisco and Macedo, Caio F. B. and Vicente, Rodrigo and Cardoso, Vitor",
    title = "{Extreme-Mass-Ratio Inspirals in Ultralight Dark Matter}",
    eprint = "2312.06767",
    archivePrefix = "arXiv",
    primaryClass = "gr-qc",
    doi = "10.1103/PhysRevLett.133.121404",
    journal = "Phys. Rev. Lett.",
    volume = "133",
    number = "12",
    pages = "121404",
    year = "2024"
}

@article{Speeney:2024mas,
    author = "Speeney, Nicholas and Berti, Emanuele and Cardoso, Vitor and Maselli, Andrea",
    title = "{Black holes surrounded by generic matter distributions: Polar perturbations and energy flux}",
    eprint = "2401.00932",
    archivePrefix = "arXiv",
    primaryClass = "gr-qc",
    doi = "10.1103/PhysRevD.109.084068",
    journal = "Phys. Rev. D",
    volume = "109",
    number = "8",
    pages = "084068",
    year = "2024"
}

@article{Zhang:2024ugv,
    author = "Zhang, Chao and Fu, Guoyang and Dai, Ning",
    title = "{Detecting dark matter halos with extreme mass-ratio inspirals}",
    eprint = "2401.04467",
    archivePrefix = "arXiv",
    primaryClass = "gr-qc",
    doi = "10.1088/1475-7516/2024/04/088",
    journal = "JCAP",
    volume = "04",
    pages = "088",
    year = "2024"
}

@article{Macedo:2024qky,
    author = "Macedo, Caio F. B. and Rosa, Jo\~ao Lu\'\i{}s and Rubiera-Garcia, Diego",
    title = "{Optical appearance of black holes surrounded by a dark matter halo}",
    eprint = "2402.13047",
    archivePrefix = "arXiv",
    primaryClass = "gr-qc",
    doi = "10.1088/1475-7516/2024/07/046",
    journal = "JCAP",
    volume = "07",
    pages = "046",
    year = "2024"
}

@article{Ravanal:2024odh,
    author = "Ravanal, Yuri and G\'omez, Gabriel and Cruz, Norman",
    title = "{Scalar field dark matter around charged black holes}",
    eprint = "2404.06774",
    archivePrefix = "arXiv",
    primaryClass = "astro-ph.CO",
    doi = "10.1103/PhysRevD.110.023027",
    journal = "Phys. Rev. D",
    volume = "110",
    number = "2",
    pages = "023027",
    year = "2024"
}

@article{Aurrekoetxea:2024cqd,
    author = "Aurrekoetxea, Josu C. and Marsden, James and Clough, Katy and Ferreira, Pedro G.",
    title = "{Self-interacting scalar dark matter around binary black holes}",
    eprint = "2409.01937",
    archivePrefix = "arXiv",
    primaryClass = "gr-qc",
    doi = "10.1103/PhysRevD.110.083011",
    journal = "Phys. Rev. D",
    volume = "110",
    number = "8",
    pages = "083011",
    year = "2024"
}

@article{PhysRevD.109.124056,
  title = {Gravitational radiation from hyperbolic encounters in the presence of dark matter},
  author = {Chowdhuri, Abhishek and Singh, Rishabh Kumar and Kangsabanik, Kaushik and Bhattacharyya, Arpan},
  journal = {Phys. Rev. D},
  volume = {109},
  issue = {12},
  pages = {124056},
  numpages = {22},
  year = {2024},
  month = {Jun},
  publisher = {American Physical Society},
  doi = {10.1103/PhysRevD.109.124056},
  url = {https://link.aps.org/doi/10.1103/PhysRevD.109.124056}
}

@article{Cheng:2024mgl,
    author = "Cheng, Ya-Ze and Cao, Yan and Tang, Yong",
    title = "{Effects of black hole environments on extreme mass-ratio hyperbolic encounters}",
    eprint = "2411.03095",
    archivePrefix = "arXiv",
    primaryClass = "gr-qc",
    month = "11",
    year = "2024"
}

@article{Rahman:2023sof,
    author = "Rahman, Mostafizur and Kumar, Shailesh and Bhattacharyya, Arpan",
    title = "{Probing astrophysical environment with eccentric extreme mass-ratio inspirals}",
    eprint = "2306.14971",
    archivePrefix = "arXiv",
    primaryClass = "gr-qc",
    doi = "10.1088/1475-7516/2024/01/035",
    journal = "JCAP",
    volume = "01",
    pages = "035",
    year = "2024"
}

@article{PhysRevLett.88.191301,
  title = {Dark Matter Spikes and Annihilation Radiation from the Galactic Center},
  author = {Merritt, David and Milosavljevi\ifmmode \acute{c}\else \'{c}\fi{}, Milos and Verde, Licia and Jimenez, Raul},
  journal = {Phys. Rev. Lett.},
  volume = {88},
  issue = {19},
  pages = {191301},
  numpages = {4},
  year = {2002},
  month = {Apr},
  publisher = {American Physical Society},
  doi = {10.1103/PhysRevLett.88.191301},
  url = {https://link.aps.org/doi/10.1103/PhysRevLett.88.191301}
}

@article{PhysRevD.64.043504,
  title = {Dark-matter spike at the galactic center?},
  author = {Ullio, Piero and Zhao, HongSheng and Kamionkowski, Marc},
  journal = {Phys. Rev. D},
  volume = {64},
  issue = {4},
  pages = {043504},
  numpages = {10},
  year = {2001},
  month = {Jul},
  publisher = {American Physical Society},
  doi = {10.1103/PhysRevD.64.043504},
  url = {https://link.aps.org/doi/10.1103/PhysRevD.64.043504}
}

@article{PhysRevD.78.083506,
  title = {Dark matter dynamics in the galactic center},
  author = {Vasiliev, Eugene and Zelnikov, Maxim},
  journal = {Phys. Rev. D},
  volume = {78},
  issue = {8},
  pages = {083506},
  numpages = {11},
  year = {2008},
  month = {Oct},
  publisher = {American Physical Society},
  doi = {10.1103/PhysRevD.78.083506},
  url = {https://link.aps.org/doi/10.1103/PhysRevD.78.083506}
}

@article{PhysRevLett.113.151302,
  title = {Galactic Center Gamma-Ray Excess from Dark Matter Annihilation: Is There a Black Hole Spike?},
  author = {Fields, Brian D. and Shapiro, Stuart L. and Shelton, Jessie},
  journal = {Phys. Rev. Lett.},
  volume = {113},
  issue = {15},
  pages = {151302},
  numpages = {5},
  year = {2014},
  month = {Oct},
  publisher = {American Physical Society},
  doi = {10.1103/PhysRevLett.113.151302},
  url = {https://link.aps.org/doi/10.1103/PhysRevLett.113.151302}
}

@article{PhysRevLett.115.231302,
  title = {Black Hole Window into $p$-Wave Dark Matter Annihilation},
  author = {Shelton, Jessie and Shapiro, Stuart L. and Fields, Brian D.},
  journal = {Phys. Rev. Lett.},
  volume = {115},
  issue = {23},
  pages = {231302},
  numpages = {5},
  year = {2015},
  month = {Dec},
  publisher = {American Physical Society},
  doi = {10.1103/PhysRevLett.115.231302},
  url = {https://link.aps.org/doi/10.1103/PhysRevLett.115.231302}
}

@article{PhysRevLett.95.011301,
  title = {Dark Minihalos with Intermediate Mass Black Holes},
  author = {Zhao, HongSheng and Silk, Joseph},
  journal = {Phys. Rev. Lett.},
  volume = {95},
  issue = {1},
  pages = {011301},
  numpages = {4},
  year = {2005},
  month = {Jun},
  publisher = {American Physical Society},
  doi = {10.1103/PhysRevLett.95.011301},
  url = {https://link.aps.org/doi/10.1103/PhysRevLett.95.011301}
}

@article{PhysRevD.72.103517,
  title = {New signature of dark matter annihilations: Gamma rays from intermediate-mass black holes},
  author = {Bertone, Gianfranco and Zentner, Andrew R. and Silk, Joseph},
  journal = {Phys. Rev. D},
  volume = {72},
  issue = {10},
  pages = {103517},
  numpages = {11},
  year = {2005},
  month = {Nov},
  publisher = {American Physical Society},
  doi = {10.1103/PhysRevD.72.103517},
  url = {https://link.aps.org/doi/10.1103/PhysRevD.72.103517}
}

@article{Eda:2013gg,
    author = "Eda, Kazunari and Itoh, Yousuke and Kuroyanagi, Sachiko and Silk, Joseph",
    title = "{New Probe of Dark-Matter Properties: Gravitational Waves from an Intermediate-Mass Black Hole Embedded in a Dark-Matter Minispike}",
    eprint = "1301.5971",
    archivePrefix = "arXiv",
    primaryClass = "gr-qc",
    doi = "10.1103/PhysRevLett.110.221101",
    journal = "Phys. Rev. Lett.",
    volume = "110",
    number = "22",
    pages = "221101",
    year = "2013"
}

@article{Eda:2014kra,
    author = "Eda, Kazunari and Itoh, Yousuke and Kuroyanagi, Sachiko and Silk, Joseph",
    title = "{Gravitational waves as a probe of dark matter minispikes}",
    eprint = "1408.3534",
    archivePrefix = "arXiv",
    primaryClass = "gr-qc",
    doi = "10.1103/PhysRevD.91.044045",
    journal = "Phys. Rev. D",
    volume = "91",
    number = "4",
    pages = "044045",
    year = "2015"
}

@article{PhysRevD.97.064003,
  title = {Gravitational waves with dark matter minispikes: The combined effect},
  author = {Yue, Xiao-Jun and Han, Wen-Biao},
  journal = {Phys. Rev. D},
  volume = {97},
  issue = {6},
  pages = {064003},
  numpages = {10},
  year = {2018},
  month = {Mar},
  publisher = {American Physical Society},
  doi = {10.1103/PhysRevD.97.064003},
  url = {https://link.aps.org/doi/10.1103/PhysRevD.97.064003}
}

@article{PhysRevD.102.103022,
  title = {Extreme dark matter tests with extreme mass ratio inspirals},
  author = {Hannuksela, Otto A. and Ng, Kenny C. Y. and Li, Tjonnie G. F.},
  journal = {Phys. Rev. D},
  volume = {102},
  issue = {10},
  pages = {103022},
  numpages = {9},
  year = {2020},
  month = {Nov},
  publisher = {American Physical Society},
  doi = {10.1103/PhysRevD.102.103022},
  url = {https://link.aps.org/doi/10.1103/PhysRevD.102.103022}
}

@ARTICLE{1943ApJ....97..255C,
       author = {{Chandrasekhar}, S.},
        title = "{Dynamical Friction. I. General Considerations: the Coefficient of Dynamical Friction.}",
      journal = {apj},
         year = 1943,
        month = mar,
       volume = {97},
        pages = {255},
          doi = {10.1086/144517},
       adsurl = {https://ui.adsabs.harvard.edu/abs/1943ApJ....97..255C}
}

@article{Zi:2025lio,
    author = "Zi, Tieguang and Kumar, Shailesh",
    title = "{Probing scalar field with generic extreme mass-ratio inspirals around Kerr black holes}",
    eprint = "2508.00516",
    archivePrefix = "arXiv",
    primaryClass = "gr-qc",
    month = "8",
    year = "2025"
}

@article{Ostriker_1999,
doi = {10.1086/306858},
url = {https://dx.doi.org/10.1086/306858},
year = {1999},
month = {mar},
publisher = {},
volume = {513},
number = {1},
pages = {252},
author = {Eve C. Ostriker},
title = {Dynamical Friction in a Gaseous Medium},
journal = {The Astrophysical Journal}
}

@article{Kim_2007,
doi = {10.1086/519302},
url = {https://dx.doi.org/10.1086/519302},
year = {2007},
month = {aug},
publisher = {},
volume = {665},
number = {1},
pages = {432},
author = {Hyosun Kim and Woong-Tae Kim},
title = {Dynamical Friction of a Circular-Orbit Perturber in a Gaseous Medium},
journal = {The Astrophysical Journal}
}

@article{Macedo:2013qea,
    author = "Macedo, Caio F. B. and Pani, Paolo and Cardoso, Vitor and Crispino, Lu\'\i{}s C. B.",
    title = "{Into the lair: gravitational-wave signatures of dark matter}",
    eprint = "1302.2646",
    archivePrefix = "arXiv",
    primaryClass = "gr-qc",
    doi = "10.1088/0004-637X/774/1/48",
    journal = "Astrophys. J.",
    volume = "774",
    pages = "48",
    year = "2013"
}

@article{10.1093/mnras/104.5.273,
    author = {Bondi, H. and Hoyle, F.},
    title = {On the Mechanism of Accretion by Stars},
    journal = {Monthly Notices of the Royal Astronomical Society},
    volume = {104},
    number = {5},
    pages = {273-282},
    year = {1944},
    month = {10},
    issn = {0035-8711},
    doi = {10.1093/mnras/104.5.273},
    url = {https://doi.org/10.1093/mnras/104.5.273},
    eprint = {https://academic.oup.com/mnras/article-pdf/104/5/273/8072203/mnras104-0273.pdf},
}

@BOOK{1983bhwd.book.....S,
       author = {{Shapiro}, Stuart L. and {Teukolsky}, Saul A.},
        title = "{Black holes, white dwarfs and neutron stars. The physics of compact objects}",
         year = 1983,
          doi = {10.1002/9783527617661},
       adsurl = {https://ui.adsabs.harvard.edu/abs/1983bhwd.book.....S},
      publisher = {A Wiley-Interscience Publication, New York: Wiley, 1983}
}

@article{PhysRevLett.110.221101,
  title = {New Probe of Dark-Matter Properties: Gravitational Waves from an Intermediate-Mass Black Hole Embedded in a Dark-Matter Minispike},
  author = {Eda, Kazunari and Itoh, Yousuke and Kuroyanagi, Sachiko and Silk, Joseph},
  journal = {Phys. Rev. Lett.},
  volume = {110},
  issue = {22},
  pages = {221101},
  numpages = {5},
  year = {2013},
  month = {May},
  publisher = {American Physical Society},
  doi = {10.1103/PhysRevLett.110.221101},
  url = {https://link.aps.org/doi/10.1103/PhysRevLett.110.221101}
}

@article{Yue:2018vtk,
    author = "Yue, Xiao-Jun and Han, Wen-Biao and Chen, Xian",
    title = "{Dark matter: an efficient catalyst for intermediate-mass-ratio-inspiral events}",
    eprint = "1802.03739",
    archivePrefix = "arXiv",
    primaryClass = "gr-qc",
    doi = "10.3847/1538-4357/ab06f6",
    journal = "Astrophys. J.",
    volume = "874",
    number = "1",
    pages = "34",
    year = "2019"
}

@article{Dai:2021olt,
    author = "Dai, Ning and Gong, Yungui and Jiang, Tong and Liang, Dicong",
    title = "{Intermediate mass-ratio inspirals with dark matter minispikes}",
    eprint = "2111.13514",
    archivePrefix = "arXiv",
    primaryClass = "gr-qc",
    doi = "10.1103/PhysRevD.106.064003",
    journal = "Phys. Rev. D",
    volume = "106",
    number = "6",
    pages = "064003",
    year = "2022"
}

@article{Nichols:2023ufs,
    author = "Nichols, David A. and Wade, Benjamin A. and Grant, Alexander M.",
    title = "{Secondary accretion of dark matter in intermediate mass-ratio inspirals: Dark-matter dynamics and gravitational-wave phase}",
    eprint = "2309.06498",
    archivePrefix = "arXiv",
    primaryClass = "gr-qc",
    doi = "10.1103/PhysRevD.108.124062",
    journal = "Phys. Rev. D",
    volume = "108",
    number = "12",
    pages = "124062",
    year = "2023"
}

@article{Shadykul:2024ehz,
    author = "Shadykul, Darkhan and Chakrabarty, Hrishikesh and Malafarina, Daniele",
    title = "{Intermediate mass ratio inspirals in dark matter halos}",
    eprint = "2410.18657",
    archivePrefix = "arXiv",
    primaryClass = "gr-qc",
    month = "10",
    year = "2024"
}

@article{PhysRevD.105.043009,
  title = {Measuring the dark matter environments of black hole binaries with gravitational waves},
  author = {Coogan, Adam and Bertone, Gianfranco and Gaggero, Daniele and Kavanagh, Bradley J. and Nichols, David A.},
  journal = {Phys. Rev. D},
  volume = {105},
  issue = {4},
  pages = {043009},
  numpages = {22},
  year = {2022},
  month = {Feb},
  publisher = {American Physical Society},
  doi = {10.1103/PhysRevD.105.043009},
  url = {https://link.aps.org/doi/10.1103/PhysRevD.105.043009}
}

@misc{karydas2024sharpeningdarkmattersignature,
      title={Sharpening the dark matter signature in gravitational waveforms I: Accretion and eccentricity evolution}, 
      author={Theophanes K. Karydas and Bradley J. Kavanagh and Gianfranco Bertone},
      year={2024},
      eprint={2402.13053},
      archivePrefix={arXiv},
      primaryClass={gr-qc},
      url={https://arxiv.org/abs/2402.13053}, 
}

@book{Poisson_Will_2014, place={Cambridge}, title={Gravity: Newtonian, Post-Newtonian, Relativistic}, publisher={Cambridge University Press}, author={Poisson, Eric and Will, Clifford M.}, year={2014}}

@misc{our_github_repo_nonlinear_memo,
  author       = {Rishabh Kumar Singh},
  title        = {Github-nonlinear-memory: Nonlinear-GW-memory-in-astrophysical-environment},
  year         = {2025},
  howpublished = {\url{https://github.com/rks-circle/Nonlinear-GW-memory-in-astrophysical-environment}},
  note         = {Accessed: August 8, 2025}
}

@article{Usseglio:2025iwt,
    author = "Usseglio, Davide and Trestini, David and Chowdhuri, Abhishek",
    title = "{Quasi-Keplerian parametrization for compact binaries on hyperbolic orbits in scalar-tensor theories at second post-Newtonian order}",
    eprint = "2504.13829",
    archivePrefix = "arXiv",
    primaryClass = "gr-qc",
    doi = "10.1103/mfg1-t3l3",
    month = "4",
    year = "2025"
}

@article{Cho:2021onr,
    author = "Cho, Gihyuk and Dandapat, Subhajit and Gopakumar, Achamveedu",
    title = "{Third order post-Newtonian gravitational radiation from two-body scattering: Instantaneous energy and angular momentum radiation}",
    eprint = "2111.00818",
    archivePrefix = "arXiv",
    primaryClass = "gr-qc",
    doi = "10.1103/PhysRevD.105.084018",
    journal = "Phys. Rev. D",
    volume = "105",
    number = "8",
    pages = "084018",
    year = "2022"
}

@article{PhysRevD.111.083041,
  title = {Probing dark matter halo profiles with multiband observations of gravitational waves},
  author = {Tahelyani, Divya and Bhattacharyya, Arpan and Sengupta, Anand S.},
  journal = {Phys. Rev. D},
  volume = {111},
  issue = {8},
  pages = {083041},
  numpages = {14},
  year = {2025},
  month = {Apr},
  publisher = {American Physical Society},
  doi = {10.1103/PhysRevD.111.083041},
  url = {https://link.aps.org/doi/10.1103/PhysRevD.111.083041}
}

@article{PhysRevD.84.064023,
  title = {Importance of including small body spin effects in the modelling of extreme and intermediate mass-ratio inspirals},
  author = {Huerta, E. A. and Gair, Jonathan R.},
  journal = {Phys. Rev. D},
  volume = {84},
  issue = {6},
  pages = {064023},
  numpages = {21},
  year = {2011},
  month = {Sep},
  publisher = {American Physical Society},
  doi = {10.1103/PhysRevD.84.064023},
  url = {https://link.aps.org/doi/10.1103/PhysRevD.84.064023}
}

@article{PhysRevD.66.044002,
  title = {Zoom and whirl: Eccentric equatorial orbits around spinning black holes and their evolution under gravitational radiation reaction},
  author = {Glampedakis, Kostas and Kennefick, Daniel},
  journal = {Phys. Rev. D},
  volume = {66},
  issue = {4},
  pages = {044002},
  numpages = {33},
  year = {2002},
  month = {Aug},
  publisher = {American Physical Society},
  doi = {10.1103/PhysRevD.66.044002},
  url = {https://link.aps.org/doi/10.1103/PhysRevD.66.044002}
}

@article{Kumar_2024,
doi = {10.1088/1475-7516/2024/10/047},
url = {https://dx.doi.org/10.1088/1475-7516/2024/10/047},
year = {2024},
month = {oct},
publisher = {IOP Publishing},
volume = {2024},
number = {10},
pages = {047},
author = {Kumar, Shailesh and Singh, Rishabh Kumar and Chowdhuri, Abhishek and Bhattacharyya, Arpan},
title = {Exploring waveforms with non-GR deviations for extreme mass-ratio inspirals},
journal = {Journal of Cosmology and Astroparticle Physics}
}

@article{PhysRevD.100.024009,
  title = {Observational signature of the logarithmic terms in the soft-graviton theorem},
  author = {Laddha, Alok and Sen, Ashoke},
  journal = {Phys. Rev. D},
  volume = {100},
  issue = {2},
  pages = {024009},
  numpages = {5},
  year = {2019},
  month = {Jul},
  publisher = {American Physical Society},
  doi = {10.1103/PhysRevD.100.024009},
  url = {https://link.aps.org/doi/10.1103/PhysRevD.100.024009}
}

@article{PhysRevD.46.4304,
  title = {Hereditary effects in gravitational radiation},
  author = {Blanchet, Luc and Damour, Thibault},
  journal = {Phys. Rev. D},
  volume = {46},
  issue = {10},
  pages = {4304--4319},
  numpages = {0},
  year = {1992},
  month = {Nov},
  publisher = {American Physical Society},
  doi = {10.1103/PhysRevD.46.4304},
  url = {https://link.aps.org/doi/10.1103/PhysRevD.46.4304}
}

@article{Blanchet:1997jj,
    author = "Blanchet, Luc",
    title = "{Gravitational wave tails of tails}",
    eprint = "gr-qc/9710038",
    archivePrefix = "arXiv",
    doi = "10.1088/0264-9381/15/1/009",
    journal = "Class. Quant. Grav.",
    volume = "15",
    pages = "113--141",
    year = "1998",
    note = "[Erratum: Class.Quant.Grav. 22, 3381 (2005)]"
}

@article{Zhang:2017jma,
    author = "Zhang, P. -M. and Duval, C. and Horvathy, P. A.",
    title = "{Memory Effect for Impulsive Gravitational Waves}",
    eprint = "1709.02299",
    archivePrefix = "arXiv",
    primaryClass = "gr-qc",
    doi = "10.1088/1361-6382/aaa987",
    journal = "Class. Quant. Grav.",
    volume = "35",
    number = "6",
    pages = "065011",
    year = "2018"
}

@article{Blanchet_2017,
doi = {10.1088/1361-6382/aa79d7},
url = {https://dx.doi.org/10.1088/1361-6382/aa79d7},
year = {2017},
month = {jul},
publisher = {IOP Publishing},
volume = {34},
number = {16},
pages = {164001},
author = {Blanchet, Luc and Le Tiec, Alexandre},
title = {First law of compact binary mechanics with gravitational-wave tails},
journal = {Classical and Quantum Gravity}
}

@article{PhysRev.166.1272,
  title = {Gravitational Radiation in the Limit of High Frequency. II. Nonlinear Terms and the Effective Stress Tensor},
  author = {Isaacson, Richard A.},
  journal = {Phys. Rev.},
  volume = {166},
  issue = {5},
  pages = {1272--1280},
  numpages = {0},
  year = {1968},
  month = {Feb},
  publisher = {American Physical Society},
  doi = {10.1103/PhysRev.166.1272},
  url = {https://link.aps.org/doi/10.1103/PhysRev.166.1272}
}

@article{Ivanov:2025ozg,
    author = "Ivanov, Mikhail M. and Li, Yue-Zhou and Parra-Martinez, Julio and Zhou, Zihan",
    title = "{Resummation of Universal Tails in Gravitational Waveforms}",
    eprint = "2504.07862",
    archivePrefix = "arXiv",
    primaryClass = "hep-th",
    reportNumber = "MIT-CTP/5863",
    month = "4",
    year = "2025"
}

@article{3jhf-vdjz,
  title = {High-post-Newtonian-order dynamical effects induced by tail-of-tail interactions in a two-body system},
  author = {Bini, Donato and Damour, Thibault and Geralico, Andrea},
  journal = {Phys. Rev. D},
  volume = {112},
  issue = {4},
  pages = {044028},
  numpages = {13},
  year = {2025},
  month = {Aug},
  publisher = {American Physical Society},
  doi = {10.1103/3jhf-vdjz},
  url = {https://link.aps.org/doi/10.1103/3jhf-vdjz}
}

@article{Bini:2025vuk,
    author = "Bini, Donato and Damour, Thibault",
    title = "{High precision black hole scattering: Tutti frutti vs worldline effective field theory}",
    eprint = "2504.20204",
    archivePrefix = "arXiv",
    primaryClass = "hep-th",
    doi = "10.1103/8ks7-2blq",
    journal = "Phys. Rev. D",
    volume = "112",
    number = "4",
    pages = "044002",
    year = "2025"
}

@article{Bhattacharyya:2024aeq,
    author = "Bhattacharyya, Arpan and Ghosh, Debodirna and Ghosh, Saptaswa and Pal, Sounak",
    title = "{Observables from classical black hole scattering in Scalar-Tensor theory of gravity from worldline quantum field theory}",
    eprint = "2401.05492",
    archivePrefix = "arXiv",
    primaryClass = "hep-th",
    doi = "10.1007/JHEP04(2024)015",
    journal = "JHEP",
    volume = "04",
    pages = "015",
    year = "2024"
}

@article{Bhattacharyya:2024kxj,
    author = "Bhattacharyya, Arpan and Ghosh, Debodirna and Ghosh, Saptaswa and Pal, Sounak",
    title = "{Bootstrapping the spinning two body problem in dynamical Chern-Simons gravity using worldline QFT}",
    eprint = "2407.07195",
    archivePrefix = "arXiv",
    primaryClass = "hep-th",
    doi = "10.1007/JHEP04(2025)175",
    journal = "JHEP",
    volume = "04",
    pages = "175",
    year = "2025"
}

@article{Miller:2025yyx,
    author = "Miller, Andrew L.",
    title = "{Gravitational wave probes of particle dark matter: a review}",
    eprint = "2503.02607",
    archivePrefix = "arXiv",
    primaryClass = "astro-ph.HE",
    month = "3",
    year = "2025"
}

@article{PhysRevD.89.104059,
  title = {Can environmental effects spoil precision gravitational-wave astrophysics?},
  author = {Barausse, Enrico and Cardoso, Vitor and Pani, Paolo},
  journal = {Phys. Rev. D},
  volume = {89},
  issue = {10},
  pages = {104059},
  numpages = {60},
  year = {2014},
  month = {May},
  publisher = {American Physical Society},
  doi = {10.1103/PhysRevD.89.104059},
  url = {https://link.aps.org/doi/10.1103/PhysRevD.89.104059}
}

@article{PhysRevD.105.024072,
  title = {Gravitational memory and compact extra dimensions},
  author = {Ferko, Christian and Satishchandran, Gautam and Sethi, Savdeep},
  journal = {Phys. Rev. D},
  volume = {105},
  issue = {2},
  pages = {024072},
  numpages = {38},
  year = {2022},
  month = {Jan},
  publisher = {American Physical Society},
  doi = {10.1103/PhysRevD.105.024072},
  url = {https://link.aps.org/doi/10.1103/PhysRevD.105.024072}
}

@article{DeLuca:2024cjl,
    author = "De Luca, Valerio and Khoury, Justin and Wong, Sam S. C.",
    title = "{Gravitational memory and soft theorems: The local perspective}",
    eprint = "2412.01910",
    archivePrefix = "arXiv",
    primaryClass = "gr-qc",
    doi = "10.1103/gbg1-mz49",
    journal = "Phys. Rev. D",
    volume = "112",
    number = "2",
    pages = "L021502",
    year = "2025"
}

@article{PhysRevD.106.064022,
  title = {Gravitational wave memory and its tail in cosmology},
  author = {Jokela, Niko and Kajantie, K. and Sarkkinen, Miika},
  journal = {Phys. Rev. D},
  volume = {106},
  issue = {6},
  pages = {064022},
  numpages = {30},
  year = {2022},
  month = {Sep},
  publisher = {American Physical Society},
  doi = {10.1103/PhysRevD.106.064022},
  url = {https://link.aps.org/doi/10.1103/PhysRevD.106.064022}
}

@article{Hait:2022ukn,
    author = "Hait, Arpan and Mohanty, Subhendra and Prakash, Suraj",
    title = "{Frequency space derivation of linear and nonlinear memory gravitational wave signals from eccentric binary orbits}",
    eprint = "2211.13120",
    archivePrefix = "arXiv",
    primaryClass = "gr-qc",
    doi = "10.1103/PhysRevD.109.084037",
    journal = "Phys. Rev. D",
    volume = "109",
    number = "8",
    pages = "084037",
    year = "2024"
}

@article{10.1111/j.1365-2966.2007.12408.x,
    author = {Barausse, E.},
    title = {Relativistic dynamical friction in a collisional fluid},
    journal = {Monthly Notices of the Royal Astronomical Society},
    volume = {382},
    number = {2},
    pages = {826-834},
    year = {2007},
    month = {12},
    issn = {0035-8711},
    doi = {10.1111/j.1365-2966.2007.12408.x},
    url = {https://doi.org/10.1111/j.1365-2966.2007.12408.x},
    eprint = {https://academic.oup.com/mnras/article-pdf/382/2/826/3434565/mnras0382-0826.pdf},
}

@article{PhysRevD.78.124020,
  title = {Model waveform accuracy standards for gravitational wave data analysis},
  author = {Lindblom, Lee and Owen, Benjamin J. and Brown, Duncan A.},
  journal = {Phys. Rev. D},
  volume = {78},
  issue = {12},
  pages = {124020},
  numpages = {12},
  year = {2008},
  month = {Dec},
  publisher = {American Physical Society},
  doi = {10.1103/PhysRevD.78.124020},
  url = {https://link.aps.org/doi/10.1103/PhysRevD.78.124020}
}

\end{document}